\documentclass[11pt,a4paper]{article}
\usepackage{cite}
\usepackage{graphicx}
\usepackage{tikz}
\usetikzlibrary{hobby,calc,shapes, positioning, backgrounds,trees, decorations, decorations.markings, decorations.pathmorphing, arrows, shapes.geometric, snakes}
\usepackage{tikz-feynman}
\usepackage{amssymb}
\usepackage{amsmath}
\usepackage{amsfonts}
\usepackage{dsfont}
\usepackage{mathtools}
\usepackage{array}
\usepackage{rotating}
\usepackage{bbold,amsfonts}
\allowdisplaybreaks

\usepackage[utf8]{inputenc}
\usepackage{bm}
\usepackage{xcolor}
\usepackage{float}
\usepackage{braket}
\usepackage{placeins}
\usepackage[height=8.8in,width=6.45in]{geometry}
\usepackage[font=small,labelfont=bf]{caption}
\usepackage[hidelinks]{hyperref}
\usepackage{booktabs,float,slashed}
\usepackage{standalone}
\usepackage{caption}
\usepackage{subcaption}
\usepackage{makecell}
\usetikzlibrary{decorations.pathreplacing}
\numberwithin{equation}{section}
\newcommand{\vev}[1]{{\left\langle #1 \right\rangle}}

\newcommand{\beq}{\begin{equation}}
\newcommand{\eeq}{\end{equation}}

\DeclareMathOperator{\Tr}{Tr}
\DeclareMathOperator{\tr}{tr}

\newcommand{\ii}{\mathrm{i}}

\makeatletter
\newcommand*{\letterdef@}{}
\newcommand*{\letterdef}[3]{%
	\def\letterdef@##1{\expandafter\newcommand\csname #1\endcsname{#2{##1}}}%
	\@tfor\@tempa :=#3\do{\expandafter\letterdef@\expandafter{\@tempa}}}
\makeatother
\letterdef{c#1} {\mathcal}{ABCDEFGHIJKLMNOPQRSTUVWXYZ} % \cX = \mathcal{X}
\letterdef{rm#1}{\mathrm} {dDeimM} % \rmX = \mathrm{X} for X in {dDmM}

\newcommand{\gym}{g_{_{\rm YM}}}

\def\ie{\begin{equation}\begin{aligned}}
\def\fe{\end{aligned}\end{equation}}

\DeclareMathAlphabet{\mathbb}{U}{msb}{m}{n} %math fonts like R^4 look much better

\DeclareMathAlphabet{\mathbb}{U}{msb}{m}{n} %math fonts like R^4 look much better

\begin{document}

\begin{titlepage}

\begin{flushright}
%\scriptsize 
{\small QMUL-PH-24-12}
\end{flushright}

\vspace*{10mm}
\begin{center}
{\LARGE \bf 
All-loop heavy-heavy-light-light correlators  \\
\vspace*{1.6mm}  in $\mathcal{N}=4$ super Yang--Mills theory}

\vspace*{15mm}

{\Large Augustus Brown, Francesco Galvagno, and Congkao Wen}

\vspace*{8mm}
	
Centre for Theoretical Physics, Department of Physics and Astronomy, \\
Queen Mary University of London, London, E1 4NS, UK
			\vskip 0.3cm
			
	{\small
		E-mail:
		\texttt{a.a.x.brown,f.galvagno,c.wen@qmul.ac.uk}
	}
\vspace*{0.8cm}
\end{center}

\begin{abstract}

\vspace*{0.3cm}

We study Heavy-Heavy-Light-Light (HHLL) correlators $\langle \mathcal{H} \mathcal{H} \mathcal{O}_2 \mathcal{O}_2 \rangle$ in $\mathcal{N}=4$ super Yang-Mills theory with $SU(N)$ gauge group at generic $N$. The light operator $\mathcal{O}_2$ is the dimension two superconformal primary in the stress tensor multiplet and $\mathcal{H}$ is a general half-BPS superconformal primary operator with dimension (or $R$-charge) $\Delta_{\mathcal{H}}$. We consider the large-charge 't Hooft limit, where $\Delta_{\mathcal{H}} \rightarrow \infty$ with fixed 't Hooft-like coupling $\lambda:=\Delta_{\mathcal{H}}\, g_{_{\rm YM}}^2$. We show that the $L$-loop contribution to the HHLL correlators in the leading large-charge limit is universal for any choice of the heavy operator $\mathcal{H}$, given as $\lambda^L \sum_{\ell=0}^{L} \Phi^{(\ell)} \Phi^{(L-\ell)}$ with an $SU(N)$ colour factor coefficient, where $\Phi^{(\ell)}$ is the ladder Feynman integral, which is known to all loops. The dependence on the explicit form of the heavy operator lies only in the colour factor coefficients. We determine such colour factors for several classes of heavy operators, and show that the large charge limit leads to minimal powers of $N$. For 
the special class of ``canonical heavy operators", 
one can even resum the all-loop ladder integrals and determine the correlators at finite $\lambda$. Furthermore, upon integrating over the spacetime dependence the resulting integrated HHLL correlators agree with the existing results derived from supersymmetric localisation. Finally, as an application of the all-loop analytic results, we derive exact expressions for the structure constants of two heavy operators and the Konishi operator, finding intriguing connections with the integrated HHLL correlators. 

\end{abstract}
\end{titlepage}
\setcounter{tocdepth}{2}
\tableofcontents

\section{Introduction}
\label{sec:intro}

The maximally supersymmetric $\cN=4$ super Yang-Mills theory (SYM) \cite{Brink:1976bc}   
has played a primary role in the context of Quantum Field Theory in four dimensions, thanks to the possibility of determining large sets of observables in a fully analytic way. Of particular significance for this paper is the four-point correlator of half-BPS superconformal primaries, which have been extensively studied in both the strong \cite{Goncalves:2014ffa,Rastelli:2016nze,Alday:2017xua,Caron-Huot:2018kta,Alday:2019nin,Drummond:2020dwr,Abl:2020dbx,Aprile:2020mus,Huang:2021xws, Drummond:2022dxw, Alday:2023mvu} and the weak coupling regime 
\cite{Eden:2011we, Eden:2012tu,Drummond:2013nda, Bourjaily:2016evz, Caron-Huot:2021usw}. 
At the perturbative level, which will be directly relevant for this paper, remarkable progress has been made in the well-studied large-$N$ limit \cite{tHooft:1973alw}, which selects only a specific class of conformal Feynman integrals. Using extremely powerful methods developed in \cite{Eden:2011we, Eden:2012tu}, the loop integrands for these four-point correlators in $\cN=4$ SYM have been systematically determined to ten loops \cite{Bourjaily:2015bpz, Bourjaily:2016evz} (see \cite{Chicherin:2015edu, Caron-Huot:2021usw} for the correlators of half-BPS operators with generic charges and \cite{Fleury:2019ydf} for beyond the planar limit at four loops), although evaluating the Feynman integrals may still be a highly challenging task \cite{Drummond:2013nda}. Further discussions on the recent developments about four-point correlators in $\cN=4$ SYM may be found in a nice review \cite{Heslop:2022xgp}, which also contains additional references.

In achieving these impressive results, the 't Hooft limit (\textit{i.e.} a double scaling limit where $N\to\infty$ while keeping $\gym^2 N$ fixed, where $\gym$ is the Yang-Mills coupling) has played a crucial role in simplifying the perturbative analysis at weak coupling. It is natural to investigate other possible parameters in the theory that may also simplify the perturbative expansions and lead to analytic results at high loop orders. In the past few years, the ideal candidate has been found in the context of Conformal Field Theories (CFT) with a large global charge, where the properties of large-charge operators have been initially studied using Effective Field Theory (EFT) techniques \cite{Hellerman:2015nra,Alvarez-Gaume:2016vff, Monin:2016jmo, Banerjee:2017fcx} (see the review \cite{Gaume:2020bmp} for additional references). In order to obtain information about the microscopic details of the large charge expansion, it is possible to combine the large charge parameter with a weak coupling expansion, by realising a 't Hooft-like double scaling limit of the type: $Q \to\infty$ with $g^{\#} Q$ fixed, where $Q$ generically denotes the charge, and $g$ is the parameter of the perturbative expansion. This idea was originally derived in \cite{Bourget:2018obm} in the context of extremal correlators in $\cN=2$ superconformal theories in the presence of BPS operators with large conformal dimensions (or equivalently, the charges of R-symmetry). In this setup the perturbative expansion could be organised in terms of the large-charge double scaling limit, leading to several results for extremal correlators using supersymmetric localisation techniques \cite{Beccaria:2018xxl,Grassi:2019txd,Beccaria:2020azj}, which for rank-1 theories matched the expectations from large-charge EFTs \cite{Hellerman:2017sur,Hellerman:2018xpi,Hellerman:2021yqz,Hellerman:2021duh}. The idea of organising the perturbative series in terms of a large charge 't Hooft-like coupling has then been further exploited for non-SUSY scalar QFTs \cite{Arias-Tamargo:2019kfr,Arias-Tamargo:2019xld,Badel:2019khk}, combining this approach with a semiclassical analysis \cite{Badel:2019oxl,Giombi:2020enj,Giombi:2022gjj}.
 Therefore it is natural to investigate the effect of the large charge 't Hooft limit on the perturbative series for the simplest four-dimensional gauge theory $\cN=4$ SYM, in comparison with the standard large-$N$ 't Hooft limit.

In $\cN=4$ SYM, this idea was first applied to a class of Heavy-Heavy-Light-Light (HHLL) correlators with $SU(N)$ gauge group in the context of the integrated correlators\cite{Paul:2023rka,Brown:2023why} (see also \cite{Caetano:2023zwe}). The four-point correlation functions of the form $\langle \cH \cH \mathcal{O}_2 \mathcal{O}_2 \rangle$ were considered, where $\mathcal{O}_2$ is the half-BPS superconformal primary operator with dimension two in the stress tensor multiplet and the heavy operators $\cH$ are certain half-BPS superconformal primary operators with a large conformal dimension (or, equivalently, R-symmetry charge). 
The integrated version of the HHLL correlators, where the spacetime dependence of the correlators is integrated over a certain supersymmetric invariant integration measure \cite{Binder:2019jwn}, can be computed by supersymmetric localisation \cite{Pestun:2007rz}. For certain choices of heavy operators, the 
integrated HHLL correlators were studied in \cite{Paul:2023rka,Brown:2023why} in the large-charge 't Hooft expansion. It was even possible to study the large-charge expansions with a finite (complexified) Yang-Mills coupling $\tau$ and investigate the non-perturbative effects and their modular properties. It was found that the results take a remarkably similar form to the fixed-$\tau$ large-$N$ expansions of integrated correlators of four operators with fixed charges \cite{Chester:2019jas, Dorigoni:2021bvj, Dorigoni:2021guq,Dorigoni:2022cua}.

Understanding the (unintegrated) HHLL correlators in $\cN=4$ SYM is of course much more challenging, since localisation techniques cannot directly apply. The HHLL correlator for the special case of the $SU(2)$ gauge group was considered in \cite{Caetano:2023zwe}.
The choice of the heavy operator $\cH$ is fixed by the $SU(2)$ gauge group to be the multitrace operator $(\cO_2)^p$ with $p\rightarrow \infty$. In the large-charge 't Hooft limit, (\textit{i.e.} $p \rightarrow \infty$ and $\lambda=2p g^2_{_{\rm YM}}$ being fixed), the HHLL correlator may be interpreted as a two-point function of the light operators in a heavy background generated by the large-charge operators. Following the results from \cite{Giombi:2020enj}, such an interpretation was employed in \cite{Caetano:2023zwe} to express the $SU(2)$ HHLL correlator in the leading large-charge 't Hooft limit in a closed form, arising as a resummation of conformal ladder integrals \cite{Broadhurst:2010ds}.

These results for the HHLL correlator with $SU(2)$ gauge group and for the $SU(N)$ integrated HHLL correlators indicate that the large-charge expansion leads to remarkable simplifications of $\cN=4$ SYM in this limit. In this paper we study the HHLL correlators $\langle \cH \cH \mathcal{O}_2 \mathcal{O}_2 \rangle$ in $\mathcal{N}=4$ SYM with a generic $SU(N)$ gauge group, where $\cH$ can be any half-BPS superconformal primary operators with a large conformal dimension or charge $\Delta_{\cH}$.
We follow a perturbative approach in the weak Yang-Mills coupling regime, considering the large-charge 't Hooft double scaling limit:
\begin{equation}\label{eq:doublescaling_def}
    \Delta_\cH\to\infty\, , ~~~\rm with~~~ \lambda=\Delta_\cH \,\gym^2~~~ {\rm fixed}~,
\end{equation}
which implies that the dimension $\Delta_\cH$ is the largest parameter in the game. Hence, all our results are valid for $\Delta_{\cH} \gg N^2$; the examples we will study are the cases where $N$ is fixed to be a finite value.  We study the Feynman integrals at each perturbative order, employing the chiral Lagrangian insertion method for constructing loop integrands \cite{Eden:2011we, Eden:2012tu} as well as supersymmetric non-renormalisation theorems \cite{Baggio:2012rr}, which impose strong constraints on the possible Feynman diagrams. Exploiting some large-charge combinatorial arguments on the chiral Lagrangian insertions, we are able to identify the class of Feynman integrals contributing at the leading order in the large-charge limit. In full generality we show that the dynamical part $\cT_\cH(u, v)$ of the $\langle \cH \cH \mathcal{O}_2 \mathcal{O}_2 \rangle$ correlators to all loops can be expressed as 
\ie \label{eq:SUN-intro}
\cT_\cH(u, v) =  \sum_{L=1}^{\infty} d_{\cH, N; L} \,  \frac{(-a)^L}{ u }\sum_{\ell=0}^{L}   \Phi^{(\ell)}(u, v)  \Phi^{(L-\ell)}(u, v) \, ,
\fe
where $u, v$ are conformal cross ratios, $a= \lambda/(4\pi^2)$ with $\lambda$ defined in \eqref{eq:doublescaling_def}, and $\Phi^{(\ell)}(u, v) $ is the $\ell$-loop ladder Feynman integral (and we set $\Phi^{(0)}(u, v)=1$). The analytic expression for $\Phi^{(\ell)}(u,v)$ is known to all loop orders \cite{Usyukina:1993ch} and is given in \eqref{eq:ladderL}. We remark that the spacetime dependent part of \eqref{eq:SUN-intro} is universal for any choice of the heavy operators $\cH$. 

The coefficients $d_{\cH, N; L}$ depend on the precise form of $\cH$ as well as the $SU(N)$ gauge group. These coefficients are essentially the $SU(N)$ colour factors associated with the corresponding Feynman diagrams that are leading in the large charge expansion. We determine them for concrete examples of heavy operators, showing that the large-charge combinatorial argument is associated with the chiral Lagrangian insertions that minimise the power of $N$. Hence at each loop order, the large-charge limit selects the class of Feynman diagrams inserted in the maximally non-planar way in the colour space. The diagrammatic results for $d_{\cH, N; L}$ match precisely with the expectations from the integrated HHLL correaltors computed using supersymmetric localisation \cite{Paul:2023rka,Brown:2023why}. This interpretation of the large-charge 't Hooft limit at the level of Feynman diagrams confirms the findings of \cite{Beccaria:2020azj}, where it has been shown that the analogous class of maximally non-planar Feynman diagrams captures the leading large charge limit of chiral/anti-chiral two-point correlators in $\cN=2$ SCFTs.

When specifying fixed values of $N$, we identify a particular class of heavy operators, denoted here as ``canonical heavy operators", for which the colour factor coefficients $d_{\cH, N; L}$ are proportional to $c^L$ for some constant $c$ for any loop order $L$. It is straightforward to see that for $d_{\cH, N; L} \sim c^L$, the HHLL correlators $\cT_{\cH}(u, v)$ given in \eqref{eq:SUN-intro} can be recast as 
\ie \label{eq:THUV-fin-spe2}
\cT_{\cH}(u, v) =   {\beta \over u } \left[ \Bigg(   \sum_{\ell=0}^{\infty} (-c\,a)^\ell  \, \Phi^{(\ell)}(u, v) \Bigg)^2 -1 \right] \, , 
\fe
where $\beta$ is some overall constant. Remarkably, the all-loop ladder Feynman integrals can be resummed \cite{Broadhurst:2010ds}. Therefore, for all the ``canonical heavy operators", we have access of the HHLL correlators even in the finite coupling regime. We denote this class of operators as ``canonical'', since it suggests the canonical direction of taking the large charge limit, where the HHLL correlators simplify drastically. 
The case of the $SU(2)$ gauge group was studied in \cite{Caetano:2023zwe}, where the only possible half-BPS heavy operator is $(\cO_2)^p$, for which $d_{\cH, 2; L} =2^{2-L}$. Therefore it represents the simplest example of the ``canonical heavy operators". However, for higher rank gauge groups, $(\cO_2)^p$ is not canonical anymore, as we show in \eqref{eq:coffO2p}. Hence the problem of finding the ``canonical'' large charge direction for any rank $N$ seems a very non-trivial problem. In this paper we make some progress, finding the canonical operators for the $SU(3)$ gauge group, see \eqref{eq:O03p} and the discussion afterwards.
%As we mentioned the results in \cite{Caetano:2023zwe} were obtained using large-charge EFT. Our analysis provides a direct diagrammatic understanding for the $SU(2)$ result of \cite{Caetano:2023zwe} and extends it to the most general HHLL correlators when considering gauge groups with higher ranks and generic forms of the heavy operators. 

Hence the ``canonical heavy operators" are a very special class of heavy operators. We also consider more generic heavy operators, for which the HHLL correlators cannot be expressed in the form of \eqref{eq:THUV-fin-spe2}. To be concrete, we consider a class of generic heavy operators, including those studied in \cite{Paul:2023rka, Brown:2023why} in the context of integrated correlators. The existing known results for the integrated HHLL correlators \cite{Paul:2023rka, Brown:2023why} also allow us to further verify the all-loop expression; indeed we find that the integrated version of $\cT_{\cH}(u, v)$ given in \eqref{eq:SUN-intro} precisely matches with the results in \cite{Paul:2023rka, Brown:2023why}, where the integrated correlators were computed using supersymmetric localisation. 

Finally, we analyse the OPE limits of our results. From the HHLL correlators, we can straightforwardly read off the relevant CFT data. In particular, the HHLL correlators predict the structure constants $C_{\cH \cH \cK}$ of two heavy operators and the Konishi operator to all loops. These all-loop expressions can in fact be resummed, and lead to exact results for the structure constants, which can be further expanded in the large-coupling regime. Intriguingly, we find that the all-loop expressions of the structure constants in the large-charge 't Hooft limit take a very similar form to the expressions of the corresponding integrated HHLL correlators \cite{Paul:2023rka, Brown:2023why}, and we give precise relations between $C_{\cH \cH \cK}$ and the integrated correlators. 

The rest of the paper is organised as follows. In section \ref{sec:HHLL} we review some general properties of the four-point correlators of half-BPS superconformal primary operators in $\cN=4$ SYM. Some of the main results of the paper are contained in section \ref{sec:integrand}, where we show that the HHLL correlators at any loops are given by particularly simple Feynman integrals, with a single unfixed colour factor coefficient at a given loop order. We also analyse the OPE limits of the proposed all-loop expression, which confirm the consistency of the results and allow us to derive the structure constants of two heavy operators and the Konishi operator. In section \ref{sec:colourfactor}, we provide a general procedure for fixing the colour factor coefficients by studying the relevant Feynman diagrams we determined in the previous section. We compute explicitly the colour factors for some specific examples of heavy operators.  In section \ref{sec:localisation}, we review the results of the integrated HHLL correlators, from which we can also obtain the colour factors, and we find perfect agreement with the results from Feynman diagrams in section \ref{sec:colourfactor}. We further discuss the intriguing relation between integrated HHLL correlators with the structure constants $C_{\cH \cH \cK}$, and obtain exact expressions for $C_{\cH \cH \cK}$ for some examples of heavy operators and explore their weak and strong coupling expansions.  Finally, the summary and the discussion about future directions are in section \ref{sec:conclusion}.  The paper also includes three appendices: appendix \ref{app:color} provides some basic group theory formulas that are needed for the computation of the colour factors in section \ref{sec:colourfactor}. In appendix \ref{app:N=1SUSY}, we compute the one and two-loop contributions to the HHLL correlators for arbitrary charges (\textit{i.e.} not in the large-charge limit) using $\cN=1$ superspace formalism. We then show that in the large-charge limit the superspace results match with the proposed formula \eqref{eq:SUN-intro} up to two loops. Appendix \ref{app:integrated} provides details of constructing heavy operators and reports some perturbative results for some examples of integrated HHLL correlators with a generic $SU(N)$ gauge group and generic charges.

\section{Perturbative four-point correlators}
\label{sec:HHLL}

In this section, we introduce the four-point correlators of half-BPS operators in $\cN=4$ SYM that we study in this paper:
\begin{align} \label{eq:four-pt1}
    \langle \cH(x_1, Y_1) \cH(x_2, Y_2) \cO_2(x_3, Y_3) \cO_2(x_4, Y_4) \rangle \, ,
\end{align}
where $\cO_2$ is taken to be the single-trace superconformal primary operator obtained as the following gauge invariant combinations of the six scalar fields $\Phi^I$,
\begin{equation}\label{O20prime}
 \cO_p(x, Y)  = {1\over p} Y_{I_1} \cdots Y_{I_p} \Tr\left( \Phi^{I_1}(x) \cdots \Phi^{I_p}(x) \right)~,
\end{equation}
with $p=2$. Here $x$ is spacetime position of the operator and the  $Y_I$, with $I=1,\dots 6$, denotes the $SO(6)$ R-symmetry null vector. The operators $\cH$ are some generic half-BPS superconformal primaries, 
which are combinations of single- and multi-trace operators defined as\footnote{Throughout this paper, we will sometimes use the shorthand notation $(\cO_m)^p$, which should be understood as $\cO_{m,\ldots, m}$ with $p$ traces, and similarly $(\cO_m)^p(\cO_n)^q:=\cO_{m,\ldots, m, n, \ldots, n}$, etc..}
\ie  \label{eq:sin-tr}
   \cO_{p_1, \cdots, p_n}(x, Y) = {p_1 \cdots p_n\over p}  \cO_{p_1}(x, Y) \cdots  \cO_{p_n}(x, Y) \, .
\fe
We denote the dimension (or equivalently $SO(6)_R$ charge) of the operator $\cH$  as $\Delta_{\cH}$,  which eventually will be taken to be large when we consider the large-charge limit and we will refer $\cO_2$ as the light operator and $\cH$ as the heavy operator, when we consider the large-charge limit. 

Before studying four-point correlators in the large-charge limit, we begin by considering the superconformal symmetry constraints on the correlators of generic half-BPS operators. In general, the four-point correlators in \eqref{eq:four-pt1} can be expressed as
\begin{equation}\label{eq:4pt22pp}
    \langle \cH(x_1, Y_1) \cH(x_2, Y_2) \cO_2(x_3, Y_3) \cO_2(x_4, Y_4) \rangle = \cG_{\rm free}(x_i,Y_i) + \cI_4(x_i,Y_i) \, (d_{12})^{\Delta_{\cH}-2}~ \cT_{\cH}(u,v) ~,
\end{equation}
where the prefactor $\cI_4(x_i,Y_i)$ is completely fixed by the superconformal symmetry \cite{Eden:2000bk, Nirschl:2004pa}, which takes the following form, 
\ie 
\cI_4(x_i,Y_i) =\cN_{\cH} {(z-\alpha)(z-\bar \alpha)(\bar z-\alpha)(\bar z-\bar \alpha) \over z {\bar z} (1-z)(1-\bar z)} (d_{13})^2 (d_{24})^2 \, ,
\fe
where  $d_{ij}=Y_i\cdot Y_j / x_{ij}^2$, and we have included the normalisation factor of the two-point function of the heavy operator, 
\ie 
\langle \cH(x_1, Y_1) \cH(x_2, Y_2) \rangle = \cN_{\cH}  \, (d_{12})^{\Delta_{\cH}} \, ,
\fe
which is independent of the coupling and can be computed by free-theory Wick contractions.  
In the expressions of $\cI_4(x_i,Y_i)$ and $\cT_{\cH}(u,v)$,  the spacetime cross ratios are defined as follows
\ie \label{eq:zzb}
   u &=z\, \bar{z}= {x_{12}^2 x_{34}^2 \over x_{13}^2 x_{24}^2} ~, \quad \quad v=(1-z)(1- \bar{z})= {x_{14}^2 x_{23}^2 \over x_{13}^2 x_{24}^2} ~, 
   \fe
   and the R-symmetry cross ratios are given as
   \ie
   \alpha\, \bar{\alpha}&=  {Y_1\cdot Y_2 Y_3\cdot Y_4 \over Y_1\cdot Y_3 Y_2\cdot Y_4} ~, \quad \quad (1-\alpha)(1- \bar{\alpha})=  {Y_1\cdot Y_4 Y_2\cdot Y_3 \over Y_1\cdot Y_3 Y_2\cdot Y_4}  ~.
\fe
In \eqref{eq:4pt22pp}, we have also separated out the free part of the correlator, $\cG_{\rm free}(x_i,Y_i)$, which can be computed by free-theory Wick contractions; therefore,  all the non-trivial dynamics is contained in  $\cT_{\cH}(u,v)$, often called the reduced correlator in the literature, which is the main focus of our work.

\subsection{Review of the chiral Lagrangian insertion method}
\label{sec:3.1}
The most efficient approach to compute the reduced correlator $\cT_{\cH} (u,v)$ in $\mathcal{N}=4$ SYM at weak coupling is the chiral Lagrangian insertion method \cite{Eden:2011we, Eden:2012tu}, which allows to trade the construction of $L$-loop integrands for four-point functions of half-BPS operators with tree level correlators with $L$-insertions of chiral Lagrangians. 
We will follow the discussion of \cite{Dorigoni:2021rdo, Green:2020eyj}. We begin with the following exact relation of the correlators in $\mathcal{N}=4$ SYM: 
\ie \label{eq:soft}
 \cD_{w}    \langle  \cO_1(x_1)  \ldots \cO_n (x_n) \rangle  =  \int d^4 x_{n+1}   \langle  \cO_1(x_1)  \ldots \cO_n (x_n)   \cL(x_{n+1})  \rangle \, ,
\fe
where $\cL(x)$ is the chiral Lagrangian, which is the top component of the super multiplet of $\cO_2$, being related to $\cO_2$ by the action of four supercharges, $\cL \sim Q^4 \cO_2$ \cite{Dolan:2001tt}. After imposing equations of motion, $\cL$ is given by
\begin{align}
 \label{eq:otaudef} 
 \cL =   \tr\bigg\{  -{1\over 2}  F_{\alpha\beta} F^{\alpha\beta} 
+\sqrt{2} \lambda^{\alpha A} [\phi_{AB},\lambda_\alpha^B]   - \frac18  [\phi^{AB},\phi^{CD}][\phi_{AB},\phi_{CD}] \bigg\} \, .
\end{align}
And $\cD_{w}$ denotes the $SL(2, \mathbb{Z})$ covariant derivative, which is defined as
\ie
 \cD_{w} = \tau_2\, \partial_{\tau} - i {w \over 2} \, , \qquad {\rm with} \quad \tau = \tau_1 + \ii \tau_2 = {\theta \over 2\pi} + \ii {4\pi \over \gym^2}\, , 
 \fe
where $w$ is the total bonus $U(1)_Y$ weights \cite{Intriligator:1998ig} of the $n$-point correlators of the operators $\cO_1,  \ldots, \cO_n$.  The relation \eqref{eq:soft} is valid non-perturbatively, and relates any $n$-point correlator to an $(n{+}1)$-point correlator with a chiral Lagrangian insertion. In the holographic dual type IIB superstring theory, such relation has an interesting interpretation as the soft-dilaton theorem, where the additional chiral Lagrangian (integrated over spacetime) is interpreted as a soft dilaton in the type IIB superstring theory \cite{Green:2020eyj}.\footnote{More detailed discussion of the $U(1)_Y$ weights and the relation \eqref{eq:soft} as well as its applications, especially at non-perturbative level, can be found in \cite{Dorigoni:2021rdo, Green:2020eyj}.} 

Applying the relation \eqref{eq:soft} repeatedly to the four-point correlators of our interest in \eqref{eq:4pt22pp}, we have the following relation,  
\ie \label{eq:soft-2}
& \cD_{L-1} \ldots \cD_{1}  \cD_{0}    \langle \cH(x_1, Y_1) \cH(x_2, Y_2) \cO_2(x_3, Y_3) \cO_2(x_4, Y_4) \rangle  \cr 
=&  \int d^4 x_5 \ldots  d^4 x_{L+4} \langle \cH(x_1, Y_1) \cH(x_2, Y_2) \cO_2(x_3, Y_3) \cO_2(x_4, Y_4) \cL(x_5)  \ldots \cL(x_{L+4})  \rangle \, ,
\fe
where we have used the fact that the four-point correlator has zero  $U(1)_Y$ weight since the operators are the bottom components of supermultiplets \cite{Dorigoni:2021rdo, Green:2020eyj}. Note that the derivative $\cD_{0}$ annihilates the free part of the four-point correlator, namely $\cG_{\rm free}(x_i,Y_i)$ in \eqref{eq:4pt22pp}. And importantly, due to the superconformal Ward identity, one can also factor out the same R-symmetry factor $\cI_4(x_i, Y_i) (d_{12})^{\Delta_\cH-2}$  for the $(4+\ell)$-point correlator on the RHS because the chiral Lagrangian is the highest component of the supermultiplet, which is independent of the $SO(6)_R$ vector $Y_I$. Thus, \eqref{eq:soft-2} leads to a relation for the reduced correlator $\cT_{\cH}(u,v)$, which we would like to study. 

As shown in \cite{Dorigoni:2021rdo}, when applying the relation \eqref{eq:soft-2} in the perturbation theory, it leads to a very efficient way of constructing the loop integrands for the four-point correlator \cite{Eden:2011we, Eden:2012tu}, 
\begin{align}
     \label{eq:soft-3}
& \langle \cH(x_1, Y_1) \cH(x_2, Y_2) \cO_2(x_3, Y_3) \cO_2(x_4, Y_4) \rangle \vert_{ L-{\rm loop}} \\
=&\, {(-1)^L \over L!} \int d^4 x_5 \ldots  d^4 x_{L+4} \langle \cH(x_1, Y_1) \cH(x_2, Y_2) \cO_2(x_3, Y_3) \cO_2(x_4, Y_4) \cL(x_5)  \ldots \cL(x_{\ell+4})  \rangle \vert_{\rm tree} \, . \nonumber
\end{align}
The above relation relates the $L$-loop corrections to the four-point correlator with the tree-level $(L+4)$-point correlator with $L$ additional chiral Lagrangian inserted (see also Figure \ref{fig:example_L} for a prototypical diagram). This relation implies that the perturbative expansion of the reduced correlators $\cT_{\cH}(u, v)$ can be expressed as \cite{Eden:2011we, Eden:2012tu}
\ie\label{eq:TH_exp}
\cT_{\cH}(u, v) =    \sum_{L=1}^{\infty} \left(-\frac{\gym^2}{4\pi^2}\right)^L x_{13}^2 x_{24}^2 \, F_{\cH}^{(L)}(x_i)\, .
\fe
The $L$-loop integral $F^{(L)}(x_i)$ is defined as 
\ie \label{eq:FLx}
F_{\cH}^{(L)}(x_i) = {x^2_{12}x_{13}^2x_{14}^2 x_{23}^2 x_{24}^2 x_{34}^2 \over  (\pi^2)^L } \int d^4 x_5 \ldots d^4 x_{L+4} \sum_{\alpha} d^{(\alpha)}_{\cH, N; L}\, f^{(L)}_{\alpha} (x_i)\,,
\fe
where $d^{(\alpha)}_{\cH, N; L}$ are the $L$-loop colour factors, depending on the rank of the gauge group and on the form of the $\cH$ operator, and each $f^{(L)}_{\alpha} (x_i)$ is some rational function (often called $f$-graph functions \cite{Eden:2011we, Eden:2012tu}), which takes the following general form, 
\ie \label{eq:f-graph}
f^{(L)}_{\alpha} (x_i)={P^{(L)}_{\alpha}(x_{ij}^2) \over \prod_{1\leq i<j \leq 4+L} x_{ij}^2} \, . 
\fe
The numerator $P^{(L)}_{\alpha}$ is a polynomial of $x_{ij}^2$ that has weight $(L-1)$ at each point $x_i$ and has $S_2 \times S_{2+L}$ permutation symmetry, namely it is symmetric in $\{x_1, x_2\}$ and in $\{x_3, x_4, x_5 \ldots, x_{4+L}\}$. The symmetry property in the heavy operator insertion points $\{x_1, x_2\}$ is manifest, while the enhanced permutation symmetry between the external points $x_3, x_4$ and the integration points $x_5,  \ldots, x_{4+L}$ is because the chiral Lagrangian $\cL$ and $\cO_2$ are in the same super multiplet. Equivalently, this is due to the fact that the same R-symmetry factor $\cI_4(x_i, Y_i) (d_{12})^{\Delta_\cH-2}$ (as we discussed earlier) can be factored out on both sides of the equation \eqref{eq:soft-2} \cite{Eden:2011we, Eden:2012tu}.

After this brief review, in the next section, we will utilise the chiral Lagrangian insertion method to derive the Feynman integrals for the HHLL correlators to all-loop orders in the large-charge 't Hooft limit. 

\section{All-loop integrands for Heavy-Heavy-Light-Light correlators}
\label{sec:integrand}

In this section, we define the large-charge 't Hooft limit of the HHLL correlators. Using the chiral Lagrangian insertion method reviewed in section~\ref{sec:HHLL} and combining it with certain supersymmetric non-renormalisation theorems, remarkably, we are able to determine the perturbative loop integrands for the HHLL correlators to all-loop orders in the large-charge 't Hooft limit. 

\subsection{Large charge 't Hooft limit}

We begin by specifying the regime in the parameter space that we investigate in this work. We consider a limit very similar to the 't Hooft limit where the usual role of $N$ is replaced by the dimension or charge of the operator, $\Delta_\cH$. More specifically, we take the following double scaling limit between $\Delta_{\cH}$ and the Yang-Mills coupling \cite{Bourget:2018obm} \footnote{In the following, sometimes we also denote $\Delta_\cH = h$ to simplify the notation.}:
\begin{equation}\label{eq:doublescaling_def2}
    \Delta_{\cH} \rightarrow \infty~, ~~  \rm with ~~~ \lambda =\Delta_{\cH} \, g_{_{\rm YM}}^2~~\rm fixed~,
\end{equation}
where we denote (here and in the following) $\lambda$ as the large-charge 't Hooft coupling. Our discussion is valid for any $\Delta_{\cH} \gg N^2$. As a first consequence of this limit, we see that the structure of the heavy operators is fixed to be multi-trace because the dimensions of the operators are larger than the rank of the gauge group. In this section, we do not need the explicit form of $\cH$, as long as the dimension $\Delta_\cH$ is the largest parameter. We will return to the possible choices of heavy operators in section~\ref{sec:heavy_choice}.

The implications of the large-charge 't Hooft limit \eqref{eq:doublescaling_def2} on the reduced correlator $\cT_{\cH}(u, v)$ were first studied in the context of integrated correlators in \cite{Paul:2023rka, Brown:2023why}. It has been shown that the integrated HHLL correlators in $\cN=4$ SYM admit a topological expansion in the charge space (similar to the usual 't Hooft expansion in the colour space). As we will see later, this property still holds for the (unintegrated) HHLL correlators $\cT_{\cH}(u, v)$, such that
\ie
\cT_{\cH}(u,v)  = \sum_{g=0}^{\infty} {1\over \Delta_{\cH}^{g} }\, \cT^{(g)}_{\cH}(u,v)  \, .
\fe
Similarly to the usual 't Hooft large-$N$ expansion, each genus contribution  has a perturbative expansion in the large-charge 't Hooft coupling $\lambda$, 
\ie \label{eq:TH_genusexp}
\cT^{(g)}_{\cH}(u,v)  = \sum_{L=1}^{\infty} (-a)^L\, \mathcal{T}^{(g)}_{\cH; L}(u, v) \, , \quad {\rm with} \quad a = \frac{\lambda}{4\pi^2}\, , 
\fe
where each $\mathcal{T}^{(g)}_{\cH; L}(u, v)$ is obtained by combining a sum of Feynman integrals in terms of the $f$-graph functions as in the form of \eqref{eq:f-graph}, with appropriate colour factors. 

In this paper, we will focus on the leading genus-zero contribution at the large charge limit. From now on, to simplify the notation we will drop the superscript ``$(0)$" and denote $\cT^{(0)}_{\cH}(u,v)$ simply as $\cT_{\cH}(u,v)$, and $\cT^{(0)}_{\cH; L}(u,v)$ as $\cT_{\cH; L}(u,v)$. Compared with the general expression \eqref{eq:TH_exp} for fixed charge, we will find that the $L$-loop contribution to $\cT_{\cH}(u,v)$ in the large-charge 't Hooft limit can be expressed in terms of a single class of $f$-graph functions, where the colour factor $d_{\cH, N; L}$ is completely factored out, namely
\ie\label{eq:factor_TI}
\cT_{\cH; L}(u, v) = d_{\cH, N; L}\, \mathcal{I}^{(L)}(u, v)\,. 
\fe
The other implication of this formula is that the spacetime part of the reduced correlator $\mathcal{I}^{(L)}(u, v)$ is universal, and does not depend on the detailed multi-trace structure of the heavy operators $\cH$, as long as the large charge limit \eqref{eq:doublescaling_def2} holds. In terms of $f$-graph functions, using \eqref{eq:TH_exp} and \eqref{eq:FLx}  $\mathcal{I}^{(L)}(u, v)$ can be expressed as
\ie \label{eq:defcI}
\mathcal{I}^{(L)}(u, v) = {x^2_{12}x_{13}^4 x_{14}^2 x_{23}^2 x_{24}^4 x_{34}^2 \over  (\pi^2)^L } \int d^4 x_5 \ldots d^4 x_{L+4}  \, f^{(L)}_{\alpha} (x_i) \, ,
\fe
where we have removed the summation over the index $\alpha$ because, as we will see, only  a particular $f$-graph function contributes to the HHLL correlators. This fact is also related to the factorisation of the colour factors in \eqref{eq:factor_TI}. 
The dependence on the precise form of $\cH$ resides in the colour factors $d_{\cH, N; L}$ which will be computed in section~\ref{sec:colourfactor}.
In the next subsection we will show that the factorisation \eqref{eq:factor_TI} holds in the large-charge 't Hooft limit, and we will derive the spacetime Feynman integral part $\mathcal{I}^{(L)}(u, v)$ or equivalently determine the function $f^{(L)}_{\alpha} (x_i)$. 

\subsection{Diagrammatic analysis using chiral Lagrangian insertion}

\begin{figure}
    \centering
    \begin{tikzpicture}
      % Define the coordinates for the three vertical points
          \coordinate (x7) at (0, 1.);
    \coordinate (x8) at (0,-1.);
    \coordinate (x5) at (0,0.2);
    %\coordinate (x6) at (0,-1/2);
    \coordinate (x4) at (0, -2);
    \coordinate (x3) at (0, 2);
    
    % Define the coordinates for the left and right points
    \coordinate (x1) at (-2, 0);
    \coordinate (x2) at (2, 0);
    
    % Draw the red dots
    \fill[blue] (x1) circle (2pt);
    \fill[blue] (x2) circle (2pt);
    \fill[blue] (x3) circle (2pt);
    \fill[blue] (x4) circle (2pt);
    \fill[red] (x5) circle (2.5pt);
    %\fill[red] (x6) circle (2.5pt);
        \fill[red] (x7) circle (2.5pt);
    \fill[red] (x8) circle (2.5pt);

    \fill[blue] (-2,1.2) circle (2pt);
    \node at (-2,0.95) {.};
    \node at (-2,0.85) {.};
    \node at (-2,0.75) {.};
    \fill[blue] (-2,0.5) circle (2pt);
    \fill[blue] (-2,0.25) circle (2pt);
    \fill[blue] (-2,-0.25) circle (2pt);
    \fill[blue] (-2,-0.5) circle (2pt);
    \node at (-2,-0.95) {.};
    \node at (-2,-0.85) {.};
    \node at (-2,-0.75) {.};
    \fill[blue] (-2,-1.2) circle (2pt);

    \fill[blue] (2,1.2) circle (2pt);
    \node at (2,0.95) {.};
    \node at (2,0.85) {.};
    \node at (2,0.75) {.};
    \fill[blue] (2,0.5) circle (2pt);
    \fill[blue] (2,0.25) circle (2pt);
    \fill[blue] (2,-0.25) circle (2pt);
    \fill[blue] (2,-0.5) circle (2pt);
    \node at (2,-0.95) {.};
    \node at (2,-0.85) {.};
    \node at (2,-0.75) {.};
    \fill[blue] (2,-1.2) circle (2pt);

    \draw (-2,0) to (-1.7,0.15);
    \draw (-2,0) to (-1.7,0.05);
    \draw (-2,0) to (-1.7,-0.05);
    
    \draw (-2,0.25) to (-1.7,0.3);
    \draw (-2,0.25) to (-1.7,0.4);

    \draw (-2,0.5) to (-1.7,0.55);
   \draw (-2,0.5) to (-1.7,0.65);

    \draw (-2,1.2) to (-1.7,1.3);
    \draw (-2,1.2) to (-1.7,1.4);
    \draw (-2,1.2) to (-1.7,1.2);
    \draw (-2,1.2) to (-1.7,1.1);

    \draw (-2,-0.25) to (-1.7,-0.3);
    \draw (-2,-0.25) to (-1.7,-0.4);

    \draw (-2,-0.5) to (-1.7,-0.55);
   \draw (-2,-0.5) to (-1.7,-0.65);

    \draw (-2,-1.2) to (-1.7,-1.3);
    \draw (-2,-1.2) to (-1.7,-1.4);
    \draw (-2,-1.2) to (-1.7,-1.2);

    \draw (2,0) to (1.7,0.15);
    \draw (2,0) to (1.7,0.05);
    \draw (2,0) to (1.7,-0.05);
    
    \draw (2,0.25) to (1.7,0.3);
    \draw (2,0.25) to (1.7,0.4);

    \draw (2,0.5) to (1.7,0.55);
    \draw (2,0.5) to (1.7,0.65);

    \draw (2,1.2) to (1.7,1.3);
    \draw (2,1.2) to (1.7,1.4);
    \draw (2,1.2) to (1.7,1.2);
     \draw (2,1.2) to (1.7,1.1);

    \draw (2,-0.25) to (1.7,-0.3);
    \draw (2,-0.25) to (1.7,-0.4);

    \draw (2,-0.5) to (1.7,-0.55);
    \draw (2,-0.5) to (1.7,-0.65);

    \draw (2,-1.2) to (1.7,-1.3);
    \draw (2,-1.2) to (1.7,-1.4);
    \draw (2,-1.2) to (1.7,-1.2);
    
    \draw (x5) to (0, 0.5);
    \draw (x5) to (0, -0.1);
    \draw (x5) to (0.3, 0.2);
    \draw (x5) to (-0.3, 0.2);
     \draw (x7) to (0, 0.7);
    \draw (x7) to (0, 1.3);
    \draw (x7) to (0.3, 1);
    \draw (x7) to (-0.3, 1);
    \draw (x8) to (0, -0.7);
    \draw (x8) to (0, -1.3);
    \draw (x8) to (0.3, -1);
    \draw (x8) to (-0.3, -1);

    \draw (x3) to (0.2,1.6);
     \draw (x3) to (-0.2,1.6);

    \draw (x4) to (0.2,-1.6);
     \draw (x4) to (-0.2,-1.6);
    
 \node at (0,-0.3) {.};
\node at (0,-0.4) {.};
    \node at (0,-0.5) {.};

     \node at (-2.4,0) {$x_1$}; \node at (2.4,0) {$x_2$};
      \node at (0,2.4) {$x_3$}; \node at (0,-2.4) {$x_4$};
\end{tikzpicture}  
    \caption{A graphical representation of the $(L+4)$-point correlator arising from the chiral Lagrangian insertion method, to be contracted with free propagators. The insertions at points $x_1$ and $x_2$ are generic multi-trace half-BPS operators, the insertions at $x_3$ and $x_4$ are the two light operators. We have distinguished the insertions of the chiral Lagrangians at points $x_5\dots x_{4+L}$ by red dots. The chiral Lagrangian insertions are represented with four-scalar vertices, even though this is not the most general case, see \eqref{eq:otaudef}. However, for this paper, we will see that the diagrams dominating in the large-charge limit arise purely from the four-scalar vertices.}
    \label{fig:example_L}
\end{figure}
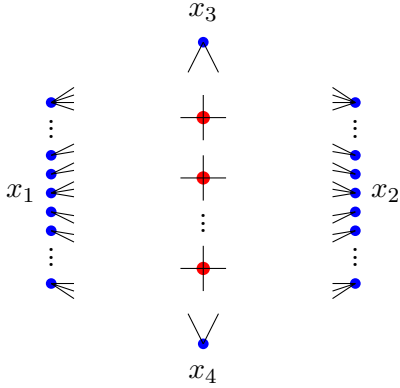

This subsection is devoted to determining the class of $f$-graph functions \eqref{eq:f-graph} that contribute to the HHLL correlators in the large-charge planar limit using the chiral Lagrangian insertion method.
In order to clarify our arguments mainly based on combinatorial reasons, we integrate the chiral Lagrangian insertion method with a graphical approach. We represent the $L$-loop contribution to the correlators with the insertion of $L$ chiral Lagrangians as in Figure~\ref{fig:example_L}.

\subsubsection{Supersymmetric non-renormalisation theorems}

Before analysing the possible diagrams or Feynman integrals that may contribute in the large-charge limit, we would like to take into account all the constraints on the diagrams from $\cN=4$ maximal supersymmetry, that allow us to exclude a priori certain classes of diagrams. As we mentioned, the chiral Lagrangian is related to superconformal primary $\cO_2$ by supersymmetry via $\cL \sim Q^4 \cO_2$ \cite{Dolan:2001tt}. This allows one to derive supersymmetric non-renormalisation theorems \cite{Baggio:2012rr}, which in particular imply that any correlators with fewer than four half-BPS insertions and any non-zero number of chiral Lagrangians should vanish, namely
\ie \label{eq:0-diagram}
\langle O_1(x_1, Y_1) \cdots O_n(x_n, Y_n)  \cL(x_{n+1})  \ldots \cL(x_{n+L}) \rangle=0\, ,  \quad {\rm if} \quad n \leq 3 \quad {\rm and} \quad L \geq 1 \, .
\fe
It is worth noting that the vanishing of these correlators combined with the relation \eqref{eq:soft} immediately implies the fact that two- and three-point functions of half-BPS operators do not receive any quantum corrections \cite{Lee:1998bxa, Baggio:2012rr}. It is important to note that the supersymmetric non-renormalisation theorems \eqref{eq:0-diagram} apply to any half-BPS insertions, with  
\ie \label{eq:non-gauge}
O_i(x, Y) = \Phi_{I}(x) \cdots \Phi_{J}(x)\, Y^{I} \cdots  Y^{J}  
\fe
being any collection of fundamental scalar fields inserted at a single point $x$ and not necessarily gauge invariant. In particular, \eqref{eq:0-diagram} implies that any diagrams with two (or more) fundamental fields from a single chiral Lagrangian connected to a single half-BPS operator do not contribute (or more precisely it will be cancelled by other diagrams). 
%This argument allows to exclude any contributions that contain a combination like \eqref{eq:0-diagram} as a subdiagram.
%which is not necessarily gauge invariant.  also apply if we do not trace over the gauge group for the half BPS operators, because $O_i(x, Y)$ are half BPS whether we trace over the gauge group or not. 
%The supersymmetric non-renormalisation theorems \eqref{eq:0-diagram} impose strong constraints on the allowed diagrams using the chiral Lagrangian insertion method \eqref{eq:soft-3}.  
Let us clarify this point by considering a schematic example showing the class of diagrams that are not allowed by \eqref{eq:0-diagram}, see below: 
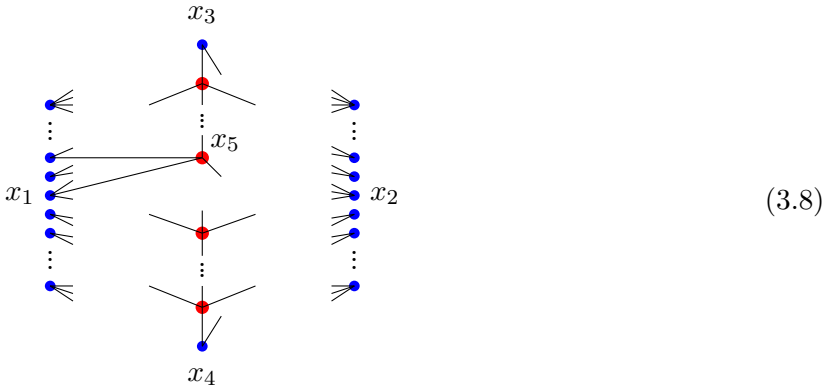
\ie \label{eq:non-allowed}
\begin{tikzpicture}
      % Define the coordinates for the three vertical points
          \coordinate (x7) at (0, 4/3+0.15);
    \coordinate (x8) at (0,-4/3-0.15);
    \coordinate (x5) at (0,1/2);
    \coordinate (x6) at (0,-1/2);
    \coordinate (x4) at (0, -2);
    \coordinate (x3) at (0, 2);
    
    % Define the coordinates for the left and right points
    \coordinate (x1) at (-2, 0);
    \coordinate (x2) at (2, 0);
    
    % Draw the red dots
    \fill[blue] (x1) circle (2pt);
    \fill[blue] (x2) circle (2pt);
    \fill[blue] (x3) circle (2pt);
    \fill[blue] (x4) circle (2pt);
    \fill[red] (x5) circle (2.5pt);
    \fill[red] (x6) circle (2.5pt);
        \fill[red] (x7) circle (2.5pt);
    \fill[red] (x8) circle (2.5pt);

    \fill[blue] (-2,1.2) circle (2pt);
    \node at (-2,0.95) {.};
    \node at (-2,0.85) {.};
    \node at (-2,0.75) {.};
    \fill[blue] (-2,0.5) circle (2pt);
    \fill[blue] (-2,0.25) circle (2pt);
    \fill[blue] (-2,-0.25) circle (2pt);
    \fill[blue] (-2,-0.5) circle (2pt);
    \node at (-2,-0.95) {.};
    \node at (-2,-0.85) {.};
    \node at (-2,-0.75) {.};
    \fill[blue] (-2,-1.2) circle (2pt);

    \fill[blue] (2,1.2) circle (2pt);
    \node at (2,0.95) {.};
    \node at (2,0.85) {.};
    \node at (2,0.75) {.};
    \fill[blue] (2,0.5) circle (2pt);
    \fill[blue] (2,0.25) circle (2pt);
    \fill[blue] (2,-0.25) circle (2pt);
    \fill[blue] (2,-0.5) circle (2pt);
    \node at (2,-0.95) {.};
    \node at (2,-0.85) {.};
    \node at (2,-0.75) {.};
    \fill[blue] (2,-1.2) circle (2pt);

    \draw (-2,0) to (-1.7,0.2);
    \draw (-2,0) to (-1.7,-0.05);
    
    \draw (-2,0.25) to (-1.7,0.3);
    \draw (-2,0.25) to (-1.7,0.4);

    \draw (-2,0.5) to (-1.7,0.63);
 %   \draw (-2,0.5) to (-1.7,0.65);

    \draw (-2,1.2) to (-1.7,1.3);
    \draw (-2,1.2) to (-1.7,1.4);
    \draw (-2,1.2) to (-1.7,1.2);
    \draw (-2,1.2) to (-1.7,1.1);

    \draw (-2,-0.25) to (-1.7,-0.3);
    \draw (-2,-0.25) to (-1.7,-0.4);

    \draw (-2,-0.5) to (-1.7,-0.55);
   \draw (-2,-0.5) to (-1.7,-0.65);

    \draw (-2,-1.2) to (-1.7,-1.3);
    \draw (-2,-1.2) to (-1.7,-1.4);
    \draw (-2,-1.2) to (-1.7,-1.2);

    \draw (2,0) to (1.7,0.15);
    \draw (2,0) to (1.7,0.05);
    \draw (2,0) to (1.7,-0.05);
    
    \draw (2,0.25) to (1.7,0.3);
    \draw (2,0.25) to (1.7,0.4);

    \draw (2,0.5) to (1.7,0.55);
    \draw (2,0.5) to (1.7,0.65);

    \draw (2,1.2) to (1.7,1.3);
    \draw (2,1.2) to (1.7,1.4);
    \draw (2,1.2) to (1.7,1.2);
     \draw (2,1.2) to (1.7,1.1);

    \draw (2,-0.25) to (1.7,-0.3);
    \draw (2,-0.25) to (1.7,-0.4);

    \draw (2,-0.5) to (1.7,-0.55);
    \draw (2,-0.5) to (1.7,-0.65);

    \draw (2,-1.2) to (1.7,-1.3);
    \draw (2,-1.2) to (1.7,-1.4);
    \draw (2,-1.2) to (1.7,-1.2);
    
    \draw (x1) to (x5);
    \draw (-0.7,-0.25) to (x6);
    \draw (x5) to (0, 4/5);
    \draw (x5) to (1/4, 1/4);
 %   \draw (x5) to (x2);
   \draw (x6) to (0, -4/5);
    \draw (x6) to (0.7,-0.25);

    \draw (x3) to (0.25,1.6);
    \draw (x6) to (0,-1/2+0.3);

    \draw (x4) to (0.25,-1.6);
 %   \draw (x5) to (0,1/2-0.3);
    
    \draw (x7) to  (x3);
    \draw (x7) to  (0, 6/5);
      \draw (x7) to  (0.7, 1.2);
           \draw (x7) to  (-0.7, 1.2);
            \draw (x5) to  (-2, 0.5);
    \draw (x8) to  (x4);
    \draw (x8) to  (0, -6/5);
          \draw (x8) to  (0.7, -1.2);
            \draw (x8) to  (-0.7, -1.2);
    
 \node at (0,1.06) {.};
\node at (0,0.99) {.};
    \node at (0,0.92) {.};
     \node at (0,-1.06) {.};
\node at (0,-0.99) {.};
    \node at (0,-0.92) {.};

     \node at (-2.4,0) {$x_1$}; \node at (2.4,0) {$x_2$};
%     \node at (-0.3,1/2+0.2) {$x_5$};
      \node at (0.3,1/2+0.2) {$x_5$};
      \node at (0,2.4) {$x_3$}; \node at (0,-2.4) {$x_4$};
\end{tikzpicture}  
\fe
Starting from the picture represented in Figure \ref{fig:example_L}, we have highlighted the chiral Lagrangian at position $x_5$, which is connected to the half-BPS operator at the position $x_1$ with two free propagators.  
%Indeed, as specified in \eqref{eq:soft-3}, the construction of loop integrands using chiral Lagrangian insertion requires one to close the diagrams using free propagators only.  
Clearly, this part of the diagram is of the form  \eqref{eq:0-diagram}, so it vanishes due to maximal supersymmetry. This property also follows from a different point of view: such a diagram would naively generate double poles of the form $1/(x^2_{15})^2$, which should not be there in any loop integrands by a simple OPE analysis \cite{Eden:2012tu}. The correct allowed form of the loop integrands is displayed in the expression \eqref{eq:f-graph}, which has only simple poles $1/x^2_{ij}$. 

\subsubsection{The integrands at one and two loops}

After excluding this class of illegal diagrams, we now determine the diagrams contributing to the large-charge 't Hooft limit. 
We keep $\cH$ as a general heavy operator, as long as the dimension (or charge) $\Delta_\cH = h \gg N^2$.  
We are interested in the reduced correlator $\cT_{\cH}(u, v)$, which is completely independent of the $SO(6)_R$ vectors $Y_I$. This allows us to choose some particular $SO(6)_R$ channel to simplify the discussion, especially at the level of drawing the Feynman diagrams. In particular we conveniently choose the R-symmetry channels for the heavy operators to be orthogonal to the $\cO_2$ operators, \textit{i.e.} $Y_1\cdot Y_3=Y_2\cdot Y_3=Y_1\cdot Y_4=Y_2\cdot Y_4=0$. With this choice, the light and heavy operators can be connected only via chiral Lagrangian insertions.

We begin by considering the first few loop orders as examples. At one-loop order ($O(\gym^2)$), the numerator polynomials $P^{(1)}_{\alpha}(x_{ij}^2)$ in \eqref{eq:f-graph} should have degree $1-1=0$, so it can only be a constant. The only possible integrand with the right properties is then
\ie \label{eq:f-graph1}
f^{(1)} (x_i)={1 \over \prod_{1\leq i<j \leq 5} x_{ij}^2} \, , 
\fe
which gives rise to the one-loop box integral. Diagrammatically, the one-loop contribution can be obtained by the single insertion of a chiral Lagrangian, as shown below:
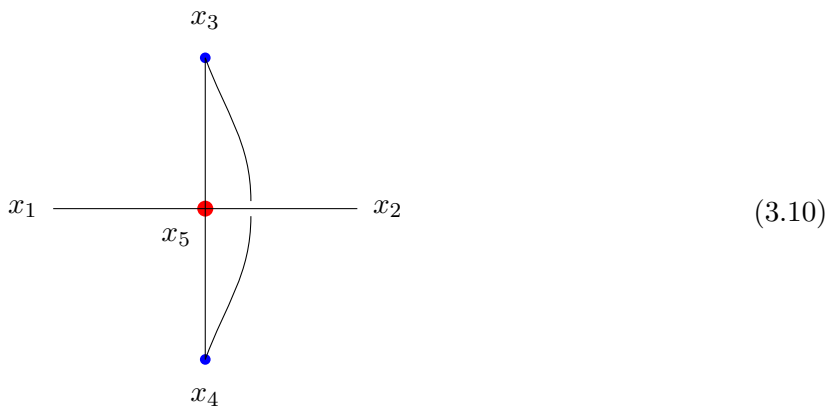
\ie \label{eq:one-loop}
\begin{tikzpicture}
    % Define the coordinates for the three vertical points
    \coordinate (x5) at (0, 0);
    \coordinate (x4) at (0, -2);
    \coordinate (x3) at (0, 2);
    
    % Define the coordinates for the left and right points
    \coordinate (x1) at (-2, 0);
    \coordinate (x2) at (2, 0);
    
    % Draw the red dots
    \fill[blue] (x3) circle (2pt);
    \fill[blue] (x4) circle (2pt);
    \fill[red] (x5) circle (3pt);
    
    % Draw curved lines connecting D and E to A, B, and C in black
    \draw[black] (x1) -- (x5);
    \draw[black] (x3) -- (x5);
    \draw[black] (x4) -- (x5);
    \draw[black] (x2) -- (x5);
    \draw (x3) to[out=-70, in=90] (0.6,0.1);
    \draw (0.6,-0.1) to[out=-90, in=70] (x4);

%    \draw (x3) to[out=-70, in=90] (0.6,0.15);
 %  \draw (0.6,-0.15) to[out=-90, in=70] (x4);
  
    \node[scale=1.] at (-2.4,0) {$x_1$}; \node[scale=1.]  at (2.4,0) {$x_2$};
     \node[scale=1.]  at (-0.38,-0.38) {$x_5$};
      \node[scale=1.]  at (0,2.5) {$x_3$}; \node[scale=1.]  at (0,-2.5) {$x_4$};
\end{tikzpicture}
\fe
where the two lines of $\cL(x_5)$ are then connected to $\cH(x_1)$ and $\cH(x_2)$: each is connected to one of the legs of $\cH(x_1)$ and $\cH(x_2)$ by the free propagator, respectively. Using the explicit form of the chiral Lagrangian \eqref{eq:otaudef}, we see that the diagram \eqref{eq:one-loop} reproduces the expected one-loop box integral \eqref{eq:f-graph1} (up to terms which are independent of the loop integration point $x_5$ that ensures the $f$-graph to have correct weights at each point $x_i$).  The insertion of \eqref{eq:one-loop} in $\vev{\cH(x_1)\cH(x_2)\cO_2(x_3)\cO_2(x_4)}$ comes with a combinatorial factor scaling with the number of legs of the heavy operators at $x_1$ and $x_2$.\footnote{The combinatorial factor should take into account the multitrace structure of the heavy operator. In particular one should consider the legs inside the same trace as indistinguishable, and the correct combinatorics would be in terms of the number of trace insertions in $\cH$ rather than the dimension of the operator. In the large charge limit, however, these two quantities scale in the same way, up to numerical factors that depends on the microscopic structure of the heavy operator $\cH$. This is irrelevant for the analysis of loop integrands of the present section, and will be instead fully considered in section \ref{sec:colourfactor} when computing the combinatorial and colour factors exactly.} This $\binom{h}{1} = h$ factor combines nicely with the Yang-Mills coupling to realise the large charge 't Hooft coupling  $\lambda= h\, \gym^2$.
The colour factor for a general number of colours $N$ at one-loop order will be computed in the next section.

We now consider the two-loop case, which is more effective for clarifying our argument of large charge diagrams. We start from listing all the possible $f$-graph functions \eqref{eq:f-graph}, as described in section \ref{sec:3.1} and following the prescription of \cite{Eden:2011we, Eden:2012tu}. For two-loop computations of the HHLL correlators,  after imposing the $S_2 \times S_{4}$ permutation symmetry, there are two distinct $f$-graphs, with weight four at each position $x_i$:
\begin{subequations}\label{eq:fgraphs_2loop}
    \begin{align}
        f^{(2)}_{1} (x_i)&= {1\over 16} {x_{12}^2 x_{34}^2 x_{56}^2 \over \prod_{1\leq i <j \leq 6} x_{ij}^2  } + S_2 \times S_4 \, , \label{eq:fgraphs_2loop_a} \\
f^{(2)}_{2}(x_i) &={1\over 4} {x_{13}^2 x_{24}^2 x_{56}^2 \over \prod_{1\leq i <j \leq 6} x_{ij}^2  } + S_2 \times S_4 \, . \label{eq:fgraphs_2loop_b}
    \end{align}
\end{subequations}
The claim is that the leading large charge limit selects only a single loop integrand of the $f$-graphs in \eqref{eq:fgraphs_2loop}, which turns out to be $ f^{(2)}_{1}(x_i)$.\footnote{See also \cite{Jiang:2023uut} for a $f$-graphs analysis to compute a special class of HHLL correlator of the form \eqref{eq:four-pt1}, where the operator $\cH$ is chosen to be the so-called dimension $N$ determinant operator for $SU(N)$ gauge group. Such an operator becomes heavy when considering the usual large-$N$ 't Hooft limit, but due to the different regime both $f^{(2)}_{1} (x_i)$ and $f^{(2)}_{2} (x_i)$ contribute.}

\begin{figure}[t]
    \centering
    \begin{tikzpicture}
      % Define the coordinates for the three vertical points
    \coordinate (x5) at (-2/3,0);
    \coordinate (x6) at (2/3,0);
    \coordinate (x4) at (0, -2);
    \coordinate (x3) at (0, 2);
    
    % Define the coordinates for the left and right points
    \coordinate (x1) at (-1.6, 0);
    \coordinate (x2) at (1.6, 0);
    
    % Draw the red dots
    \fill[blue] (x3) circle (2pt);
    \fill[blue] (x4) circle (2pt);
    \fill[red] (x5) circle (3pt);
    \fill[red] (x6) circle (3pt);

    \draw (x1) to (x2);
    \draw (x5) to (x3);
    \draw (x5) to (x4);
    \draw (x6) to (x4);
    \draw (x6) to (x3);
  
    \node[scale=1.]  at (-2.,0) {$x_1$}; \node[scale=1.]  at (2.,0) {$x_2$};
     \node[scale=1.]  at (-2/3-0.3,-0.35) {$x_5$};
      \node[scale=1.]  at (2/3+0.3,-0.35) {$x_6$};
      \node[scale=1.]  at (0,2.5) {$x_3$}; \node[scale=1.]  at (0,-2.5) {$x_4$};
      \node[scale=1.]  at (0,-3.5) {$(a)$};
\end{tikzpicture}
\hspace{1.cm}
\begin{tikzpicture}
      % Define the coordinates for the three vertical points
    \coordinate (x5) at (0,2/3);
    \coordinate (x6) at (0,-2/3);
    \coordinate (x4) at (0, -2);
    \coordinate (x3) at (0, 2);
    
    % Define the coordinates for the left and right points
    \coordinate (x1) at (-2, 0);
    \coordinate (x2) at (2, 0);
    
    % Draw the red dots
    \fill[blue] (x3) circle (2pt);
    \fill[blue] (x4) circle (2pt);
    \fill[red] (x5) circle (3pt);
    \fill[red] (x6) circle (3pt);

    \draw (-1.1,1/3) to (x5);
    \draw (-1.1,-1/3) to (x6);
    \draw (1.1,1/3) to (x5);
    \draw (1.1,-1/3) to (x6);
    \draw (x5) to (x3);
    \draw (x6) to (x4);
    \draw (x5) to (x6);

    \draw (x3) to[out=-70, in=90] (0.6,2/3-0.05);
    \draw (0.6,2/3-0.3) to (0.6,-2/3+0.3);
    \draw (0.6,-2/3+0.05) to[out=-90, in=70] (x4);

   \node[scale=1.] at (-1.6,0) {$x_1$}; 
   \node[scale=1.] at (1.6,0) {$x_2$};
     \node[scale=1.] at (-0.4,2/3+0.3) {$x_5$};
      \node[scale=1.] at (-0.4,-2/3-0.3) {$x_6$};
      \node[scale=1.] at (0,2.5) {$x_3$}; \node[scale=1.] at (0,-2.5) {$x_4$};
      \node[scale=1.] at (0,-3.5) {$(b)$};
\end{tikzpicture}
\hspace{1.cm}
\begin{tikzpicture}
      % Define the coordinates for the three vertical points
    \coordinate (x5) at (0,2/3);
    \coordinate (x6) at (0,-2/3);
    \coordinate (x4) at (0, -2);
    \coordinate (x3) at (0, 2);
    
    % Define the coordinates for the left and right points
    \coordinate (x1) at (-2, 0);
    \coordinate (x2) at (2, 0);
    
    % Draw the red dots
    \fill[blue] (x3) circle (2pt);
    \fill[blue] (x4) circle (2pt);
    \fill[red] (x5) circle (3pt);
    \fill[red] (x6) circle (3pt);

    \draw (-1.1,1/3) to (x5);
    \draw (-1.1,-1/3) to (x6);
    \draw (1.1,1/3) to (x5);
    \draw (1.1,-1/3) to (x6);
    \draw (x5) to (x3);
    \draw (x6) to (x4);

    \draw (x3) to[out=-70, in=90] (0.4,2/3+0.05);
    \draw (0.4,2/3-0.25) to[out=-90, in=70] (x6);

    \draw (x5) to[out=180+70, in=90] (-0.4,-2/3+0.25);
    \draw (-0.4,-2/3) to[out=-90, in=180-70] (x4);

   \node[scale=1.] at (-1.6,0) {$x_1$}; 
   \node[scale=1.] at (1.6,0) {$x_2$};
     \node[scale=1.] at (-0.4,2/3+0.25) {$x_5$};
      \node[scale=1.] at (0.4,-2/3-0.25) {$x_6$};
      \node[scale=1.] at (0,2.5) {$x_3$}; \node[scale=1.] at (0,-2.5) {$x_4$};
      \node[scale=1.] at (0,-3.5) {$(c)$};
\end{tikzpicture} 
    \caption{Classes of two-loop diagrams. Figure $(a)$ is a typical representative of the class of diagrams coming with a $\binom{h}{1}$ combinatorial factor, hence subleading in the large charge limit. Diagrams $(b)$ and $(c)$ have a $\binom{h}{2}$ combinatorial factor and maximise the large-charge combinatorics at two loops. Thanks to the two lines connecting $x_3$ and $x_4$, diagram $(b)$ should come with a relative factor of $2$ compared to $(c)$.}
    \label{fig:twoloops}
\end{figure}
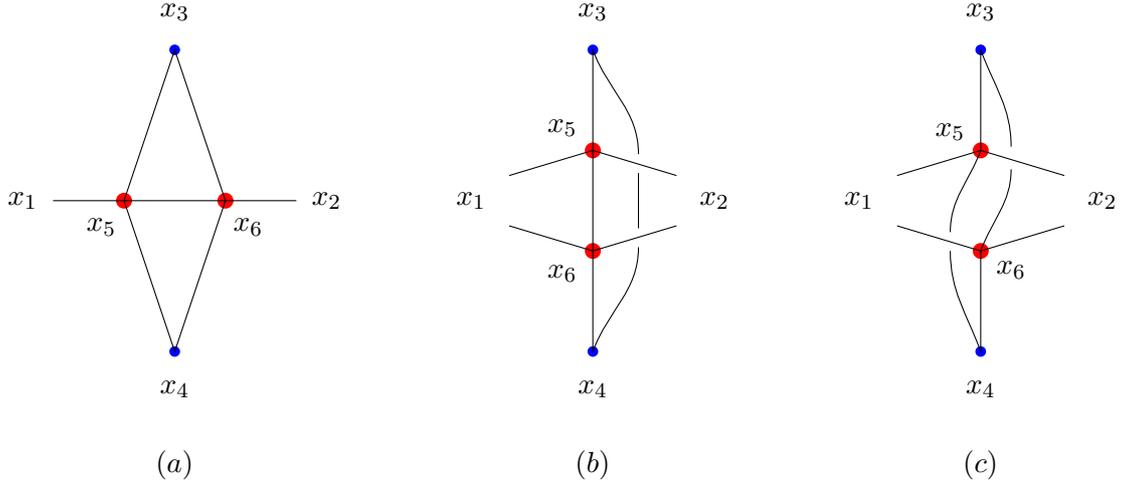

The key element to single out the relevant class of diagrams in the large-charge limit is the combinatorial factor, which is less trivial than the one-loop case. In Figure~\ref{fig:twoloops}, we show a few typical types of diagrams with the insertions of two chiral Lagrangians.  Diagrams like the one represented in diagram $(a)$ of Figure~\ref{fig:twoloops} can be connected to points $x_1, x_2$ (where the heavy operators sit, see Figure~\ref{fig:example_L}) with a combinatorial factor of $\binom{h}{1}=h$, as in the one-loop case. 
Diagrams $(b)$ and $(c)$, on the other hand, are connected to the points $x_1, x_2$ with a $\binom{h}{2} \sim h^2$ combinatorial factor, and so they both dominate in the large-$h$ limit. Once again, such large-charge scaling allows to form precisely the large-charge 't Hooft coupling $\lambda^2 = (h\, g^2_{_{\rm YM}})^2$ as defined in \eqref{eq:doublescaling_def2} for the two-loop contribution.
From these considerations we see that to maximise the charge at each perturbative order we need to connect the maximal number of legs from chiral Lagrangians to the heavy operators, paying attention to avoid diagrams like \eqref{eq:non-allowed}, which are not allowed by the supersymmetric non-renormalisation theorems. This also means that only the four-scalar vertex in the chiral Lagrangian \eqref{eq:otaudef} should be used because only the combination of these vertices maximise the number of lines being connected to the heavy operators.

We now consider the Feynman integrands associated with the diagrams $(b)$ and $(c)$ in Figure~\ref{fig:twoloops}. As we pointed out, all the lines drawn in the diagrams are scalar propagators, so we can directly read off the associated loop integrands at two loops from the diagrams. The diagram $(b)$ in Figure~\ref{fig:twoloops} generates the following two-loop integrand
\ie \label{eq:2l-ladder}
I^{(2)}_{(b)} \sim  {1 \over x_{15}^2 x^2_{16} x_{25}^2 x^2_{26} x_{56}^2 x_{35}^2 x_{46}^2 } \, . 
\fe
The diagram $(c)$ in Figure~\ref{fig:twoloops} leads to another type of two-loop integrand:
\ie \label{eq:1l-ladderS}
I^{(2)}_{(c)} \sim  {1 \over x_{15}^2 x_{25}^2 x_{35}^2 x_{45}^2 x^2_{16} x^2_{26}x^2_{36}  x_{46}^2 } \, .
\fe
This diagrammatic analysis is schematic, which does not specify the coefficients of these Feynman integrals. To impose the enhanced permutation symmetry as discussed at the end of subsection \ref{sec:3.1}, we complete the above expressions by multiplying them with appropriate factors of $x_{ij}^2$ with $i,j<5$ (importantly without introducing higher-order poles), so that they have weight four in all $x_i$. We then apply the $S_2 \times S_{4}$ permutation symmetry acting separately on $x_{1}, x_2$ and on $x_3, x_4, x_5, x_6$. After some trivial algebra, one can realise that the combination of diagrams $(b)$ and $(c)$ actually belongs to a single class of the two-loop $f$-graph functions, \textit{i.e.} $f^{(2)}_{1}$ in \eqref{eq:fgraphs_2loop_a}.  

The fact that diagrams $(b)$ and $(c)$ in Figure~\ref{fig:twoloops} belong to a single $f$-graph function implies that they have the same colour factor. This important
point will also be proved by a direct colour factor computation in section~\ref{sec:colourfactor}. Then $d_{\cH,N;2}$ can be factored out as in \eqref{eq:factor_TI}. In this section we concentrate on computing the spacetime part $\mathcal{I}^{(L)}(u,v)$ of the two-loop correlator, while $d_{\cH,N;2}$ will be explicitly computed in section~\ref{sec:colourfactor} for some choices of $\cH$.

After some simplification, we conveniently rewrite $f^{(2)}_{1}$  as
\ie \label{eq:2l-ladder2}
f^{(2)}_{1} = {1 \over 8} \frac{1}{ \prod_{i=3}^{6} x^2_{i,i+1} x^2_{1i} x^2_{2i}}\,+ S_{4} \, ,      
\fe
where $x_{7}$ should be identified as $x_3$, the prefactor $1/8$ is the symmetry factor, and we have dropped the action of $S_2$ on $x_1, x_2$ since it only generates another overall symmetry factor. Now we can write down the two-loop integral, using the definition \eqref{eq:defcI} applied on the $f^{(2)}_{1}$ integrand given in \eqref{eq:2l-ladder2}:\footnote{As we discussed above, compared to \eqref{eq:FLx} we can factor out the $d_{\cH,N;2}$ colour factor, according to the large-charge factorisation \eqref{eq:factor_TI}.}
\ie \label{eq:Fladder2}
\cI^{(2)}(u, v) = {x^2_{12}x_{13}^4x_{14}^2 x_{23}^2 x_{24}^4 x_{34}^2 \over (\pi^2)^2 } \int d^4 x_5 d^4 x_{6}\, f^{(2)}_{1} (x_i) \, .
\fe
%\ie \label{eq:Fladder2}
%F^{(2)}(x_i) = {x^2_{12}x_{13}^2x_{14}^2 x_{23}^2 x_{24}^2 x_{34}^2 \over (\pi^2)^2 } \int d^4 x_5 d^4 x_{6}\, f^{(2)}_{1} (x_i) \, .
%\fe
This Feynman integral can be recast in simple combinations of well-known ladder Feynman integrals. The $L$-loop ladder integral is defined as 
\ie \label{eq:ladderL0}
\Phi^{(L)}(u, v) =  \int {d^4 x_5 \over \pi^2} \ldots {d^4 x_{L+4}  \over \pi^2} \frac{x_{34}^2  \, (x_{12}^2)^{L}}{ \prod_{i=4}^{L+4} x^2_{i,i+1} \prod_{i=5}^{L+4} x^2_{1i} x^2_{2i}} \, , 
\fe
where we have identified $x_{L+5}:=x_3$. Famously, the ladder integral is known to any loop orders in terms of polylogarithms \cite{Usyukina:1993ch}:  
\ie \label{eq:ladderL}
\Phi^{(L)}(u, v) = {u \over  z -\bar{z}} \sum_{r=0}^L {(-1)^r (2L-r)! \over r! (L-r)! L! } \log^r (v) \left( \text{Li}_{2L-r}(1-z) - \text{Li}_{2L-r}(1-\bar{z}) \right)  \, ,
\fe
where the cross ratios are defined in \eqref{eq:zzb}. 
Therefore, the two-loop integral $F^{(2)}(x_i)$ in \eqref{eq:Fladder2} can be rewritten in a very simple form, purely in terms of ladder integrals, such that the spacetime part of the reduced correlator $\cI^{(2)}(u,v)$ from the large-charge expansion \eqref{eq:factor_TI} at two-loops reads
\ie \label{eq:F2loop}
\mathcal{I}^{(2)}(u,v)= {1  \over u} \left[ \left(  \Phi^{(1)}(u, v) \right)^2 +2  \Phi^{(2)}(u, v)  \right] \, .
\fe 
Hence the diagrams $(b)$ and $(c)$ in Figure~\ref{fig:twoloops} maximising the combinatorial factor at large charge generate the simple two-loop result \eqref{eq:F2loop}. A posteriori, such a result can also be derived from the structure of the diagrams $(b)$ and $(c)$ in terms of scalar propagators; diagram $(b)$ has the conformal structure of the two-loop ladder diagram, and diagram $(c)$ is essentially the square of the one-loop ladder integral drawn in \eqref{eq:one-loop}. The relative factor of two comes from the different symmetry factors in the chiral Lagrangian insertion.

From the above discussion, we see that only a particular two-loop $f^{(2)}$-graph function contributes in the large-charge 't Hoot limit. As we will see shortly in the next subsection, the fact that only a particular type of $f$-graph function contributes in the large-charge 't Hooft limit holds even at higher loops. As in the two-loop case, such an $f$-graph function at $L$ loops generates products of ladder integrals, generalising \eqref{eq:F2loop}. 
Therefore, we will denote this special $f$-graph as $f_{\rm ladder}$. 

An additional check of the two-loop result \eqref{eq:F2loop} can be found in appendix \ref{app:N=1SUSY}, where we report the two-loop computation for $\cT_\cH(u,v)$ following the results of \cite{DAlessandro:2005fnh} obtained with Feynman diagram computations via $\cN=1$ superspace formalism. In this setting, one needs to specify the choice of the heavy operator, so we revisit the results of \cite{DAlessandro:2005fnh} (obtained at fixed conformal dimensions) in the case of HHLL correlators for $\cH = (\cO_2)^p$.  In this way, we explicitly verify that taking the large-charge limit (\textit{i.e.} $p\rightarrow \infty$) indeed also leads to \eqref{eq:F2loop}.

\subsubsection{HHLL correlators at all loops}

We can now generalise our argument to higher-loop orders. The large charge combinatorial argument for the two-loop case applies to the higher-loop contributions analogously. The diagrams dominating in the large-charge limit must have the maximal number of lines from the chiral Lagrangians connecting to the heavy operators (once again, up to the allowed conditions coming from the supersymmetric non-renormalisation theorems), in order to maximise the combinatorial factor. The generalisation of the diagrams $(b), (c)$ in Figure~\ref{fig:twoloops} to the higher-loop case takes the following typical form:
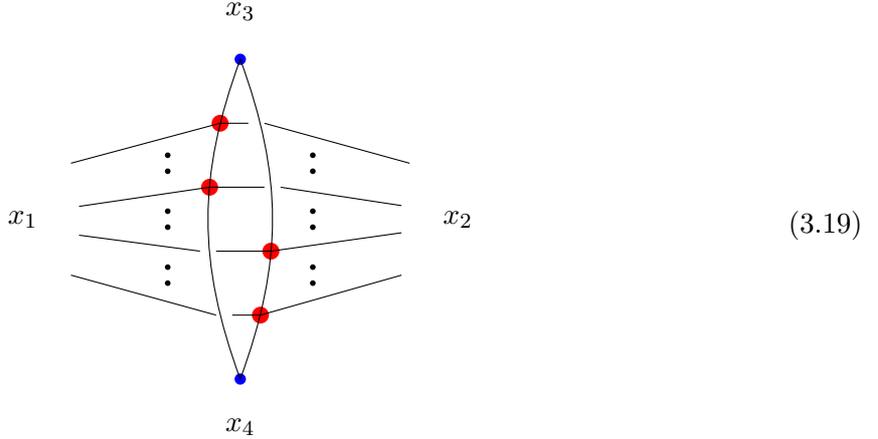
\ie \label{eq:L-loops}
\begin{tikzpicture}[scale=1.06]
    \coordinate (x5) at (-0.25,1.2);
    \coordinate (x6) at (-0.38,0.4);
    \coordinate (x7) at (0.38,-0.4);
    \coordinate (x8) at (0.25,-1.2);
    \coordinate (a1) at (-2.1,0.7);
    \coordinate (a2) at (-2,0.16);
    \coordinate (a3) at (-2,-0.2);
    \coordinate (a4) at (-2.1,-0.7);
    \coordinate (b1) at (2.1,0.7);
    \coordinate (b2) at (2,0.17);
    \coordinate (b3) at (2,-0.17);
    \coordinate (b4) at (2,-0.7);
    \coordinate (x4) at (0, -2);
    \coordinate (x3) at (0, 2);
    
    % Draw the red dots
    \fill[blue] (x3) circle (2pt);
    \fill[blue] (x4) circle (2pt);
    \fill[red] (x5) circle (3pt);
    \fill[red] (x6) circle (3pt);
    \fill[red] (x7) circle (3pt);
    \fill[red] (x8) circle (3pt);

    \draw (x3) to [out=-110, in=110] (x4); \draw (x4) to [out=70, in=290] (x3);
    \draw (a1) to (x5); \draw (x5) to (0.1,1.2); \draw (0.3,1.2) to (b1);
    \draw (a2) to (x6); \draw (x6) to (0.3,0.4); \draw (0.5,0.4) to (b2);
    \draw (b3) to (x7); \draw (x7) to (-0.3,-0.4); \draw (-0.5,-0.4) to (a3);
    \draw (b4) to (x8); \draw (x8) to (-0.1,-1.2); \draw (-0.3,-1.2) to (a4);
 %  \fill[black] (-0.85,0) circle (1.1 pt);
  \fill[black] (-0.9,0.1) circle (1 pt);
   \fill[black] (-0.9,-0.1) circle (1 pt);
  \fill[black] (-0.9,0.8) circle (1 pt);
   \fill[black] (-0.9,0.6) circle (1 pt);
     \fill[black] (-0.9,-0.6) circle (1 pt);
   \fill[black] (-0.9,-0.8) circle (1 pt);
     \fill[black] (0.9,0.1) circle (1 pt);
   \fill[black] (0.9,-0.1) circle (1 pt);
  \fill[black] (0.9,0.8) circle (1 pt);
   \fill[black] (0.9,0.6) circle (1 pt);
     \fill[black] (0.9,-0.6) circle (1 pt);
   \fill[black] (0.9,-0.8) circle (1 pt);
       \node[scale=1.]  at (-2.7,0) {$x_1$}; \node[scale=1.]  at (2.7,0) {$x_2$};
    \node[scale=1.]  at (0,2.6) {$x_3$}; \node[scale=1.]  at (0,-2.6) {$x_4$};
\end{tikzpicture}  
\fe
where each chiral Lagrangian has two lines attached to the two heavy operators, and two lines connected to other chiral Lagrangians or the light operators at $x_3$ and $x_4$. This is the only class of diagrams behaving as $\binom{h}{L} \sim (h)^L$ in the large-$h$ limit for a diagram with $L$ chiral Lagrangian insertions.  Again, such large-charge scaling reproduces the expected 't Hooft-like coupling $(h\, g^2_{_{\rm YM}})^L = \lambda^L$ at $L$ loops.
Similarly to the two-loop case, the distribution of the chiral Lagrangians on the two lines connecting $x_3$ and $x_4$ determines the relative symmetry factors of the diagrams. Interpreting this point from a graphical perspective, when the red dots in \eqref{eq:L-loops} are equally distributed between the two legs, the diagram has a relative symmetry factor of $\tfrac{1}{2}$.
Also, as in two-loop case, the whole class of diagrams depicted in \eqref{eq:L-loops} carries a unique colour factor $d_{\cH,N;L}$, that can be factored out as in \eqref{eq:factor_TI}. This will be shown in section~\ref{sec:colourfactor}. We can also see this fact by realising that the integrands represented in Fig. \eqref{eq:L-loops} arise from a single $f$-graph function, as we will show below. 

We now read off the class of Feynman integrands arising from diagrams like \eqref{eq:L-loops}, as in the two-loop case.
All the lines are scalar propagators, so the loop integration points $x_5,  \ldots, x_{L+4}$ cannot appear in the numerators of the $f$-graph functions (or they must be cancelled with the denominator of \eqref{eq:f-graph}). The same must happen for $x_3, x_4$ due to the $S_{L+2}$ symmetry that relates $x_3, x_4$ with the integration points. From these considerations, the $f$-graph  should generate the following Feynman integrands (and only these):
\ie \label{eq:large-L-loop}
I_{\rm ladder, \ell}^{(L)}(x_i) &\sim  {1 \over  \left( x_{35}^2 x_{4, \ell+4}^2 \prod_{i=5}^{\ell+3} x^2_{i,i+1} \right) \left( x_{3, \ell+5}^2 x_{4, L+4}^2 \prod_{i=\ell+5}^{L+3} x^2_{i,i+1} \right) \left(  \prod_{i=5}^{L+4} x^2_{1i} x^2_{2i}  \right) }  \, .
\fe
Similarly to the two-loop case, these integrands are products of two ladder integrals of $\ell$ and $L-\ell$ loops, for $\ell \leq L$. 

As we commented and similarly to the two-loop case, this class of integrands is associated with a unique $f$-graph function, after imposing the $S_2 \times S_{L+2}$ permutation symmetry. This leads to the conclusion that the loop integrand relevant for the $L$-loop contribution to the HHLL correlator must take the following form:
\ie \label{eq:large-c2}
f_{\rm ladder}^{(L)}(x_i) = {1 \over 2(L{+}2)} \frac{(x_{12}^2)^{L-2}}{ \prod_{i=3}^{L+4} x^2_{i,i+1} x^2_{1i} x^2_{2i}}\,+ S_{L+2} \, , 
\fe
where again $x_{L+5}:=x_3$ and the numerator $(x_{12}^2)^{L-2}$ ensures the correct conformal weights for $x_1, x_2$. Here we have summed over all the permutations of $x_3, x_4, \ldots, x_{L+4}$ and $1/(2(L{+}2))$  is the symmetry factor.  For $L=1$ and $L=2$, the expression \eqref{eq:large-c2} reduces to  the one-loop and two-loop result \eqref{eq:f-graph1} and \eqref{eq:2l-ladder2}, respectively. And more generally, it correctly gives rise to $I_{\rm ladder, \ell}^{(L)}(x_i)$ given in \eqref{eq:large-L-loop} for all possible choices of $\ell$, with the relative factor $1/2$ for $\ell=L/2$ and $L$ even, as we pointed out earlier. 

We may also express $f_{\rm ladder}^{(L)}(x_i)$ in the form of \eqref{eq:f-graph}, with the numerator 
\ie \label{eq:Pladder}
P_{\rm ladder}^{(L)}(x_{ij}^2) = \big(x_{12}^2\big)^{L-1} \widetilde{P}_{\rm ladder}^{(L)}(x_{ij}^2) \, , 
\fe
where $\tilde{P}_{\rm ladder}^{(L)}(x_{ij}^2)$ is a sum of products of $x_{ij}^2$ with $3\leq i, j \leq L+4$, and is given by 
\ie \label{eq:tPL}
\widetilde{P}_{\rm ladder}^{(L)}(x_{ij}^2)= \frac{1}{2(L{+}2)}   \left(\prod_{5 \leq j   \leq L+3} \prod_{3 \leq i \leq j-2 } x_{ij}^2 \right) \left(\prod_{4 \leq k \leq L+2} x_{k, L+4}^2 \right)  + S_{L+2} \, . 
\fe
Each $x^2_{ij}$ in $\widetilde{P}_{\rm ladder}^{(L)}$  maximally appears once, so that they cancel out the corresponding terms in the denominator of \eqref{eq:f-graph}, since only scalar Feynman integrals (\textit{i.e.} the loop integration points do not appear in the numerator)  can contribute to the HHLL correlators in the leading large-charge limit.  
 
In summary, we find that the $L$-loop contribution to the HHLL correlators at the large charge limit is given by a single combination of Feynman integrals defined in \eqref{eq:defcI} with the integrand $f_{\rm ladder}^{(L)}(x_i)$ given in \eqref{eq:large-c2}, namely:
\ie \label{eq:FladderL}
\cI^{(L)}(u, v) = {x^2_{12}x_{13}^4 x_{14}^2 x_{23}^2 x_{24}^4 x_{34}^2 \over (\pi^2)^L } \int d^4 x_5 \ldots d^4 x_{L+4}\, f^{(L)}_{\rm ladder} (x_i) \, .
\fe
As in the two loop case \eqref{eq:F2loop}, in \eqref{eq:FladderL} one can recognise a sum of particular combinations of the ladder integrals given in \eqref{eq:ladderL0} and \eqref{eq:ladderL}: 
\ie \label{eq:FL}
\mathcal{I}^{(L)}(u,v) =  {1 \over u} \sum_{\ell=0}^{L}   \Phi^{(\ell)}(u, v)  \Phi^{(L-\ell)}(u, v)   \, ,
\fe
where for $\Phi^{(0)}(u, v)$ it should be understood as $\Phi^{(0)}(u, v)=1$. 

In conclusion, considering the full loop expansion \eqref{eq:TH_genusexp} and plugging the above result in \eqref{eq:factor_TI}, we can write down in full generality the result for the reduced HHLL correlator at the leading genus zero in the large-charge 't Hooft limit: 
\ie \label{eq:THUV-fin}
\cT_{\cH}(u, v) =  \sum_{L=1}^{\infty} d_{\cH, N; L} \,  {(-a)^L \over u}\sum_{\ell=0}^{L}   \Phi^{(\ell)}(u, v)  \Phi^{(L-\ell)}(u, v)  \, , 
\fe
where we recall $a=\lambda/(4\pi^2) = \Delta_{\cH} \, g^2_{_{\rm YM}}/(4\pi^2)$. The dependence on the explicit form of the heavy operators $\cH$ lies only in the colour factor coefficients $d_{\cH, N; L}$. They will be determined in the next section where we will specify some explicit examples of heavy operators.

\subsubsection{Resummation properties for canonical heavy operators}
We add some considerations on the possible resummation properties of the result \eqref{eq:THUV-fin}, in order to make some contact with the $SU(2)$ result from \cite{Caetano:2023zwe} and to outline how resummation can be performed in the higher rank case for the ``canonical heavy operators", which we will introduce. 

We may recast the expression \eqref{eq:THUV-fin} in a more suggestive form, 
\ie \label{eq:THUV-fin2}
\cT_{\cH}(u, v) =  \sum_{L=1}^{\infty} d_{\cH, N; L} \, {a^L \over u }\Bigg(   \sum_{\ell=0}^{\infty} (-a)^\ell  \Phi^{(\ell)}(u, v) \Bigg)^2 \, \Bigg{\vert}_{a^L}  \, , 
\fe
where the subscript $a^L$ means taking the coefficient of $a^L$. Remarkably, the sum of the all-loop ladder integrals in the parenthesis of \eqref{eq:THUV-fin2} can be resummed \cite{Broadhurst:2010ds}. The finite coupling result after the resummation can be expressed in a variety of ways that are not particularly interesting for our goals. Instead, a particular representation of the resummation that will be later used for the conformal block expansion in the heavy-light channel is given by \cite{Fleury:2016ykk, Caetano:2023zwe}
\ie \label{eq:resum-form}
\sum_{\ell=0}^{\infty} (-a)^\ell  \Phi^{(\ell)}(u, v)= {u \over  \sqrt{v}} \sum_{r=1}^{\infty} {r \, e^{-\sigma \sqrt{r^2+4 a}} \over  \sqrt{r^2+4 a}} {\sin(r \varphi) \over \sin(\varphi)} \, ,
\fe
where $e^{i \varphi} = \sqrt{z/\bar{z}}$ and $e^{-\sigma} = \sqrt{z \bar{z}}$.

The resummed formula is particularly useful for the case where the colour factor coefficients $d_{\cH, N; L}$ take a particularly simple form. As we mentioned, $d_{\cH, N; L}$ depends on the choice of $\cH$, and in general is some complicated function of $N$ and $L$, as we will see in section~\ref{sec:colourfactor} and section~\ref{sec:localisation}. However, choosing a proper $\cH$ such that $d_{\cH, N; L}\sim c^L$, where $c$ is a general numerical constant, would allow to simply rescale the coupling $a$ and use the resummation formula \eqref{eq:resum-form} as follows
\ie \label{eq:THUV-fin-spe}
\cT_{\cH}(u, v) =   {\beta \over u } \left[ \Bigg(   \sum_{\ell=0}^{\infty} (-c\,a)^\ell  \, \Phi^{(\ell)}(u, v) \Bigg)^2 -1 \right] \, ,
\fe
where $\beta$ can be an irrelavant overall constant.
We will call the heavy operators that have coefficients $d_{\cH, N; L}\sim c^L$ as ``canonical heavy operators". 
Let us consider the simplest example.
In the case of $SU(2)$ gauge group, there is a unique choice for heavy operator $\cH=(\cO_2)^p$ with $p\rightarrow \infty$. In this case $d_{\cH, 2; L}=2^{2-L}$  \cite{Caetano:2023zwe}, so $(\cO_2)^p$ is a canonical heavy operator for $SU(2)$.  Due to the resummation, one may study the strong coupling limit of the HHLL correlators of the canonical heavy operators, as discussed in \cite{Caetano:2023zwe} for the $SU(2)$ gauge group. One of the interesting features is that in the strong coupling expansion, $\cT_{\cH}(u, v)$, as given in \eqref{eq:THUV-fin-spe}, is exponentially decayed. However, this is not a general feature for heavy operators whose colour factor coefficients are not in this form, and this property strongly depends on the rank of the gauge group. This has already been seen for the integrated HHLL correlators, in general they contain both power series and exponentially decayed terms in the strong coupling expansion \cite{Brown:2023why}.

\subsection{The OPE analysis}

After determining the general structure of the all-loop contribution to the HHLL correlators, we will now analyse the OPE properties of $\cT_{\cH}(u, v)$ and determine non-trivial CFT data from our results. 

\subsubsection{The $s$-channel OPE analysis}

A significant property of the $L$-loop integrand for the HHLL correlators is the factor $(x_{12}^2)^{L-1}$ in \eqref{eq:Pladder}, which is the highest power of $x_{12}^2$ the $f^{(L)}$-graph function could have. This property is closely related to the $s$-channel OPE limit, for which $x_1 \rightarrow x_2, x_3 \rightarrow x_4$, or $u\rightarrow 0, v \rightarrow 1$. The dominant non-BPS operator exchanged in the $s$-channel OPE is the Konishi operator $ \cK$, which requires the reduced correlator  $\cT_{\cH}(u, v)$ to behave as
\ie \label{eq:s-ope}
\cT_{\cH}(u, v) \sim \widetilde{C}_{\cH \cH \cK}\, \widetilde{C}_{\cO_2 \cO_2 \cK}\, u^{\gamma_{\cK}/2} \, , 
\fe
where $\gamma_{\cK}$ is the anomalous dimension of the Konishi operator, and $\widetilde{C}_{\cH \cH \cK}$ and $\widetilde{C}_{\cO_2 \cO_2 \cK}$ are the OPE coefficients. The anomalous dimension is given by \cite{Fiamberti:2007rj,Fiamberti:2008sh}
\ie \label{eq:Konishig}
\gamma_{\cK} = 3 a_t -3 a_t^2+ {21 \over 4} a_t^3+ \ldots \, ,
\fe
where $a_t$ is the usual 't Hooft coupling $a_t =N g_{_{\rm YM}}^2/(4\pi^2)$. Importantly, the formula \eqref{eq:Konishig} for the first three loops as shown is valid for general $N$ due to the absence of the non-planar contribution up to this order. The crucial point is that, when expressing \eqref{eq:Konishig} in terms of the large-charge `t Hooft coupling $\lambda =\Delta_{\cH}\, g_{_{\rm YM}}^2 $, the anomalous dimension is suppressed in the large-charge planar limit:
\ie
\gamma_{\cK} = \sum_{L=1}^{\infty} \Delta_{\cH}^{-L}  a^L  \gamma^{(L)}_{\cK} \, ,
\fe
with $a=\lambda/(4\pi^2)$. We then have 
\ie
u^{\gamma_{\cK}/2} = 1+ \frac{a }{\Delta_{\cH}}   \gamma_{\cK}^{(1)}  \log u +{a^2 \over 2\Delta_{\cH} ^2}   \left((\gamma_{\cK}^{(1)})^2 \log
   ^2(u)+2 \gamma_{\cK}^{(2)} \log
   (u)\right)  + \ldots \, ,
\fe
and $\gamma^{(L)}_{\cK}$ can be read off from \eqref{eq:Konishig}. For example, $\gamma^{(1)}_{\cK}=3N$, which in fact is the only term that is relevant for the following discussion.  

We now consider the OPE coefficients in \eqref{eq:s-ope}. For the same reason as the anomalous dimension we discussed above, we see that only the free theory part of $\widetilde{C}_{\cO_2 \cO_2 \cK}$ (denoted as $\widetilde{C}^{(0)}_{\cO_2 \cO_2 \cK}$), which is independent of the Yang-Mills coupling, is not suppressed in the large-charge limit. We can analyse the charge dependence of $\widetilde{C}_{\cH \cH \cK}$ by chiral Lagrangian insertions. Using an analysis similar to the one given in the previous section for the HHLL correlators, we find that it must behave as
\ie
\widetilde{C}_{\cH \cH \cK} = \Delta_{\cH} \widetilde{C}^{(0)}_{\cH \cH \cK}  + \sum_{L=1}^{\infty}  a^L \, \widetilde{C}^{(L)}_{\cH \cH \cK} + O(1/\Delta_{\cH})  \, . 
\fe
The same structure has also been obtained in \cite{Caetano:2023zwe} for the $SU(2)$ gauge group.

Putting everything together, the $s$-channel OPE limit given in \eqref{eq:s-ope} immediately implies that in the perturbative expansion, we have
\ie \label{eq:s-ope2}
\cT_{\cH}(u, v) \sim  {a \over 2}   \,     C^{(0)}_{\cH \cH \cK}  \gamma^{(1)}_{\cK}   \log(u) + \sum_{L=1}^{\infty}  a^L\,    C^{(L)}_{\cH \cH \cK} \, ,
\fe
 where we have omitted subleading terms in $1/\Delta_{\cH}$ and ${C}_{\cH \cH \cK}=\widetilde{C}_{\cH \cH \cK} \, \widetilde{C}^{(0)}_{\cO_2 \cO_2 \cK}$ (namely we reabsorb the tree level structure constant $\widetilde{C}^{(0)}_{\cO_2 \cO_2 \cK}$ into ${C}_{\cH \cH \cK}$ as defined in \eqref{eq:s-ope}). 
The expression \eqref{eq:s-ope2} shows that $\cT_{\cH}(u, v)$ is suppressed in the $s$-channel OPE, except at one loop (namely it is only singular at one loop order and is constant at higher loops). We will see that the expected large-charge $f$-graph functions with numerator $P^{(L)}_{\alpha}(x_{ij}^2)$ of the structure of \eqref{eq:Pladder} have precisely the required property due to its highest power in $x_{12}^2$, which is indeed suppressed in the limit of $x_1 \rightarrow x_2$ for $L>1$.  
It is worth mentioning that at low-loop orders the requirement of suppressing $s$-channel OPE is in fact enough to fully fix the loop integrands. For example, all possible two-loop integrands are listed in \eqref{eq:fgraphs_2loop}, and we see that only the one given in \eqref{eq:fgraphs_2loop_a} has the correct $s$-channel OPE properties.  

We now analyse explicitly the all-loop expression of the HHLL correlators $\cT_{\cH}(u,v)$ given in \eqref{eq:THUV-fin} in the $s$-channel OPE limit, in order to check its consistency with \eqref{eq:s-ope2} to all orders. Moreover, the analysis also enables us to directly deduce the OPE coefficients for two heavy operators and a Konishi operator, $C_{\mathcal{H} \mathcal{H} \mathcal{K}}$, at any loop orders.   In the limit $u\rightarrow 0, v\rightarrow 1$ (or equivalently $z \rightarrow 0, \bar{z} \rightarrow 0$) only the terms of $\ell=0$ and $\ell=L$ in the sum of \eqref{eq:FL} contribute, because all the other terms are proportional to $u$. To see this explicitly, the ladder integrals $\Phi^{(L)}(u, v)$ defined in \eqref{eq:ladderL} when $u\rightarrow 0, v\rightarrow 1$ behave as
\ie
\Phi^{(1)}(u, v) &\sim  u \left( 2 - \log(u) \right)\,  , 
\fe
for $L=1$, and 
\ie
\Phi^{(L)}(u, v) &\sim u \binom{2 L}{L} \zeta (2 L-1)  \, ,
\fe
for $L \geq 2$. Therefore, 
\ie \label{eq:THUV-fin3}
\cT_{\cH}(u, v) \sim  2a \, d_{\cH, N; 1} (\log(u)-2) +  2\sum_{L=2}^{\infty} \left(-a\right)^L \,  d_{\cH, N; L} \,    \binom{2L}{L} \zeta (2L-1)  \, .  
\fe
We see the expression \eqref{eq:THUV-fin3} for $\cT_{\cH}(u, v)$ in the $s$-channel OPE limit has indeed the required form \eqref{eq:s-ope2}. 

The comparison of  \eqref{eq:THUV-fin3} with \eqref{eq:s-ope2} also allows to read off the structure constants $C_{\cH \cH \cK}$. We find for  $L=0, 1$, 
\ie \label{eq:OPE-coff0}
C^{(0)}_{\cH \cH \cK} ={4 \over \gamma^{(1)}_\cK}  d_{\cH, N; 1}  =  {4 d_{\cH, N; 1} \over 3N} \, , \qquad
C^{(1)}_{\cH \cH \cK} =-4  d_{\cH, N; 1}   \, , 
\fe
where we have used $\gamma^{(1)}_\cK = 3N$. Analogously for $L \geq 2$ we get:
\ie \label{eq:OPE-coff}
C^{(L)}_{\cH \cH \cK} =   2\, d_{\cH, N; L}  \, (-1)^L     \binom{2L}{L} \zeta (2L-1)  \, .
\fe
In the next sections, we will fix the colour factor coefficients $d_{\cH, N; L}$ for various examples of heavy operators, which then fully determine the OPE coefficients $C_{\cH \cH \cK}$. 
For example, for the simplest $SU(2)$ gauge group, for which the heavy operator can only be $\cH=(\cO_2)^p$ with $p\rightarrow \infty$, we have $d_{\cH, 2; L}=2^{2-L}$ as we will see in the next section. The structure constant $C_{\cH \cH \cK}$ for the $SU(2)$ gauge group was also obtained in \cite{Caetano:2023zwe}.\footnote{The structure constant $C_{\cH \cH \cK}$ in \cite{Caetano:2023zwe} is divided by the two-point function of the Konishi operators, which we do not include here.} 

\subsubsection{The $t$-channel OPE analysis}

We now move on to consider the $t$-channel OPE expansion, for which the OPE is between heavy and light operators that leads to the following conformal block expansion of the correlators, 
\ie
\cT_{\cH}(u, v) = \sum_{\Delta, S} \sum_i \Big| C_{\cO_2 \, \cH \, \cO^{(i)}_{\Delta, S}} \Big|^2 \, \cG_{\Delta, S}(u, v)\, , 
\fe
where $\{\Delta, S\}$ are the dimension and the spin of the operators that are exchanged in the OPE expansion and the index $i$ denotes any possible degeneracy (for example, in the zero coupling, it is easy to see the existence of the degeneracy for a generic $SU(N)$ gauge group). 
Importantly, the four-dimensional conformal blocks of the heavy-light OPE are known to simplify to a Gegenbauer polynomial $C_S^{(1)}(\cos(\varphi))$  \cite{Jafferis:2017zna}, leading to
\ie 
\cG_{\Delta, S}(u, v) =  e^{-\sigma(\Delta - \Delta_{\cH})} {\sin (S+1)\varphi \over \sin \varphi }  \, ,
\fe
where $e^{i \varphi} = \sqrt{z/\bar{z}}$ and $e^{-\sigma} = \sqrt{z \bar{z}}$ and we have omitted terms that are subleading in the large charge. Hence, the HHLL correlators can be expanded as
\ie \label{eq:Tus-t}
\cT_{\cH}(u, v) = \sum_{\Delta, S} \sum_i |C_{\cO_2 \, \cH \, \cO^{(i)}_{\Delta, S}} |^2  e^{-\sigma(\Delta - \Delta_{\cH})} {\sin (S+1)\varphi \over \sin \varphi } \, . 
\fe 
The $\varphi$-dependence appearing in this particular form is indeed manifest for $\cT_{\cH}(u, v)$ when expressed in the form \eqref{eq:THUV-fin2}. Using the resummed formula \eqref{eq:resum-form}, it was shown in \cite{Caetano:2023zwe},\footnote{We have corrected the summation range on $r$ in the reference \cite{Caetano:2023zwe}; the summation dummy variable was denoted as $\bold{a}$ in the reference, which sums up to $\left \lceil {S+1 \over 2 } \right \rceil$ rather than $S+1$ as in \eqref{eq:resummation}.} 
\ie \label{eq:resummation}
{1 \over u}  \Bigg(  \sum_{\ell=0}^{\infty} (-a)^\ell  \Phi^{(\ell)}(u, v) \Bigg)^2 =\sum_{S=0}^{\infty} \sum_{n=0}^{\infty} \sum^{S+1}_{r=1} C_{S, n, r}  \, e^{-\sigma (\Delta_{S, n, r}-\Delta_{\cH})} {\sin(S+1) \varphi \over \sin \varphi }\, , 
\fe
where the coefficient $C_{S, n, r}$ and the dimension $\Delta_{S, n, r}$ are given by
\ie
C_{S, n, r} &= {(r+n)(S+2+n-r) \over \sqrt{(r+n)^2+4a} \sqrt{(S+2+n-r)^2+4a}} \, , \cr 
\Delta_{S, n, r} &= \Delta_{\cH} +\sqrt{(r+n)^2+4a} + \sqrt{(S+2+n-r)^2+4a} \, .
\fe
We therefore should match
\ie \label{eq:pertur}
\left( \sum_{\Delta} \sum_i |C_{\cO_2 \, \cH \, \cO^{(i)}_{\Delta, S}} |^2  e^{-\sigma \, \Delta } \right) \Bigg \vert_{a^L} = d_{\cH, N; L} \left( \sum_{n=0}^{\infty} \sum^{S+1}_{r=1} C_{S, n, r}  \, e^{-\sigma \, \Delta_{S, n, r} }  \right) \Bigg  \vert_{a^L} \, .
 \fe
As we commented earlier, in the case of the canonical heavy operators, where  $d_{\cH, N; L} \sim c^L$ for some constant $c$, the resummation formula \eqref{eq:resummation} also applies to the HHLL correlators to all loops, by simply rescaling the coupling constant $a$ by a factor of $c$.  For such cases, we can then immediately identify the OPE coefficient as  $C_{S, n, r}$ for any coupling $a$, as was done in \cite{Caetano:2023zwe} for $SU(2)$ (again in this case $d_{\cH, 2; L}=2^{2-L}$). More generally, for the non-canonical heavy operators, we may solve the above relation \eqref{eq:pertur} perturbatively, as indicated in \eqref{eq:pertur}. We leave the detailed analysis as a future direction.

\section{Determining the colour factors}
\label{sec:colourfactor}

In this section, we determine the colour factor coefficients $d_{\cH,N;L}$ of the class of diagrams represented in \eqref{eq:L-loops} that in the previous section were shown to dominate in the large-charge limit. Using the $f$-graph functions (and the $S_2 \times S_{2+L}$ symmetry), we have seen in section~\ref{sec:integrand} that each $L$-loop order all the diagrams \eqref{eq:L-loops} enjoy a unique colour factor $d_{\cH,N;L}$. Here we will show this important fact by the direct computation of the colour factors associated with these diagrams. Then, when explicitly computing $d_{\cH,N;L}$ as function of $N$ for specific examples of $\cH$, we show that computing the colour factors for general $SU(N)$ makes it clear that the leading order in the charge corresponds to the minimal power in $N$ for any loop order; hence the leading charge selects the class of Feynman diagrams inserted in the maximally non-planar way in the colour space.  

This phenomenon appears to be a very general feature of the large-charge limit of gauge theories. Indeed, taking the leading order in large charge maximises the number of graph edges, and then requires the genus of the Riemann surface associated to each Feynman diagram to be maximal at each perturbative order. Such behaviour has already been verified in various classes of $\cN=2$ SCFTs for the large charge behaviour of extremal correlators \cite{Beccaria:2020azj}, and we are seeing the same phenomenon for four-point correlators in $\mathcal{N}=4$ SYM. %It would be very interesting to further explore this idea for other theories and observables.

To compute the colour factors, the $\vev{\cH \cH \cO_2\cO_2}$ four-point function is defined by the colour tensors of the light operator $\cO_2$ and the heavy operator $\cH$:
\begin{equation}\label{eq:def_color1}
\begin{split}
  \cO_2 (x_i) = \frac{1}{2} \delta^{c_1c_2} (\varphi_{c_1} \varphi_{c_2})(x_i)~,~~~~~~ \cH(x_j) = X_{\cH}^{a_1 \dots a_{h}} (\varphi_{a_1} \dots \varphi_{a_{h}})(x_j)~,
  \end{split}
\end{equation}
where ${\varphi} = \phi \cdot Y$ and $X_{\cH}$ represents the general colour tensor associated with the heavy operator with dimension $\Delta_\cH = h$. The colour tensor of $\cO_2$ simply descends from the normalisation of $SU(N)$ generators as in \eqref{eq:trTT}, whereas $X_{\cH}$ in general depends on the multitrace structure of the heavy operator. The chiral Lagrangian insertion method allows us to construct loop integrands from free-theory tree-level diagrams. All the tree diagrams must be closed in a connected way using free propagators:
\begin{equation}
    \vev{\varphi^a (x_1)\, \varphi^b (x_2)} = \delta^{ab} d_{12} \, . 
\end{equation}
In this section we concentrate on the colour part, so each free propagator is associated with a colour delta function.
The colour tensor for $\cO_2$ is also just a delta function, hence it plays no role in determining the colour factor of the whole correlator, especially in the large-charge limit. Therefore, the tree-level colour factor relevant for this discussion is given by the Wick contraction of the colour tensors from the heavy operators:
\begin{equation}\label{eq:tree_contraction}
    \cG_{\textrm{free}} (N) = X_{\cH} \cdot X_{\cH}~,
\end{equation}
which will be the normalisation for the colour factor at any loop level. The computation of $\cG_{\textrm{free}} (N)$ for a given operator $\cH$ is a purely combinatorial problem, depending on $N$ and the multitrace structure of $\cH$.  

\subsection{Universal color factors at any loop order}\label{sec:colour_1}

After this setup, we are ready to outline the general procedure for computing the coefficients $d_{\cH, N; L}$ appearing in the HHLL correlators as in \eqref{eq:THUV-fin}. In particular we show with a simple argument the crucial property that all the diagrams at a given loop order $L$, \textit{i.e.} the class represented in \eqref{eq:L-loops}, enjoy a unique colour factor $d_{\cH, N; L}$.

We emphasised that the quantum corrections to the HHLL correlators in the large-charge limit arise from the four-scalar vertices inside the chiral Lagrangian definition \eqref{eq:otaudef}, therefore each chiral Lagrangian insertion comes with a colour factor
\begin{equation}\label{eq:color_ff_vertex}
 \gym^2 f^{a b e}\, f^{c d e}~.
\end{equation}
Hence the colour tensor at $L$ loops comes from combining $L$ colour vertices \eqref{eq:color_ff_vertex} to be contracted with the colour tensors \eqref{eq:def_color1} via the delta functions from tree level propagators.

For the class of Feynman diagrams that are leading in the large charge limit, see eq. \eqref{eq:L-loops}, the chiral Lagrangian insertions define a chain iteration of the vertices \eqref{eq:color_ff_vertex}, to be closed with the color tensors of the light operators $\cO_2(x_3)$ and $\cO_2(x_4)$. Since these colour tensors are simple delta functions (equivalent to any tree level propagator), see \eqref{eq:def_color1}, all the diagrams in \eqref{eq:L-loops} enjoy the same colour factor. Thus, after closing these iterated vertices with the delta functions defining the light operators, we simply obtain a closed loop defining a colour trace over the adjoint representation of $SU(N)$ (see \eqref{eq:adjTr_example} for some low-$L$ examples):
\ie \label{eq:Tr_adj}
\begin{tikzpicture}[baseline={([yshift=-0.8ex]current bounding box.center)}]
    \coordinate (x5) at (-0.25,1.2);
    \coordinate (x6) at (-0.38,0.4);
    \coordinate (x7) at (0.38,-0.4);
    \coordinate (x8) at (0.25,-1.2);
    \coordinate (a1) at (-0.8,1.2);
    \coordinate (a2) at (-0.8,0.4);
    \coordinate (a3) at (-0.8,-0.4);
    \coordinate (a4) at (-0.8,-1.2);
    \coordinate (b1) at (0.8,1.2);
    \coordinate (b2) at (0.8,0.4);
    \coordinate (b3) at (0.8,-0.4);
    \coordinate (b4) at (0.8,-1.2);
    \coordinate (x4) at (0, -2);
    \coordinate (x3) at (0, 2);
    
    % Draw the red dots
    \fill[blue] (x3) circle (2pt);
    \fill[blue] (x4) circle (2pt);
    \fill[red] (x5) circle (3pt);
    \fill[red] (x6) circle (3pt);
    \fill[red] (x7) circle (3pt);
    \fill[red] (x8) circle (3pt);

    \draw (x3) to [out=-110, in=110] (x4); \draw (x4) to [out=70, in=290] (x3);
    \draw (a1) to (x5); \draw (x5) to (0.1,1.2); \draw (0.3,1.2) to (b1);
    \draw (a2) to (x6); \draw (x6) to (0.3,0.4); \draw (0.5,0.4) to (b2);
    \draw (b3) to (x7); \draw (x7) to (-0.3,-0.4); \draw (-0.5,-0.4) to (a3);
    \draw (b4) to (x8); \draw (x8) to (-0.1,-1.2); \draw (-0.3,-1.2) to (a4);
  %  \draw[dotted] (-1.1,-0.2) to (-1.1,-0.6); \draw[dotted] (1.1,-0.2) to (1.1,-0.6);
  
      \node at (-1.1,1.2) {$a_1$}; \node at (1.1,1.2) {$b_1$};
      \node at (-1.1,0.4) {$a_2$}; \node at (1.1,0.4) {$b_2$};
      \node at (-1.1,-1.2) {$a_L$}; \node at (1.1,-1.2) {$b_L$};
      %\node at (-1.1,-0.4) {$a_3$}; \node at (1.1,-0.4) {$b_3$};
       \fill[black] (-1.1,-0.2) circle (0.8pt);
       \fill[black] (-1.1,-0.4) circle (0.8pt);
       \fill[black] (-1.1,-0.6) circle (0.8pt);
         \fill[black] (1.1,-0.2) circle (0.8pt);
       \fill[black] (1.1,-0.4) circle (0.8pt);
       \fill[black] (1.1,-0.6) circle (0.8pt);
     % \node at (0,2.5) {$x_3$}; \node at (0,-2.5) {$x_4$};
\end{tikzpicture} \,\,\,\,  = \, \Tr_{\textrm{Adj}} \left(  T^{a_1} T^{b_1} \dots T^{a_L} T^{b_L}\right) \, . 
\fe
This ensures that the colour factor is obtained from a unique trace over the matter representation of the theory, which is the adjoint representation in the case of $\cN=4$ SYM. Notice that this is the same argument as \cite{Beccaria:2020azj} for two-point extremal correlators in $\cN=2$ SCFTs. In that case, the matter content allows for a more general representation $\cR$ of the gauge group due to lower supersymmetry. As argued in the previous section, the diagrams \eqref{eq:Tr_adj} dominate in the large-charge 't Hooft-like limit. 

We then need to insert the subdiagram displayed in \eqref{eq:Tr_adj} in the four-point function, whose tree level in the colour space is defined by \eqref{eq:tree_contraction}. To do so, we conveniently open $2L$ colour indices in $X_\cH \cdot X_\cH$, to be contracted with \eqref{eq:Tr_adj} in a fully symmetrised way.  
In general, we can then define the partially contracted colour tensor of order $m$ as
\begin{equation} \label{eq:Ym}
    Y_{(m)}^{a_1\dots a_m \, b_1\dots b_m} = X_\cH^{a_1\dots a_m c_{m+1}\dots c_{2p}} X_\cH^{b_1\dots b_m c_{m+1}\dots c_{2p}}~,
\end{equation}
where repeated indices are summed over. $Y_{(L)}$ is then a colour tensor with $2L$ open indices that can then be contracted with the colour factor due to the insertion of the chiral Lagrangians, given by \eqref{eq:Tr_adj}. In general $Y_{(L)}$ carries some $h$-dependent combinatorial factors, taking care of the multitrace structure of the heavy operators $\cH$. Following the argument around \eqref{eq:L-loops} in order to recover the large-charge 't Hooft coupling $\lambda=h\, \gym^2$ at $L$-loop we select the part of $Y_{(L)}$ that scales as $h^L$, schematically $Y_{(L)} \sim h^L Y_{(L),\, \rm max}+O(h^{L-1})$. We will further clarify this point when applying it on specific examples, see eqs. \eqref{eq:Y2} and \eqref{eq:Yell}.

Putting everything together, the $L$-loop colour factor $d_{\cH, N; L}$ at leading order in the large-charge limit is computed via the following general formula\footnote{The factors of 2 in the denominator arise from the definition of the large charge coupling $a$ in \eqref{eq:TH_genusexp}.}:
\begin{equation}\label{eq:color_general}
    d_{\cH, N; L} = \frac{1}{2^{2L-1}} \frac{1}{X_\cH \cdot X_\cH} \Tr_{\textrm{Adj}} \left(  T^{a_1} T^{b_1} \dots T^{a_L} T^{b_L}\right)  Y_{(L), \, \rm max}^{(a_1\dots a_L \, b_1\dots b_L)} ~,
\end{equation}
where the bracket $( \ldots )$ applied on the indices of $Y_{(L)}$ denotes symmetrisation summing over all independent permutations, and we have normalised the result by the two-point function \eqref{eq:tree_contraction}. 

The next step is to finally choose the heavy operators $\cH$ and explicitly compute the colour factors $ d_{\cH, N; L}$ as functions of $N$. To this aim, we need to discuss the different classes of heavy operators with dimension $\Delta_\cH$.

\subsection{Classes of heavy operators}\label{sec:heavy_choice}

The large-charge 't Hooft scaling \eqref{eq:doublescaling_def2} imposes $\Delta_\cH$ as the largest parameter in the theory, and in particular $\Delta_\cH\gg N^2$. Hence $\cH$ are necessarily multi-trace operators. Indeed, when $N$ is fixed, any $\tr \Phi^M$ for $M\geq N$ can be recast in a combination of multitrace objects $\tr \Phi^{m_1} \dots \tr \Phi^{m_\ell}$ such that $m_i\leq N$. For example, when $N=2$, the single-trace operator with dimension four is related to a double-trace operator via the relation $\cO_4 = 1/2 (\cO_2)^2$. For large conformal dimension $\Delta_\cH$, there is clearly a huge degeneracy in the choice of the possible heavy operators, which then need to be properly organised.

In general, it is convenient to organise the half-BPS superconformal primary operators in terms of the following tower of operators \cite{Gerchkovitz:2016gxx,Brown:2023cpz, Brown:2023why}
\begin{equation}
    \label{eq:def-tower}
\cO^{(i)}_{p|M} = (\cO_2)^p \cO^{(i)}_{0|M}\, ,
\end{equation}
where $(\cO_2)^p$ denotes the multi-trace operator containing $p$ copies of $\cO_2$, defined in \eqref{O20prime}, and $\cO^{(i)}_{0|M}$ is a linear combination of  half-BPS operators of dimension $M$, and the index $i$ denotes possible degeneracy. 

Let us discuss some general properties of the operators in \eqref{eq:def-tower}. There are several reasons why it is convenient to explicitly separate out the operator $\cO_2$. Being part of the stress tensor (as well as the Lagrangian) supermultiplet, $\cO_2$ enjoys several special properties. For example, in supersymmetric localisation computations for the integrated correlators, the insertion of $\cO_2$'s can be generated simply by taking derivatives of $\tau$'s and $\bar\tau$'s  \cite{Gerchkovitz:2016gxx}.  The construction of $\cO^{(i)}_{0|M}$ is recursive, and is reviewed in appendix~\ref{sec:OpM}. The important point, as shown in the definition of $\cO^{(i)}_{0|M}$ given in \eqref{eq:defOm}, is that the starting operator of the recursive definition (denoted as $B^{(i)}_M$) excludes the operator $\cO_2$.

Because $\cO^{(i)}_{0|M}$ is of dimension $M$, then the operator $\cO^{(i)}_{p|M}$ has dimension $2p+M$. The large-charge limit can be achieved by taking $p\rightarrow \infty$ in  $\mathcal{O}^{(i)}_{p|M}$ with $M$ being fixed, or $M \rightarrow \infty$ with $p$ being fixed, or more generally both being large. 
The limit of $p\rightarrow \infty$ with $M$ being fixed is the main case we will consider in this paper. In this case we treat $\cO_2$ specially, which leads to simplifications. In particular, the colour factors $d_{p|M\,,N;L}$ in this case can be written in closed forms for different choices of $M$. 
Moreover, this is also the limit that has been studied in \cite{Brown:2023why} in the context of integrated correlators.  As we will review in section~\ref{sec:localisation}, the integrated correlators obey Laplace difference equations \cite{Paul:2022piq, Brown:2023why}, which allow one to recursively relate integrated correlators for $\cO^{(i)}_{p|M}$ with different values of $p$. Therefore the properties of the integrated correlators at large $p$ (with fixed $M$) are essentially governed by the Laplace difference equations. 

Another large-charge limit to consider is to take  $M \rightarrow \infty$ with $p$ being fixed in $\cO^{(i)}_{p|M}$. As we commented earlier, when $M$ is large there is a huge degeneracy in defining the operators $\cO^{(i)}_{0|M}$. We will leave a general study of the heavy operators of this type for future work. 
However, this task simplifies when restricting to the gauge group $SU(N)$ with some specific values of $N$. For instance, for the $SU(2)$ gauge group the only possible heavy operator is $\cH = (\cO_2)^p = \cO_{p|0}$, which is what was studied in  \cite{Caetano:2023zwe}. 
This can also be understood by the fact that for $SU(2)$, $\cO^{(i)}_{0|M}$ can only be the identity operator (or $M=0$). 
For the $SU(3)$ gauge group the situation is more interesting. All the operators must be built out of $\cO_2$ and $\cO_3$. In this case, $\cO^{(i)}_{0|M}$ is in fact unique for a given $M$ and the explicit form is given in \eqref{eq:O03p}.  We will consider  this type of heavy operators in more detail in section \ref{sec:canonical2} using results of integrated correlators.

We now compute the colour coefficients using the general formula \eqref{eq:color_general} for some explicit examples of the heavy operators, choosing in particular $\cH=\cO^{(i)}_{p|M}$ with $p\rightarrow \infty$ and $M$ fixed. We will denote the colour factor coefficients as $d^{(i)}_{p|M, N; L}$, which in general are some rational functions of $N$ and $L$.   

\subsection{Example: maximal multitrace operator}\label{sec4:exampleO2}
We first consider the simplest operator in the class of heavy operators of \eqref{eq:def-tower}, namely $\cO_{p|0} = (\cO_2)^p$. This operator has conformal dimension $\Delta_\cH = 2p$ and it is usually referred to as maximal multi-trace operator. Due to the trace normalization \eqref{eq:trTT}, its colour tensor can be written as a symmetrised combination of delta functions: 
\begin{equation}\label{eq:def_color_Op}
    X_{\cH}^{a_1 \dots a_{2p}} = \frac{1}{2^{p}} \delta^{(a_1a_2} \dots \delta^{a_{2p-1}a_{2p})}~.
\end{equation}
As we mentioned in the previous subsection, this operator is special that the contraction of such a colour tensor with tree level propagators as defined in \eqref{eq:tree_contraction} can be written as a closed form for any $p$ and any $N$ \cite{Gerchkovitz:2016gxx,Bourget:2018obm,Beccaria:2018xxl}:
\begin{equation}\label{eq:Gtree}
    \cG_{\mathrm{free}}(N) =  p!  \frac{\Gamma(\alpha+p)}{\Gamma(\alpha)}~, \quad {\rm with} \quad \alpha= \frac{N^2-1}{2}~.
\end{equation}
We now perform the contraction with the $L$-loop colour factor as described in \eqref{eq:color_general} using $Y_{(L)}$ defined in \eqref{eq:Ym}. %This can be achieved by partially contracting the two $X_\cH$ tensors while leaving $m$ pairs of indices free, with $m\leq L$. We can then define the partially contracted colour tensor of order $m$ as
%\begin{equation}
 %   Y_{(m)}^{a_1\dots a_m \, b_1\dots b_m} = X_\cH^{a_1\dots a_m c_{m+1}\dots c_{2p}} X_\cH^{b_1\dots b_m c_{m+1}\dots c_{2p}}~,
%\end{equation}
%where repeated indices are summed over. 
At one loop, we need a unique partially contracted tensor, which reads:
\begin{equation}\label{eq:Y1}
    Y_{(1)}^{a_1 b_1} = \binom{p}{1}\, p! \;\frac{\Gamma \left(\alpha+p\right)}{\Gamma \left(\alpha+1\right)}\,\delta^{a_1 b_1}~,
\end{equation}
where the $p$-dependent binomial factor comes from opening the pair of indices from one out of $p$ traces, and the $N$-dependent factor can be derived analogously to the free theory result \eqref{eq:Gtree}. 
The full one-loop colour factor then comes from the contraction with a single Lagrangian insertion, following the picture from \eqref{eq:one-loop}. In particular, the $\binom{p}{1}$ term precisely corresponds to the combinatorial factor necessary to realise the proper large charge scaling. Indeed, the scaling in $p$ combines with the coupling $\gym^2$ to create the large charge 't Hooft coupling $\lambda=2p \gym^2$, and after factoring out the two-point function normalisation \eqref{eq:Gtree} we find the following normalised colour factor at one-loop order: 
\begin{equation}
    d_{p|0,\,N;1} = \frac{1}{2} \frac{\Gamma (\alpha)}{\Gamma (\alpha+1)}\Tr_{\rm Adj} (T^aT^a) = N~,
\end{equation}
where we have used some basic properties of colour tensors, see \eqref{eq:adjTr_example}.  This coefficient represents the $L=1$ result of the general expression \eqref{eq:color_general} for the choice $\cH=\cO_{p|0}$.
This unique colour factor trivially matches the expectations coming from Feynman diagram computations.

\begin{figure}[t]
    \centering
    \begin{tikzpicture}
      % Define the coordinates for the three vertical points
    \coordinate (x5) at (0,0.6);
    \coordinate (x6) at (0,0.2);
    \coordinate (x4) at (0, -2);
    \coordinate (x3) at (0, 2);
    
    % Define the coordinates for the left and right points
    \coordinate (x1a) at (-2, 1.6);
     \coordinate (x1b) at (-2, 0.4);
     \coordinate (x1c) at (-2, -0.4);
     \coordinate (x1d) at (-2, -1.6);
    \coordinate (x2a) at (2, 1.6);
     \coordinate (x2b) at (2, 0.4);
     \coordinate (x2c) at (2, -0.4);
     \coordinate (x2d) at (2, -1.6);

    % Draw the red dots
    \fill[blue] (x3) circle (2pt);
    \fill[blue] (x4) circle (2pt);
    \fill[red] (x5) circle (3pt);
    \fill[red] (x6) circle (3pt);
    \fill[blue] (x1a) circle (2pt);
     \fill[blue] (x1b) circle (2pt);
      \fill[blue] (x1c) circle (2pt);
       \fill[blue] (x1d) circle (2pt);
       \fill[blue] (x2a) circle (2pt);
     \fill[blue] (x2b) circle (2pt);
      \fill[blue] (x2c) circle (2pt);
       \fill[blue] (x2d) circle (2pt);

        \fill[black] (-1.8,-1.2) circle (0.8pt);
       \fill[black] (-1.8,-1) circle (0.8pt);
       \fill[black] (-1.8,-0.8) circle (0.8pt);
        \fill[black] (1.8,-1.2) circle (0.8pt);
       \fill[black] (1.8,-1) circle (0.8pt);
       \fill[black] (1.8,-0.8) circle (0.8pt);
        \fill[black] (1.8,1.2) circle (0.8pt);
       \fill[black] (1.8,1) circle (0.8pt);
       \fill[black] (1.8,0.8) circle (0.8pt);
        \fill[black] (-1.8,1.2) circle (0.8pt);
       \fill[black] (-1.8,1) circle (0.8pt);
       \fill[black] (-1.8,0.8) circle (0.8pt);

    \draw (x1a) to[out=90, in=20] (-1.6,1.8);
     \draw (x1a) to[out=-90, in=-20] (-1.6,1.4);
      \draw (x1b) to[out=90, in=20] (-1.6,0.6);
     \draw (x1b) to[out=-90, in=-20] (-1.6,0.2);
     \draw (x1c) to[out=90, in=20] (-1.6,-0.2);
     \draw (x1c) to[out=-90, in=-20] (-1.6,-0.6);
     \draw (x1d) to[out=90, in=20] (-1.6,-1.4);
     \draw (x1d) to[out=-90, in=-20] (-1.6,-1.8);

    \draw (x2a) to[out=90, in=160] (1.6,1.8);
     \draw (x2a) to[out=-90, in=-160] (1.6,1.4);
    \draw (x2b) to[out=90, in=160] (1.6,0.6);
     \draw (x2b) to[out=-90, in=-160] (1.6,0.2);
     \draw (x2c) to[out=90, in=160] (1.6,-0.2);
     \draw (x2c) to[out=-90, in=-160] (1.6,-0.6);
     \draw (x2d) to[out=90, in=160] (1.6,-1.4);
     \draw (x2d) to[out=-90, in=-160] (1.6,-1.8);
     
    \draw (-1.1,0.6) to (x5);
    \draw (-1.1,0.2) to (x6);
    \draw (1.1,0.6) to (x5);
    \draw (1.1,0.2) to (x6);
    \draw (x5) to (x3);
    \draw (x6) to (x4);
    \draw (x5) to (x6);

    \draw (x3) to[out=-70, in=90] (0.6,2/3+0.05);
    \draw (0.6,2/3-0.6) to (0.6,-2/3);
    \draw (0.6,-2/3) to[out=-90, in=70] (x4);

   \node[scale=0.8] at (-1.1,0.8) {$a_1$};
   \node[scale=0.8] at (-1.1,0.) {$a_2$};
   \node[scale=0.8] at (1.1,0.8) {$b_1$};
   \node[scale=0.8] at (1.1,0.) {$b_2$};
   \node[scale=1] at (0,-2.8) {$(a)$};
\end{tikzpicture}
\hspace{2cm}
    \begin{tikzpicture}
      % Define the coordinates for the three vertical points
    \coordinate (x5) at (0,0.6);
    \coordinate (x6) at (0,-0.2);
    \coordinate (x4) at (0, -2);
    \coordinate (x3) at (0, 2);
    
    % Define the coordinates for the left and right points
    \coordinate (x1a) at (-2, 1.6);
     \coordinate (x1b) at (-2, 0.4);
     \coordinate (x1c) at (-2, -0.4);
     \coordinate (x1d) at (-2, -1.6);
    \coordinate (x2a) at (2, 1.6);
     \coordinate (x2b) at (2, 0.4);
     \coordinate (x2c) at (2, -0.4);
     \coordinate (x2d) at (2, -1.6);

    % Draw the red dots
    \fill[blue] (x3) circle (2pt);
    \fill[blue] (x4) circle (2pt);
    \fill[red] (x5) circle (3pt);
    \fill[red] (x6) circle (3pt);
    \fill[blue] (x1a) circle (2pt);
     \fill[blue] (x1b) circle (2pt);
      \fill[blue] (x1c) circle (2pt);
       \fill[blue] (x1d) circle (2pt);
       \fill[blue] (x2a) circle (2pt);
     \fill[blue] (x2b) circle (2pt);
      \fill[blue] (x2c) circle (2pt);
       \fill[blue] (x2d) circle (2pt);

        \fill[black] (-1.8,-1.2) circle (0.8pt);
       \fill[black] (-1.8,-1) circle (0.8pt);
       \fill[black] (-1.8,-0.8) circle (0.8pt);
        \fill[black] (1.8,-1.2) circle (0.8pt);
       \fill[black] (1.8,-1) circle (0.8pt);
       \fill[black] (1.8,-0.8) circle (0.8pt);
        \fill[black] (1.8,1.2) circle (0.8pt);
       \fill[black] (1.8,1) circle (0.8pt);
       \fill[black] (1.8,0.8) circle (0.8pt);
        \fill[black] (-1.8,1.2) circle (0.8pt);
       \fill[black] (-1.8,1) circle (0.8pt);
       \fill[black] (-1.8,0.8) circle (0.8pt);

    \draw (x1a) to[out=90, in=20] (-1.6,1.8);
     \draw (x1a) to[out=-90, in=-20] (-1.6,1.4);
      \draw (x1b) to[out=90, in=20] (-1.6,0.6);
     \draw (x1b) to[out=-90, in=-20] (-1.6,0.2);
     \draw (x1c) to[out=90, in=20] (-1.6,-0.2);
     \draw (x1c) to[out=-90, in=-20] (-1.6,-0.6);
     \draw (x1d) to[out=90, in=20] (-1.6,-1.4);
     \draw (x1d) to[out=-90, in=-20] (-1.6,-1.8);

    \draw (x2a) to[out=90, in=160] (1.6,1.8);
     \draw (x2a) to[out=-90, in=-160] (1.6,1.4);
    \draw (x2b) to[out=90, in=160] (1.6,0.6);
     \draw (x2b) to[out=-90, in=-160] (1.6,0.2);
     \draw (x2c) to[out=90, in=160] (1.6,-0.2);
     \draw (x2c) to[out=-90, in=-160] (1.6,-0.6);
     \draw (x2d) to[out=90, in=160] (1.6,-1.4);
     \draw (x2d) to[out=-90, in=-160] (1.6,-1.8);
     
    \draw (-1.1,0.6) to (x5);
    \draw (-1.1,-0.2) to (x6);
    \draw (1.1,0.6) to (x5);
    \draw (1.1,-0.2) to (x6);
    \draw (x5) to (x3);
    \draw (x6) to (x4);
    \draw (x5) to (x6);

    \draw (x3) to[out=-70, in=90] (0.6,2/3+0.05);
    \draw (0.6,2/3-0.2) to (0.6,-2/3+0.6);
    \draw (0.6,-2/3+0.35) to (0.6,-2/3);
    \draw (0.6,-2/3) to[out=-90, in=70] (x4);

   \node[scale=0.8] at (-1.1,0.8) {$a_1$};
   \node[scale=0.8] at (-1.1,-0.4) {$a_2$};
   \node[scale=0.8] at (1.1,0.8) {$b_1$};
   \node[scale=0.8] at (1.1,-0.4) {$b_2$};
   \node[scale=1] at (0,-2.8) {$(b)$};
\end{tikzpicture}
    \caption{Two-loop colour factor and the two ways to insert it in the diagram as described by \eqref{eq:Y2}. In $(a)$ the insertion comes with a $\binom{p}{1}$ combinatorics and maximises the scaling in $N$. In $(b)$ the combinatorial factor is $\binom{p}{2}$ which maximises the scaling in large charge and minimises the power in $N$.}
    \label{fig:2loop_choices}
\end{figure}
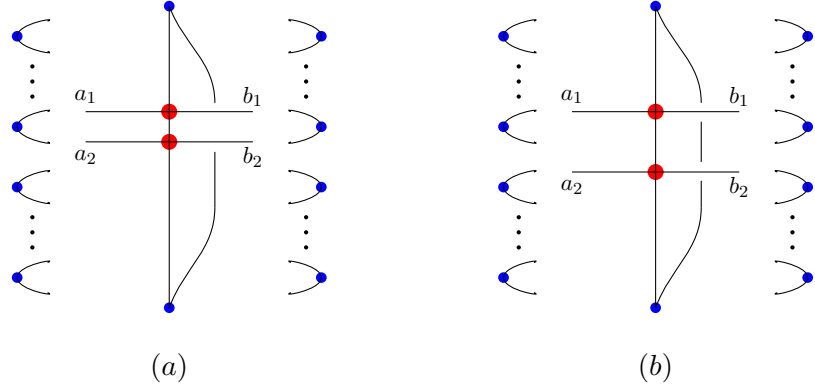

We now move to the two-loop order, and we consider only the class of diagrams that contributes at leading order in large charge limit, according to the discussion around Figure~\ref{fig:twoloops}. Such diagrams have four open indices, %to be contracted with either those in \eqref{eq:Y2_1} or in \eqref{eq:Y2_2}
so we also need a partially contracted colour tensor with four open indices. For combinatorial reasons one must separate the case where the four open indices come all from a single pair of traces (with a $\binom{p}{1}$ combinatorial factor) or from two pairs of traces (with a $\binom{p}{2}$ combinatorial factor):
\begin{subequations} \label{eq:Y2}
    \begin{align}
        Y_{(2)}^{a_1a_2b_1b_2} &= \frac{1}{2} \binom{p}{1}\, p! \;\frac{\Gamma \left(\alpha+p\right)}{\Gamma \left(\alpha+1\right)}\,\delta^{a_1 a_2} \delta^{b_1 b_2} \label{eq:Y2_1} \\  &~~~+ \binom{p}{2}\, p! \;\frac{\Gamma \left(\alpha+p\right)}{\Gamma \left(\alpha+2\right)}\,\left(\delta^{a_1 a_2} \delta^{b_1 b_2}+\delta^{a_1 b_1} \delta^{a_2 b_2}+\delta^{a_1 b_2} \delta^{b_1 a_2}\right)~.\label{eq:Y2_2}
    \end{align}
\end{subequations}
The two possibilities are depicted in Figure \ref{fig:2loop_choices}.
The combinatorial binomial factor shows that the contribution in \eqref{eq:Y2_2} is the leading term at the large-charge limit, corresponding to the contribution which is spread among the largest number of traces. Also, from the example in Figure~\ref{fig:2loop_choices} we see that the distribution among the maximal number of traces minimises the closed loops in the colour space, so that it minimises the power of $N$ in the colour factor.  The double chiral Lagrangian insertion brings the expected colour factor
\begin{equation}
    f^{a_1 c d} f^{b_1 d e} f^{a_2 e f} f^{b_2 f c} = \Tr_{\text{Adj}} (T^{a_1}T^{b_1} T^{a_2}T^{b_2})~,
\end{equation}
to be contracted with \eqref{eq:Y2_2}.
After factorizing away the free theory normalisation \eqref{eq:Gtree} the total two-loop colour factor at leading large charge order is given by
\begin{equation}
     d_{p|0,\,N;2} =\frac{1}{8} \frac{\Gamma(\alpha)}{\Gamma(\alpha+2)} \Tr_{\text{Adj}} (T^{(a_1}T^{a_1} T^{a_2}T^{a_2)})~,
\end{equation}
where the repeated indices are contracted and the adjoint trace can be computed using standard $SU(N)$ techniques, see \eqref{eq:Adj_traces_explicit}.

The generalization to higher orders is rather straightforward. At $L$ loops, we consider the partially contracted colour tensor with $2L$ open indices, which contains several terms with different combinatorial factors, in analogy with \eqref{eq:Y2}, depending on how many traces get involved in the contraction. We only keep the part that maximises the combinatorial factor:
\begin{equation}
    Y_{(L)}^{a_1 \dots a_L b_1 \dots b_L} = \binom{p}{L}\, p! \;\frac{\Gamma \left(\alpha+p\right)}{\Gamma \left(\alpha+L\right)}\, \delta^{(a_1a_2}\dots \delta^{a_{L-1}a_L}\delta^{b_1b_2}\dots \delta^{b_{L-1}b_L)} + \ldots~,\label{eq:Yell}
\end{equation}
where we have omitted the lower order terms in the large-$p$ limit, since this part of the colour tensor reproduces the correct $p^L$ scaling. For the same reasons of Figure~\ref{fig:2loop_choices}, spreading the insertion among the maximal number of traces in order to maximises the charge has also the effect of minimising the power of $N$ in the colour factor.
The contraction of the colour tensor \eqref{eq:Tr_adj} coming from $L$ chiral Lagrangians with \eqref{eq:Yell} gives a colour factor:
\begin{equation}\label{eq:Tadj}
    \Tr_{\text{Adj}} (T^{(a_1}T^{a_1} \dots T^{a_L}T^{a_L)})~,
\end{equation}
so the general $L$-loop colour factor normalised by the two-point function \eqref{eq:Gtree} is
\begin{equation}\label{eq:color_Lloop}
    d_{p|0,\,N;L} = \frac{1}{2^{2L-1}} \frac{\Gamma(\alpha)}{\Gamma(\alpha+L)} \Tr_{\text{Adj}} (T^{(a_1}T^{a_1} \dots T^{a_L}T^{a_L)}) ~,
\end{equation}
where the trace is totally symmetrised.
The colour coefficients \eqref{eq:color_Lloop} can then be explicitly evaluated using the recursive formulas shown in appendix \ref{app:color}, and they are rational functions of $N$. Some examples for low values of $L$ are shown as follows: 
\begin{equation} \label{eq:colour-factor1}
   \begin{split}
        &\hspace{1cm}d_{p|0,\,N;1} = N ~,~~~ d_{p|0,\,N;2} = \frac{5 N^2}{2^2 (N^2+1)}~,~~~d_{p|0,\,N;3} = \frac{35 N^3}{2^4 \left(N^2+1\right)(N^2+3)}~, \\
        &d_{p|0,\,N;4} = \frac{21 N^2 \left(7 N^2+2\right)}{2^5 \left(N^2+1\right) \left(N^2+3\right) \left(N^2+5\right)}~,~~~d_{p|0,\,N;5} =\frac{693 N^3}{2^6 \left(N^2+3\right) \left(N^2+5\right) \left(N^2+7\right)}~.
   \end{split}
\end{equation}
Notably, when $N=2$, the above expressions for the coefficients simplify dramatically, which lead to $d_{p|0,\,N; L}=2^{2-L}$, as we mentioned earlier. 
As we will show in the next section, the above results agree with those obtained from the integrated HHLL correlators using supersymmetric localisation.

\subsection{Other types of heavy operators}\label{sec4:exampleOMO2}

We can extend the analysis beyond $\cO_{p|0}=(\cO_2)^p$ to the cases when $ \cH=\cO^{(i)}_{p|M}$ as defined in appendix~\ref{sec:OpM} and references \cite{Brown:2023cpz, Brown:2023why}, with $p\to\infty$ the dimension $M$ of $\cO^{(i)}_{0|M}$ being fixed. This class of operators has conformal dimension $\Delta_\cH=2p+M$, and its colour tensor is a generalisation of \eqref{eq:def_color_Op}:
\begin{equation}\label{eq:def_color_OMp}
    X_{p|M}^{a_1 \dots a_{2p+M}} = \frac{1}{2^{p}} \delta^{(a_1a_2} \dots \delta^{a_{2p-1}a_{2p}} X_M^{a_{2p+1} \dots a_{2p+M})}~,
\end{equation}
where $X_M$ are the colour tensors generically defining the operators $\cO^{(i)}_{0|M}$ described in appendix~\ref{sec:OpM},  see for example:
\begin{equation}\label{eq:X34tensors}
    X_3^{a_1a_2a_3} = \tr T^{(a_1}T^{a_2}T^{a_3)}~,~~~~X_4^{a_1a_2a_3a4} = \tr T^{(a_1}T^{a_2}T^{a_3}T^{a_4)}-\frac{2N^2-3}{N(N^2+1)}\tr T^{(a_1}T^{a_2}\tr T^{a_3}T^{a_4)}~.
\end{equation}
Similarly to $\cO_{p|0}$ the problem of fully contracting two $\cO^{(i)}_{p|M}$ with tree-level propagators has a closed-form solution for any $p$ and $N$, which simply generalizes \eqref{eq:Gtree}:
\begin{equation}\label{eq:OpM_tree}
    \cG_{\rm free} (N) = X_M \cdot X_M~ p!~ \frac{\Gamma(\alpha+p+M)}{\Gamma(\alpha + M)}~,
\end{equation}
where $\alpha$ is defined in \eqref{eq:Gtree} and $X_M \cdot X_M$ is the free theory Wick contraction for $\cO^{(i)}_{0|M}$ operators. We compute some examples of $X_M \cdot X_M$ in appendix~\ref{app:color}, see eq. \eqref{eq:X34computation}. 

The quantum corrections to the four-point function are computed analogously to the $\cO_{p|0}$ case. In particular, the steps to select the leading large charge contributions follow the same argument, which is independent of the choice of $\cO^{(i)}_{0|M}$ since we are keeping $M$ fixed.
In order to maximise $p$ at $L$-loop order, the multiple Lagrangian insertions must be spread over the maximal number of traces within the $(\cO_2)^p$ part, by selecting only the contributions with a $\binom{p}{L}$ combinatorial factor. The general $L$-loop colour factor can be computed as: 
\begin{equation}\label{eq:OpM_color_l}
     d_{p|M,\,N;L} = \frac{1}{2^{2L-1}} \frac{1}{X_M\cdot X_M}  \frac{\Gamma(\alpha+M)}{\Gamma(\alpha+M+L)} \Tr_{\text{Adj}} (T^{(a_1}T^{a_1} \dots T^{a_L}T^{a_L}) X_M^{a_{L+1}\dots a_{L+M}} X_M^{a_{L+1}\dots a_{L+M})} ~,
\end{equation}
where again the colour tensors are fully symmetrised and fully contracted. As before, these colour contractions are performed using the techniques from appendix~\ref{app:color}. Let us write explicitly some examples for some $M$ and some $L$: 
\begin{subequations}
    \begin{align}
    &d_{p|M,\,N;1} = N~, \\
    &d_{p|3,\,N;2} = \frac{N^2 \left(5 N^2+67\right)}{2^2 \left(N^2+5\right) \left(N^2+7\right)}~,~~~~~~~~~~ d_{p|3,\,N;3} = \frac{35 N^3 \left(N^2+23\right)}{2^4 \left(N^2+5\right) \left(N^2+7\right) \left(N^2+9\right)}~, \\
    &d_{p|4,\,N;2} = \frac{ \left(5 N^6+96 N^4-5 N^2+144\right)}{2^2 \left(N^2+1\right) \left(N^2+7\right) \left(N^2+9\right)} ~,~~~ d_{p|4,\,N;3} = \frac{5 N \left(7 N^6+240 N^4-367 N^2+1080\right)}{2^4 \left(N^2+1\right) \left(N^2+7\right) \left(N^2+9\right) \left(N^2+11\right)}~, \\
    &d_{p|5,\,N;2} = \frac{5 \left(N^6+28 N^4+43 N^2+144\right)}{2^2 \left(N^2+5\right) \left(N^2+9\right) \left(N^2+11\right)} ~,~~~ d_{p|5,\,N;3} = \frac{5 N\! \left(7 N^6+328 N^4-515 N^2+5220\right)}{2^4\left(N^2+5\right) \left(N^2+9\right) \left(N^2+11\right) \left(N^2+13\right)}~.
\end{align}
\end{subequations}
%\begin{align}
%    &d_{p|M,\,N;1} =2 N~, \\
 %   &d_{p|3,\,N;2} = \frac{2 N^2 \left(5 N^2+67\right)}{\left(N^2+5\right) \left(N^2+7\right)}~,~~~ d_{p|3,\,N;3} = \frac{70 N^3 \left(N^2+23\right)}{\left(N^2+5\right) \left(N^2+7\right) \left(N^2+9\right)}~, \\
 %   &d_{p|4,\,N;2} = \frac{2 \left(5 N^6+96 N^4-5 N^2+144\right)}{\left(N^2+1\right) \left(N^2+7\right) \left(N^2+9\right)} ~,~~~ d_{p|4,\,N;3} = \frac{10 N \left(7 N^6+240 N^4-367 N^2+1080\right)}{\left(N^2+1\right) \left(N^2+7\right) \left(N^2+9\right) \left(N^2+11\right)}~, \\
  %  &d_{p|5,\,N;2} = \frac{10 \left(N^6+28 N^4+43 N^2+144\right)}{\left(N^2+5\right) \left(N^2+9\right) \left(N^2+11\right)} ~,~~~ d_{p|5,\,N;3} = \frac{10 N \left(7 N^6+328 N^4-515 N^2+5220\right)}{\left(N^2+5\right) \left(N^2+9\right) \left(N^2+11\right) \left(N^2+13\right)}~.
%\end{align}
We have also checked that this idea holds for higher dimensional operators (with $M\geq 6$), where there is a degeneracy, as outlined in \eqref{eq:SPO_dim6}. The colour factors for the first loop levels in these cases read:
\begin{subequations}
\begin{align}
    &d^{(1)}_{p|6,\,N;2} = \frac{5 N^8+214 N^6+1225 N^4+1724 N^2+2880}{2^2\left(N^2+11\right) \left(N^2+13\right) \left(N^4+15 N^2+8\right)}  ~,\\ &d^{(1)}_{p|6,\,N;3} = \frac{5 N \left(7 N^8+458 N^6+623 N^4+11872 N^2+19296\right)}{2^4\left(N^2+11\right) \left(N^2+13\right) \left(N^2+15\right) \left(N^4+15 N^2+8\right)}~, \\
     &d^{(2)}_{p|6,\,N;2} = \frac{\left(5 N^{10}+249 N^8+4163 N^6+20091 N^4+54980 N^2+10368\right)}{2^2\left(N^2+7\right) \left(N^2+11\right) \left(N^2+13\right) \left(N^4+15 N^2+8\right)} ~, \\&d^{(2)}_{p|6,\,N; 3} = \frac{5 N\! \left(7 N^{12}+504 N^{10}+12286 N^8+93840 N^6+491779 N^4+532560 N^2-43488\right)}{2^4\left(N^2+3\right) \left(N^2+7\right) \left(N^2+11\right) \left(N^2+13\right) \left(N^2+15\right) \left(N^4+15 N^2+8\right)}~.
\end{align}
\end{subequations}
The above results once again agree with the localisation computation, which we will discuss in the following section and some results are listed in the appendix \ref{app:integrated}. 

\section{Integrated HHLL correlators} \label{sec:localisation}

It has been shown that the HHLL correlators, when integrating out the spacetime dependence, can be computed exactly as functions of the complexified coupling constant $\tau$ \cite{Paul:2023rka, Brown:2023why, Brown:2023cpz}. The integrated correlators in general are defined as \cite{Binder:2019jwn}  
\begin{equation} \label{eq:def-integation}
\cG(\tau,\bar\tau)=I_2\left[\cT(u,v)\right]=-{2\over \pi} \int_0^{\infty} dr \int^{\pi}_0 d\theta {r^3 \sin^2 \theta \over u \, v} \cT(u,v)~,
\end{equation}
where $v=r^2, u=1-2r \cos(\theta) r +r^2$. The integrated correlators in $\mathcal{N}=4$ SYM can be shown to be related to the partition function of $\cN=2^*$ SYM on $S^4$, which can be computed using supersymmetric localisation \cite{Pestun:2007rz}, and many exact results have been obtained for integrated correlators in the literature (see \cite{Dorigoni:2022iem} for a review and an introduction to this subject). For the purpose of this paper, we are interested in the large-charge 't Hooft limit of the integrated HHLLL correlators, to compare with the all-loop expression \eqref{eq:THUV-fin}.  

From the Feynman integral point of view, the  integrated  correlators in perturbative expansion can be most conveniently viewed as the periods of the $f$-graph Feynman integrals \cite{Wen:2022oky, Brown:2023zbr, Zhang:2024ypu}, using the fact 
\ie \label{eq:def-integation2}
    I_2\left[x_{13}^2 x_{24}^2 F^{(L)}(x_i)\right]
    &=- { 1 \over 2(\pi^2)^L} \int {d^4x_1 \ldots d^4x_{L+4} \over {\rm vol(SO(1,5))}} \sum_{\alpha} d_{\cH, N; L}^{(\alpha)}\, f_{\alpha}^{(L)}(x_i)\cr &:=- { 1 \over 2}\, \sum_{\alpha}d_{\cH, N; L}^{(\alpha)}\, \cP_{f_{\alpha}^{(L)}}\, .
\fe
For the HHLL correlators, only $f_{\rm ladder}^{(L)}(x_i)$ given in \eqref{eq:large-c2} contributes, and the period for each term in $f_{\rm ladder}^{(L)}(x_i)$ is the same (due to the permutation symmetry after the integration). The result of each period is known for any loop $L$, and it is given by \cite{Belokurov:1983km, Dorigoni:2021guq}
\ie
\binom{2 L+2}{L+1}\, \zeta (2 L+1) \, . 
\fe
Therefore, the integrated HHLL correlators using the expression \eqref{eq:THUV-fin} can be written as
\ie \label{eq:periods}
\cG_\cH (\lambda) = I_2\left[\cT_{\cH}(u, v)\right] = -{1\over 2}\sum_{L=1}^{\infty} d_{\cH, N; L}\, (-a)^L\,  (L+1) \binom{2 L+2}{L+1} \zeta (2 L+1) \, . 
\fe
Hence, computing the integrated HHLL correlators from supersymmetric localisation also allows us to deduce the coefficients $d_{\cH, N; L}$, which we computed directly from Feynman diagrams in the previous section. We will see that we find perfect agreement. 

\subsection{Integrated HHLL correlators from supersymmetric localisation}

As we commented above, the integrated correlators can be computed by supersymmetric localisation. Here we will only briefly summarise the relevant results. It has been shown in \cite{Brown:2023cpz} that, for general $N$ and $p$, the integrated four-point functions $\langle \cH \cH \cO_2 \cO_2 \rangle$ with $\cH=\cO_{p|M}^{(i)}:=(\cO_2)^p\cO_{0|M}^{(i)}$ obey the following Laplace difference equation,\footnote{More general integrated four-point functions $\langle \cH \cH' \cO_2 \cO_2 \rangle$, where $\cH$ is different from $\cH'$ have also been studied in \cite{Brown:2023cpz, Brown:2023why} and obey a similar Laplace difference equation, which generalises the equation \eqref{eq:laplace}.} 
\ie \label{eq:laplace}
\Delta_{\tau} \cG^{(i)}_{p|M}(\tau,\bar{\tau}) &= \kappa_{p+1}^{(N,M)} \bigg(\cG^{(i)}_{p+1|M}(\tau,\bar{\tau}) -\cG^{(i)}_{p|M}(\tau,\bar{\tau})\bigg) - \kappa_{p}^{(N,M)} \bigg(\cG^{(i)}_{p|M}(\tau,\bar{\tau}) -\cG^{(i)}_{p-1|M}(\tau,\bar{\tau})\bigg) \cr
&+ \frac{N^2-1}{2} \cG_{1|0}(\tau,\bar{\tau}) \, ,
\fe
where we have denoted the integrated correlator as $\cG^{(i)}_{p|M}(\tau,\bar{\tau})$ and the laplacian $\Delta_{\tau}$ is defined as $\Delta_{\tau} = 4\tau_2^2 \partial_\tau \partial_{\bar{\tau}}$,  and
\ie
\kappa_p^{(N,M)} = p \left(p+\frac{N^2+2M-3}{2} \right) \, . 
\fe
Furthermore, the source term $\cG_{1|0}(\tau,\bar{\tau})$ in the Laplace-difference equation denotes the integrated correlator of  $\langle \cO_2  \cO_2  \cO_2  \cO_2 \rangle$, whose expression is known for any $N$ and $\tau$ \cite{Dorigoni:2021guq, Dorigoni:2022cua}. The Laplace difference equation is a recursion relation: once the initial condition, \textit{i.e.} $\cG^{(i)}_{0|M}(\tau,\bar{\tau})$, is given, we may then solve $\cG^{(i)}_{p|M}(\tau,\bar{\tau})$ recursively.

For the relevance of our discussion, we focus on only the perturbative terms, for which the solution to the recursive relation  \eqref{eq:laplace} can be written as \cite{Paul:2023rka} \footnote{The extension for the full $SL(2, \mathbb{Z})$-invariant expression is also known, which can be written in terms of the non-holomorphic Eisenstein series.} 
\begin{align} \label{eq:sol-rec}
\cG^{(i)}_{p|M}(\tau_2) &= \sum_{s=2}^\infty 4(-1)^s \tau_2^{1-s}\, \pi^{\frac{1}{2}-s} (2s-1) \, \Gamma(s+\frac{1}{2} ) \,\zeta(2s-1) \cr
&\bigg[ N(N-1) \, _3F_2(2-N,s,1-s;3,2;1) \big( \,_3F_2(-p,s,1-s;1,\frac{N^2+2M-1}{2};1) -1 \big) \cr
&+2s(s-1)g^{(i)}_M(N,s)\,_3F_2(-p,s,1-s;1,\frac{N^2+2M-1}{2};1) \bigg] \, ,
\end{align}
where $\tau_2 =4\pi/g_{_{\rm YM}}^2$, and the initial data is contained in $g^{(i)}_M(N,s)$. More explicitly, $g_0(N,s) = 0$ for the operator $\cH=(\cO_2)^p$, and for general  $\cH=(\cO_2)^p \cO^{(i)}_{0|M}$, $g^{(i)}_M(N,s)$ is related to the initial condition $\cG^{(i)}_{0|M}(\tau_2)$ by
\ie
\cG^{(i)}_{0|M}(\tau_2) = \sum_{s=2}^\infty 8 (-1)^s \pi^{\frac{1}{2}-s} s (s-1) (2s-1) \tau_2^{1-s} \Gamma(s+\frac{1}{2}) \zeta(2s-1) g^{(i)}_M(N,s) \, . 
\fe
We are interested in  the large-$p$ limit of the expression \eqref{eq:sol-rec} with fixed $a =\lambda/(4\pi^2)=p\, g_{_{\rm YM}}^2/(2\pi^2)$. Considering the large-charge planar limit and using $\tau_2 =2p/(\pi\, a)$, we find
\ie \label{eq:localisation}
\cG^{(i)}_{p|M}(\lambda) &= -\sum_{L=1}^\infty  (-a)^{L} \zeta(2L+1)   \frac{2^{L+3} \, \Gamma(L+\frac{3}{2})^2 \, \Gamma(\frac{N^2+2M-1}{2})}{\pi \, \Gamma(L+1)\, \Gamma(\frac{N^2+2M-1}{2} {+} L)} \cr
&\big[N(N-1) \, _3F_2(2-N,L+1,-L;3,2;1)+ 2L(L+1)\, g^{(i)}_M(N,L+1) \big] + \ldots \, ,
\fe
where we have omitted terms that are subleading in the large-$p$ limit. Comparing the localisation result 
\eqref{eq:localisation} and the result from our all-loop expression \eqref{eq:periods}, we immediately obtain
\ie\label{c_localisation_prediction}
d_{p|M, N; L} &= \frac{ \Gamma(L+\frac{3}{2}) \Gamma(\frac{N^2+2M-1}{2})}{2^{L-2}\sqrt{\pi}  \Gamma(\frac{N^2+2M-1}{2}+L)} \cr
& \big[N(N-1) \, _3F_2(2-N,L+1,-L;3,2;1)+ 2L(L+1) g^{(i)}_M(N,L+1) \big] \, . 
\fe
%\ie\label{c_localisation_prediction}
%c_{\cO^{(i)}_{p|M}, N; L} &= \frac{4(-1)^{L} (L+1) \Gamma(L+\frac{3}{2}) \Gamma(\frac{N^2+2M-1}{2})}{\sqrt{\pi} \Gamma(\frac{N^2+2M-1}{2}+L)} \cr
%& \big[N(N-1) \, _3F_2(2-N,L+1,-L;3,2;1)+ 2L(L+1) g^{(i)}_M(N,L+1) \big] \, . 
%\fe
We will consider explicit examples of heavy operators and provide formulas for the initial conditions, $g^{(i)}_M(N,L+1)$. 

\subsubsection{Examples of heavy operators and colour factors}

Let us now specify the heavy operators and consider some concrete examples of $\cO^{(i)}_{p|M}$. For the simplest case, when $M=0$, we have $\cH=(\cO_2)^p$. In this case, $g_0(N,s) =0$, and so we have
\ie \label{eq:coffO2p0}
d_{p|0, N; L} =     \frac{ N\, \Gamma(\frac{N^2+1}{2}) \, \Gamma(L+\frac{3}{2})}{2^{L-3} \sqrt{\pi}\, (N+1)\, \Gamma(\frac{N^2-1}{2}+L)}  \, _3F_2(2-N,L+1,-L; 3,2; 1) \, . 
\fe
The result matches exactly the explicit colour factors computed from Feynman diagrams as given in \eqref{eq:colour-factor1}. For example, for $N=2,3,4$, we have   
\ie \label{eq:coffO2p}
d_{p|0, 2; L} &= 2^{2-L} \, , \qquad  
d_{p|0, 3; L}=\frac{3 \left(L^2+L+6\right) \Gamma
   \left(L+\frac{3}{2}\right)}{ 2^{L-3} \sqrt{\pi }\, \Gamma (L+4)} \, , \cr
d_{p|0, 4; L} &=   \frac{45045 \left(L^2+L+4\right)
   \left(L^2+L+18\right) \Gamma
   \left(L+\frac{3}{2}\right)}{2^{L+6} \Gamma
   \left(L+\frac{15}{2}\right)} \, . 
\fe
As we commented in the previous sections, the colour factor coefficient of the $SU(2)$ gauge group, $d_{p|0, 2; L}$, is very different from others: it has the ``canonical form", $d_{\cH, N; L} \sim c^L$ (with $c=1/2$).  However, this is not a generic feature. As we see in the examples given in \eqref{eq:coffO2p}, for the same heavy operator $\cH =(\cO_2)^p$, the colour factor coefficients are not in the form of $c^L$ for the cases with $N>2$. 

Another example of the heavy operators we would like to consider is $\cH =\cO_{p|3} :=(\cO_2)^p \cO_{0|3}$ (namely $M=3$), with $p\rightarrow \infty$. In this case, the initial condition is given by \cite{Paul:2023rka}
\begin{align} \nonumber
g_3(N,s) = \frac{N}{84 (N {+} 1) (N {+} 2)}\Bigg[ &6  \left(5 N  (s^2-s+2 )-8 (s-4) (s+3)\right) \, _3F_2(3-N,s,1-s;4,4;1) \\ 
&+14 (N-3)
   \left(s^2-s+18\right) \, _3F_2(4-N,s,1-s;4,4;1) \cr
   &-(N-3) (s-4)^2 (s+3)^2 \,
   _3F_2(4-N,s,1-s;5,5;1)\Bigg]\, .
\end{align}
Using the above formula for the initial condition, the colour factor coefficient for this case $d_{p|3, N; L}$ can then be obtained from \eqref{c_localisation_prediction}. For example, for $SU(3)$ and $SU(4)$, we have, 
\ie
d_{p|3,3;L} &=\frac{9 \left(3 L^4+6 L^3+77 L^2+74 L+120\right) \Gamma \left(L+\frac{3}{2}\right)}{ 2^{L-4} \sqrt{\pi } \, \Gamma (L+7)} \, ,  \cr 
d_{p|3,4;L} &= \frac{14549535 \left(L^6+3 L^5+64 L^4+123 L^3+655 L^2+594 L+1080\right) \Gamma
   \left(L+\frac{3}{2}\right)}{\ 2^{L+9}\, \Gamma \left(L+\frac{21}{2}\right)} \, .
\fe
Similarly, using \eqref{c_localisation_prediction}, we can obtain colour factor coefficients from the integrated correlators for the heavy operator $\cH =\cO_{p|3}=(\cO_2)^p \cO_{0|3}$ with other gauge groups, as well as for other heavy operators $\cO^{(i)}_{p|M}$ for different values of $M$. Some examples of such integrated correlators in the perturbative expansion can be found in \cite{Brown:2023cpz} and in appendix \ref{app:integrated}. 

We remark that  the colour factor coefficients we compute here using the known results of integrated HHLL correlators perfectly agree with the results obtained from the direct diagrammatic computation in section \ref{sec4:exampleOMO2}. This gives a consistency check on the results.

\subsubsection{Examples of canonical heavy operators} \label{sec:canonical2}

We see from the previous discussion that the colour factor coefficient $d_{p|0, 2; L}$ in \eqref{eq:coffO2p} for $\cH =\cO_{p|0}=(\cO_2)^p$ of the $SU(2)$ gauge group takes a very special form, $d_{p|0, 2; L} =2^{2-L}$. It is therefore a prototype of the canonical heavy operators. We will consider another example of the canonical heavy operators beyond this simplest case of $SU(2)$ gauge group. 

As we already discussed previously in section \ref{sec:heavy_choice},  the heavy operators $\cO^{(i)}_{p|M}$ with $p\rightarrow \infty$ and $M$ being fixed are not the most general form of the heavy operators. One may of course also consider $\cO^{(i)}_{p|M}$ with $p$ fixed but $M$ taken to be large. In general, such a limit is much more difficult to study. One of the complications comes from the construction of $\cO^{(i)}_{0|M}$, see appendix~\ref{app:integrated} for more details.
Even though there is an explicit procedure as shown in \eqref{eq:defOm}, which in principle allows one to recursively determine all $\cO^{(i)}_{0|M}$ (a few examples of small values of $M$ are given in \eqref{eq:example_SPOs} and \eqref{eq:SPO_dim6}), if we are interested in some large values of $M$ (especially $M \rightarrow \infty$), it is in general complicated. Taking $p\rightarrow \infty$ is relatively simpler also because the Laplace-difference equation \eqref{eq:laplace} relates the integrated correlators with different values of $p$, which allows to solve the $p$ dependence analytically as we have shown previously.

To simplify the discussion, as an example we will consider the $SU(3)$ gauge group. As we will see, this simple example already leads to some very interesting new features. The restriction to $SU(3)$ gauge group means that all the half BPS primary operators we consider are built out of $\cO_2$ and $\cO_3$. Therefore, $\cO_2$ and $\cO_3$ form a two-dimensional space, and unlike $SU(2)$, there are in principle infinite number of ways ({\it i.e.} any direction in this two-dimensional space) of taking the operators to be heavy. However, we also note that $\cO^{(i)}_{0|M}$ is still unique in this case, that is because $B_M^{(i)}$ in \eqref{eq:defOm} must be $(\cO_3)^m$ with $3m=M$ for $SU(3)$ gauge group. For $M=6$ for example, we can only have $\cO^{(2)}_{0|6}$ as given in \eqref{eq:SPO_dim6}, with $\cO_4 = 1/2 \, (\cO_{2})^2$ for $SU(3)$ gauge group, therefore it becomes
\ie \label{O06}
\cO_{0|6} = \cO_{3,3}- {1\over 24} \cO_{2,2,2}\, , 
\fe
where we have removed the degeneracy index. Below are a few more examples of higher dimensions
\ie \label{eq:O09}
\cO_{0|9} =& \, \cO_{3,3,3}- {1\over 12} \cO_{3,2,2,2} \, , \qquad  \cO_{0|12} = \cO_{3,3,3,3}- {1\over 8} \cO_{3,3,2,2,2} + {1 \over 576} \cO_{2,2,2,2,2,2} \, .
\fe
In general we will denote this class of operators as $\cO_{0|3m}$ and they can be expressed generically as
\ie \label{eq:O03p}
\cO_{0|3m} = \sum_{n } \frac{(-1)^{n_1}\, (n+1)_{\frac{n_1}{3}}}{24^{\frac{n_1}{3}} (\frac{n_1}{3})!}  (\cO_3)^n (\cO_2)^{n_1} \, , \qquad {\rm with} \quad n_1= {3(m-n) \over 2}\, ,
\fe
where the summation range over $n$ is given as $\{m, m-2, m-4, \ldots \}$. 

We now consider the heavy operator to be $\cO_{0|3m}$ with $m\rightarrow \infty$. This is a particular direction of taking $\cO^{(i)}_{p|M}$ to be heavy.  To obtain the colour factor coefficients, we compute the integrated correlators using the matrix model of supersymmetric localisation, following \cite{Paul:2023rka, Brown:2023why, Brown:2023cpz}. We find that the integrated $\langle \cO_{0|3m}\,\cO_{0|3m}\,\cO_2\,\cO_2 \rangle$ at $L$ loops can be expressed in terms of a degree-$L$ polynomial of $m$, as long as $m$ is not too small (more precisely, the polynomial ansatz is valid for $m >  L/3-1 $ at $L$ loops). Therefore it does not affect extracting the large-$m$ property. This allows us to straightforwardly determine the large-$m$ behaviour of the integrated correlators. For the first few loops, we have
\ie \label{eq:3mtau}
\cG_{0|3m}^{SU(3)}(\tau_2) =&\, \frac{54 m \zeta(3)}{\pi \tau_2}-\frac{405(m^2+3m)\zeta(5)}{2\pi^2 \tau_2^2}+\frac{315(9m^3+45m^2+74m-2)\zeta(7)}{4\pi^3 \tau_2^3} \cr
&-\frac{2835(27m^4+198m^3+537m^2+638m-32)\zeta(9)}{32\pi^4 \tau_2^4} + \dots \, . 
\fe
To the orders we list above, the polynomial ansatz is in fact valid for any $m>0$. 
From the expression given in \eqref{eq:3mtau}, we can easily obtain the the large-charge 't Hooft limit ({\it i.e.} $m \rightarrow \infty$ with $\lambda = 3m \, \gym^2$ fixed) of the integrated correlator, which is given as 
\ie
\cG_{0|3m}^{SU(3)}(\lambda) = 18 a \, \zeta(3)-\frac{45}{2} a^2 \zeta(5)+\frac{105}{4} a^3 \zeta(7) -\frac{945}{32} a^4 \zeta(9) + \dots \, ,
\fe
where $a=\lambda /(4 \pi^2) = 3m/ (\pi \tau_2)$. Now, by comparing with \eqref{eq:periods}, we find that the colour factors for the heavy operator $\cH= \cO_{0|3m}$ (with $m\rightarrow \infty$) are given by
\ie
d_{0|3m , \, 3; 1}  =3\, , \quad  d_{0|3m ,\, 3; 2}  =\frac{3}{2^2}\, , \quad d_{0|3m\,,  3; 3}  =\frac{3}{2^4}\, , \quad
d_{0|3m,\, 3; 4}  =\frac{3}{2^6}\, , \,\, \dots \, . 
\fe
The above result shows an obvious pattern for the colour factor coefficients to any loop orders,  
\ie \label{eq:cO3p}
d_{0|3m,\, 3; L} = \frac{3}{2^{2L-2}} \, .
\fe
We have verified this pattern up to $12$ loops. We see that remarkably $d_{0|3m,\, 3; L}$ is in the form of $c^L$ with $c=1/4$. Therefore $\cO_{0|3m}$ for the $SU(3)$ gauge group, like $(\cO_2)^p$ for the $SU(2)$ gauge group (with $d_{p|0,\, 2; L}$ given in the first term of \eqref{eq:coffO2p}), is a ``canonical heavy operator". Hence, the corresponding HHLL correlator can be written in the special form of \eqref{eq:THUV-fin-spe} that can be resumed using the result~\cite{Broadhurst:2010ds} and is valid for a finite 't Hooft-like coupling $\lambda$. The ultimate goal would clearly be to understand the general form of the canonical heavy operators for any $SU(N)$ gauge group. We will leave this investigation for future work.

\subsection{OPE coefficients and relation with integrated HHLL correlators}

The integrated HHLL correlators as given in \eqref{eq:periods} take a remarkably similar form as the structure constants of two heavy operators and the Konishi operator given in \eqref{eq:OPE-coff}, which we quote below,
\ie \label{eq:OPE-coff2}
C_{\cH \cH \cK} =-4d_{\cH, N; 1}\,  a+  2\sum_{L=1}^{\infty}  d_{\cH, N; L+1}  \, (-a)^{L+1}     \binom{2L+2}{L+1} \zeta (2L+1)  \, .
\fe
Comparing the above expression with \eqref{eq:periods}, we see that interestingly $C_{\cH \cH \cK}$ can be essentially identified with the integrated HHLL correlators, with the $L$-loop coefficient $d_{\cH, N; L}$ replaced by the $(L+1)$-loop coefficient (and a differential with respect to $a$)
\ie \label{eq:relations0}
\partial_a C_{\cH \cH \cK} = -4\, d_{\cH, N; 1} + 4 I_2 \left[ \cT_{\cH}(u,v) \right] \big \vert_{d_{\cH, N; L} \rightarrow d_{\cH, N; L+1}} \, .
\fe
For the canonical operators, $d_{\cH, N; L+1}=c \, d_{\cH, N; L}$, so the above relation simplifies, 
\ie  \label{eq:relations}
\partial_a C_{\cH \cH \cK} = -4\, d_{\cH, N; 1} + 4 c \, I_2 \left[ \cT_{\cH}(u,v) \right]  \, , 
\fe
which can be extended beyond the perturbative loop expansion. 
These relations allow us to understand the properties of the structure constant $C_{\cH \cH \cK}$ from the known results of the integrated correlators such as those given in \cite{Paul:2023rka, Brown:2023why}. 

Using these explicit colour factor coefficients, the formula \eqref{eq:OPE-coff2} allows us to immediately obtain the OPE coefficients. Let us begin with the heavy operator $\cH=(\cO_2)^p$, for which the colour factors are given in \eqref{eq:coffO2p0}. Just to illustrate the structures, we will consider some simple examples of $N=2,3$, with the coefficients as given in \eqref{eq:coffO2p}. For $N=2$, we then find that the OPE coefficient takes the following form 
\ie \label{eq:OPE-O2}
C^{SU(2)}_{p|0, \, p|0,\, \cK} =  
- 8a +  \sum_{L=1}^{\infty}  2^{3-L} (-a)^{L+1} \binom{2 L+2}{L+1} \zeta (2 L+1) \, .
\fe
The perturbative series is convergent and can be resummed, which leads to an exact expression
\ie \label{eq:OPE-O22}
C^{SU(2)}_{p|0, \, p|0,\, \cK} =   - 8a +   \int_0^{\infty} {dw \over \sinh^2 \left(w/(2\sqrt{2a}) \right) }   (w-2 J_1(w))\, .
\fe
We may further expand the OPE coefficient in the strong coupling regime, which is given by 
\ie \label{eq:OPE-O233}
C^{SU(2)}_{p|0, \, p|0,\, \cK} ={2 \over 3}+ 4 a \left[-1+ 2\gamma
   + \log \left(\frac{a}{2}\right)\right] + C^{SU(2),\, {\rm NP}}_{p|0, \, p|0,\, \cK} \, ,
\fe
where $C^{SU(2),\, {\rm NP}}_{p|0, \, p|0,\, \cK}$ represents exponentially decayed terms, given by 
\begin{align}
    C^{SU(2),\, {\rm NP}}_{p|0, \, p|0,\, \cK} = -\sum_{n=1}^{\infty} {(2\lambda)^{1/4} \over \pi^{3/2}\, \sqrt{n}}\, e^{- n \sqrt{2 \lambda} } \left[4 \lambda ^{1/2} +\frac{7}{2 \sqrt{2}\, n } +\frac{57}{2^6\,n^2\,
\lambda ^{1/2} }-\frac{195}{2^{9} \sqrt{2}\, n^3\, \lambda }+ \ldots \right] \, ,
\end{align}
where we have expressed the result in terms of $\lambda=2p \, g_{_{\rm YM}}^2$. Using the above expressions and the known results of the integrated HHLL correlator given in \cite{Brown:2023why},  we have verified the relation \eqref{eq:relations} (in this case $c=1/2$). In fact, the exponentially decayed term $C^{SU(2),\, {\rm NP}}_{p|0, \, p|0,\, \cK}$, due to its specific form, can be obtained from the integrated HHLL correlator using the relation \eqref{eq:relations}. 

Similarly, for the other canonical heavy operator $\cO_{0|3m}$ of $SU(3)$ gauge group, with the coefficient given in \eqref{eq:cO3p}, we find  a very similar result for the OPE coefficient
\ie \label{eq:OPE-O33}
C^{SU(3)}_{0|3m, \, 0|3m,\, \cK} =   - 12\,a +   \int_0^{\infty} {6\,dw \over \sinh^2 \left(w/(2\sqrt{a}) \right) }   (w-2 J_1(w))\, .
\fe
From the above exact expression, it is straightforward to obtain the weak coupling and strong coupling expansions. They take very similar expressions to $C^{SU(2)}_{p|0, \, p|0,\, \cK}$, so we will not display all the results here. $C^{SU(3)}_{0|3m, \, 0|3m,\, \cK}$ is related to the corresponding integrated HHLL correlator via the relation \eqref{eq:relations}, with $c=1/4$ using the result of the colour factors given in \eqref{eq:cO3p}.   

We now consider the non-canonical heavy operators. One of the simplest examples is $\cH=\cO_{p|0}=(\cO_2)^p$ (with $p\rightarrow \infty$) for the $SU(3)$ gauge group. Using the colour factor coefficients $d_{p|0, 3; L}$ given in \eqref{eq:coffO2p}, we find the structure constant  can be expressed as follows, 
\ie \label{eq:OPE-O24}
C^{SU(3)}_{p|0, \, p|0,\, \cK} &= -12a  + 6 \int_0^{\infty} { dw \over w^5 \sinh^2\left(w/\sqrt{2a}  \right) } \left[ w^6-8 \left(w^2-12\right) w^2 J_0(w){}^2 \right. \cr
& \left. -4 \left(w^4-20 w^2+96\right) w J_0(w) J_1(w)+2 \left(5
   w^4-64 w^2+192\right) J_1(w){}^2\right]\, .
\fe
The above exact expression is obtained by resumming the perturbative expansion, which is given by
\ie \label{eq:OPE-O231}
C^{SU(3)}_{p|0, \, p|0,\, \cK} =- 12a  + \sum_{L=2}^{\infty}(-2a)^L \frac{48 \left(L^2+L+6\right)  \Gamma
   \left(L+\frac{1}{2}\right) \Gamma
   \left(L+\frac{3}{2}\right) \zeta (2 L-1)}{\pi  \Gamma (L+1) \Gamma
   (L+4)} \, .
\fe
From the exact expression in \eqref{eq:OPE-O24}, we can also obtain the strong coupling expansion, it takes the following form,  
\ie \label{eq:OPE-O232}
C^{SU(3)}_{p|0, \, p|0,\, \cK} =  \frac{a}{4}  \left[-1+48
   \gamma + 24 \log \left(\frac{a}{8}\right) \right]+ C^{SU(3),\, {\rm P}}_{p|0, \, p|0,\, \cK} +  C^{SU(3),\, {\rm NP}}_{p|0, \, p|0,\, \cK} \, ,
\fe
where
\ie \label{eq:SU3-pert}
C^{SU(3),\, {\rm P}}_{p|0, \, p|0,\,\cK} =2 -12 \sum_{n=1}^{\infty} \frac{(n+1) \left(4 n^2+23\right) 
\Gamma \left(n-\frac{5}{2}\right) \Gamma
   \left(n+\frac{1}{2}\right) \Gamma
   \left(n+\frac{3}{2}\right)    \zeta (2 n+3)  }{ (2a)^{n+1/2}\, \pi ^{2n+9/2}\, \Gamma (n) } \, , 
\fe
and 
\ie \label{eq:SU3-NP}
C^{SU(3),\, {\rm NP}}_{p|0, \, p|0,\,\cK} =\pm \, i \sum_{n=1}^{\infty} e^{-n \sqrt{2\lambda}}\, {\sqrt{2\lambda} \over  n\, \pi^2} \left(-48 -{252 \, \sqrt{2} \over  n\,\lambda^{1/2}}-{2499 \over  n^2\, \lambda }-{35739 \, \sqrt{2} \over 4 \,n^3\, \lambda^{3/2} } + \ldots \right)\,. 
\fe
It is easy to see that the power series $C^{SU(3),\, {\rm P}}_{p|0, \, p|0,\,\cK}$ is asymptotic and not Borel summable. The power series is completed by the non-perturbative term displayed in \eqref{eq:SU3-NP} through the resurgence \cite{Dorigoni:2014hea,Aniceto:2018bis}.  The ambiguity lying in the $\pm \, i$ factor in the non-perturbative term $C^{SU(3),\, {\rm NP}}_{p|0, \, p|0,\,\cK}$ is a standard result of the resurgence due to the fact that the asymptotic series $C^{SU(3),\, {\rm P}}_{p|0, \, p|0,\,\cK}$  is not Borel summable, therefore defining the asymptotic series is an ambiguous process. All these properties are very similar to what has been found for the integrated correlators in the literature both in the usual large-$N$ limit \cite{Dorigoni:2021guq,Collier:2022emf, Hatsuda:2022enx, Brown:2024tru, Dorigoni:2024dhy} and in the large-charge limit \cite{Paul:2023rka, Brown:2023why}.  

Similar analysis can be applied to  $\cH=\cO_{p|0}=(\cO_2)^p$ (with $p\rightarrow \infty$) for the $SU(4)$ gauge group. Using the expression $d_{p|0, 4; L}$ given in \eqref{eq:coffO2p}, one can directly write down the perturbative expression.  However, unlike \eqref{eq:SU3-pert}, the strong coupling expansion in the $SU(4)$ case actually truncates (terms beyond $O(a^{-13/2})$ vanish). From the viewpoint of the resurgence, this case is similar to the canonical heavy operators, in that the structure constants all have truncating perturbative expansions (in the strong coupling regime), but still possess exponentially decayed non-perturbative terms. In fact, this is a general phenomenon for the integrated HHLL correlators of heavy operators $\cO^{(i)}_{p|M}$ with $p\rightarrow \infty$ for $SU(2N)$ gauge groups as shown in \cite{Brown:2023why}. As discussed in \cite{Brown:2023why}, this may be understood through ``Cheshire cat" resurgence \cite{Dunne:2016jsr, Kozcaz:2016wvy, Dorigoni:2017smz, Dorigoni:2019kux}. However, as we see in the case of $SU(3)$, the structure constant $C^{SU(3)}_{p|0, \, p|0,\, \cK}$ does contain both an asymptotic series and exponentially decayed terms in the strong-coupling expansion, which are related through the standard resurgence. Once again, this property has also been seen for the integrated HHLL correlators. As shown in \cite{Brown:2023why}, the integrated HHLL correlators of the heavy operators $\cO^{(i)}_{p|M}$ with $p\rightarrow \infty$ of $SU(2N+1)$ gauge groups contain both an asymptotic series and exponentially decayed terms in the strong-coupling expansion, exactly as we saw with $C^{SU(3)}_{p|0, \, p|0,\, \cK}$. Given their relation \eqref{eq:relations0}, we expect all these properties of the integrated HHLL correlators apply to the structure constants, as we have seen in concrete examples.

\section{Conclusion and outlook}
\label{sec:conclusion}

In this paper we have showed that at any loop order the HHLL correlators in $\mathcal{N}=4$ SYM in the large-charge 't Hooft limit can be expressed in terms of a sum of products of two ladder integrals, a class of integrals which is ubiquitous in the literature and known analytically to all loops. Therefore we have determined the HHLL correlators at any loop order, up to overall coefficients which we have further fixed for certain specific heavy operators by computing the relevant colour factors in the large-charge limit. We have compared our expressions (after integrating out the spacetime dependence) with the known results obtained from the integrated correlators using supersymmetric localisation, finding a perfect agreement. From the all-loop expressions of the HHLL correlators, we have also derived the exact expressions for the structure constants of two heavy operators and the Konishi operator, as well as other CFT data. In summary, our results show remarkable simplicity of $\mathcal{N}=4$ SYM in the large-charge limit; for the observables of four-point correlators that are studied in this paper, the large-charge limit appears to be even simpler than the usual large-$N$ 't Hooft limit, and leaves a large number of open questions.

Such a comparison between large charge and large $N$ represents one of the most intriguing points. We have already highlighted the alternating behaviour of the two limits at the level of Feynman diagrams for four-point correlators of BPS operators, also found in $\cN=2$ supersymmetric Yang-Mills theories \cite{Beccaria:2020azj}. The similarity of large-$N$ and large-charge limits have also arisen in the context of integrability \cite{Caetano:2023zwe}, where the spectrum of small deformations around the half-BPS maximal multitrace operators with a large charge follows a magnon-like dispersion relation analogous to the large-$N$ dispersion relation for non-BPS single trace operators \cite{Beisert:2005tm}. Furthermore, from localisation computations for both $\cN=2$ extremal and $\cN=4$ integrated correlators, usually written in terms of an $N\times N$ matrix model \cite{Pestun:2007rz}, several observables can be recast into an emergent  dual matrix model, where the matrix is characterised by the charges \cite{Grassi:2019txd,Beccaria:2020azj,Caetano:2023zwe}. It would clearly be very interesting to further study this ``dual'' behaviour even in less supersymmetric Yang-Mills theories.

For the HHLL correlators, we have seen that the class of canonical heavy operators, for which the colour factor coefficients $d_{\cH, N;L} \sim c^L$ (for some constant $c$ depending on the choices of the heavy operators), the all-loop expressions can be resummed using the results of \cite{Broadhurst:2010ds}. It is quite remarkable that we are able to determine a class of HHLL correlators at any value of the coupling. In the simplest case of the $SU(2)$ gauge group, the only superconformal primary heavy operator is $(\cO_2)^p$, which is a canonical heavy operator for this gauge group. As shown in \cite{Caetano:2023zwe}, the $SU(2)$ HHLL correlators can also be derived from effective field theory by interpreting the HHLL correlator as a LL correlator in the presence of a heavy background created by the heavy operators. We expect the HHLL correlators for the canonical heavy operators to be understood in this way for any fixed rank of the gauge group, hence it would be important to fully classify all the canonical heavy operators. 

For generic heavy operators (or equivalently for generic colour factor coefficients $d_{\cH, N;L}$), resumming our all-loop results is more challenging, and we have not been able to extend our results beyond the perturbative regime yet, even though the resummation for the integrated HHLL correlators as well as the structure constants of two heavy operators and the Konishi operator are relatively straightforward. Understanding the resummation properties of the HHLL correlators for generic heavy operators is clearly an important question, and we plan to continue investigating this question in future.

Recent studies of the integrated correlators show that the full non-perturbative effects of $\mathcal{N}=4$ SYM can be accessed. In the context of the HHLL correlators, it has been shown that the integrated HHLL correlators for finite $\tau$ in the large-charge expansion take universal forms in terms of modular functions \cite{Paul:2023rka, Brown:2023why}. It would therefore be very interesting to extend our analysis beyond the leading large-charge planar limit  and to include non-perturbative effects, and even to study modular properties and exact results of the HHLL correlators. 

A related question is to better understand the relations between the integrated HHLL correlators and the structure constants of two heavy operators and the Konishi operator. In the context of light operators, this same class of integrated constraints are used to find numerical bounds on the structure constants of Konishi and the stress tensor \cite{Chester:2021aun,Chester:2023ehi}. In the large charge 't Hooft limit the relation between integrated correlators and the structure constants is fully analytic, so we would like to explore whether the relations found here could be extended beyond the leading large-charge 't Hooft limit, and even non-perturbatively.
Such non-perturbative relations would be very powerful, given that the integrated HHLL correlators can be computed using supersymmetric localisation and many exact results have already been obtained in the literature. This would lead to exact results of the three-point structure constants involving the non-BPS Konishi operator.  

Heavy operators and their four-point correlators in $\cN=4$ SYM have been considered in the literature for different classes of operators and in different regimes in the parameter space (see e.g. \cite{Coronado:2018cxj, Coronado:2018ypq,Jiang:2019xdz,Aprile:2020luw, Jiang:2023uut, Brown:2024tru}). An interesting example leading to similar all-loop results is the one considered in \cite{Coronado:2018cxj, Coronado:2018ypq}. It was shown that the correlator of four special R-symmetry channels of heavy operators in $\cN=4$ SYM may also be determined to all loops, which is given by the square of the so-called octagon. These computations are performed in the usual large-$N$ 't Hooft planar limit and the dimensions of the heavy operators are much smaller than $N$, even though they are heavy (therefore it is different from the large-charge limit we consider). The octagon is expressed as the determinant of matrices whose entries are given by the ladder integrals. The size of the matrix grows as the number of loops increases, hence the loop contributions contain products of multiple ladder integrals. Moreover, it was already noticed in \cite{Brown:2023zbr} that for some particular choices of the $SO(6)_R$ vectors $Y_i$, only a subset of Feynman integrals contribute to the octagon, and this subset precisely matches with the HHLL correlators of the canonical heavy operator. It would be of great interest to clarify this relation.

Recently the large charge expansion in the presence of heavy operators has been shown to be related to Coulomb branch physics \cite{Ivanovskiy:2024vel,Cuomo:2024fuy}, thanks to the idea that large charge insertions can be associated to giving nontrivial vevs to scalar fields on the Coulomb branch at large $N$. Our results at generic $N$ could be useful to fully probe this idea, and explore different symmetry breaking patterns for the Coulomb branch of $\cN=4$ SYM.

The HHLL correlators have an interesting holographic interpretation in the context of the AdS/CFT correspondence. The heavy operators may non-trivially deform the AdS background, and the HHLL four-point correlators can be viewed as two-point functions (of the two light operators) in the non-trivial background formed by the heavy operators. In a similar context, this viewpoint has been applied to the holographic correlators in $AdS_3 \times S^3$ \cite{Galliani:2017jlg, Bombini:2017sge, Ceplak:2021wzz} and in the case of $AdS_5 \times S^5$ \cite{Buchbinder:2010ek} (once again in these cases the usual large-$N$ limit was taken).  Some further progress has also been made more recently in the case of $AdS_5 \times S^5$, where the geometry of the heavy multitrace operators has been understood \cite{Giusto:2024trt}, which may be directly relevant to our study. It would be very interesting to investigate if our results (especially for the canonical heavy operators, for which we can resum the perturbative expressions) can be extended into the regimes where the holographic descriptions are well understood. 

Given the extreme simplicity of the all-loop expressions of the HHLL correlators (especially for the canonical heavy operators), it also natural to investigate whether there are deep mathematical structures governing the results. One possibility is the positive geometry of 
the so-called correlahedron \cite{Eden:2017fow} (see also \cite{He:2024xed} for recent developments), which generalises the amplituhedron construction of on-shell scattering amplitudes in $\cN=4$ SYM \cite{Arkani-Hamed:2013jha}. The correlahedron was originally applied to the four-point function $\langle \cO_2\cO_2\cO_2\cO_2 \rangle$ in the usual large-$N$ planar limit. As we have widely discussed, the HHLL correlators appear to be even simpler, since they single out only a particular $f$-graph function at a given loop order. It would be fascinating if our all-loop results are governed by such positive geometry structures. 

\vskip 1cm
\noindent {\large {\bf Acknowledgments}}
\vskip 0.2cm
We would like to thank David Broadhurst, Paul Heslop, Yu-tin Huang, Alessandro Georgoudis, Simone Giombi,  Gregory Korchemsky, Chia-kai Kuo, Rodolfo Russo, and Yifan Wang for insightful discussions and correspondences. We would also like to thank Paul Heslop, Rodolfo Russo for the helpful comments on the draft. Finally, F.G. and C.W. thank the Galileo Galilei Institute (GGI) in Florence for hospitality and support during the workshop ``Resurgence and Modularity in QFT and String Theory". 

A.B. is supported by a Royal Society funding RF$\backslash$ERE$\backslash$210067. C.W. is supported by a Royal Society University Research Fellowship,  URF$\backslash$R$\backslash$221015. F.G. and C.W. are supported by a STFC Consolidated Grant, ST$\backslash$T000686$\backslash$1 ``Amplitudes, strings \& duality". No new data were generated or analysed during this study.

\vskip 1cm

\appendix

\section{Group theory conventions for colour computations}\label{app:color}

In this appendix, we summarise our conventions and some basic formulas for the computations of colour factors. We consider an $SU(N)$ gauge group, generated by hermitian generators $T^a$ with $a=1,\dots N^2-1$, satisfying the following Lie algebra:
\begin{equation}
    [T^a,T^b] = \ii f^{abc} T^c~.
\end{equation}
The $\mathfrak su (N)$ generators are normalised in terms of the trace of two generators in the fundamental representation as follows:
\begin{equation}\label{eq:trTT}
    \tr T^a T^b = \frac{1}{2} \delta^{ab}~.
\end{equation}
Fundamental traces of three generators define the structure constants and the totally symmetrised $d$ tensor as:
\begin{equation}\label{eq:df_symbol}
    \tr \left(\{ T^a,T^b\} T^c\right) = \frac{1}{2}d^{abc}~, ~~~~ \tr \left([T^a,T^b] T^c \right) = \frac{\ii}{2}f^{abc}~.
\end{equation}
Higher order fundamental traces can be reduced recursively to lower order traces thanks to fusion/fission identities:
\begin{align}
	\label{eq:fu_fission}
		\begin{split}
		    \tr\big(T_a M_1 T_a M_2\big) & = \frac{1}{2}\,\tr M_1\, \tr M_2
		-\frac{1}{2N}\,\tr\big(M_1 M_2\big)~,\\
		\tr\big(T_a M_1\big) ~\tr\big(T_a M_2\big) & = \frac{1}{2}\,\tr\big(M_1 M_2\big) -\frac{1}{2N}\,\tr M_1 \,\tr M_2~,
		\end{split}
\end{align}
starting with the initial conditions
\begin{equation}
    \tr \mathbf{1} = N~, ~~~~\tr T^a = 0~,~~~~ \tr T^a T^a = \frac{N^2-1}{2}~,
\end{equation}
and where $M_1$ and $M_2$ are arbitrary $N\times N$ matrices.

For example, we compute the colour part of the tree level Wick contraction $X_M\cdot X_M$ of the colour tensors in \eqref{eq:X34tensors}, defining the two-point normalisation of the operators $\cO_{0|M}$: 
\begin{equation}\label{eq:X34computation}
     X_3\cdot X_3 = \frac{3}{8N}(N^2-1)(N^2-4)~,~~~ X_4\cdot X_4 = \frac{(N^2-1)(N^2-4)(N^2-9)}{4(N^2+1)}~.
\end{equation}

\subsection{Adjoint traces}
$\cN=4$ SYM is a special gauge theory whose matter content is purely in the Adjoint representation of the gauge group $SU(N)$. Hence it is convenient to promote $d$ and $f$ symbols as defined in \eqref{eq:df_symbol} to $(N^2-1) \times (N^2-1)$ matrices:
\begin{equation}
    \ii f^{abc} = (T^a)^{bc}~,~~~d^{abc} = (D^a)^{bc}~,
\end{equation}
such that many products of structure constants $f$ arising from Feynman diagram computations can be interpreted as traces over adjoint representation, for example:
\begin{equation}\label{eq:adjTr_example}
    \begin{split}
        -f^{a_1 c d} f^{b_1 d c} &= \Tr_{\rm Adj} (T^{a_1}T^{b_1}) = N \delta^{a_1 b_1}~,~~~~~  f^{a_1 c d} f^{b_1 d e} f^{a_2 e f} f^{b_2 f c} = \Tr_{\text{Adj}} (T^{a_1}T^{b_1} T^{a_2}T^{b_2})~, \\[0.15cm]
        &-f^{a_1 c d} f^{b_1 d e} f^{a_2 e f} f^{b_2 f g} f^{a_3 g h} f^{b_3 h c} = \Tr_{\text{Adj}} (T^{a_1}T^{b_1} T^{a_2}T^{b_2} T^{a_3}T^{b_3})~.
    \end{split}
\end{equation}
Traces over the adjoint representation of multiple generators (such as those that we encounter in the main text) can be evaluated by decomposing them in terms of fundamental traces as
\begin{equation}
    \Tr_{\text{Adj}} (T^{a_1}T^{a_2}\dots T^{a_{2m}}) = \sum _{\ell=0}^{2 m} (-1)^{\ell} \binom{2 m}{\ell} \tr (T^{a_1}\dots T^{a_{\ell}}) \tr (T^{a_{\ell+1}} \dots T^{a_{2m-\ell}})~,
\end{equation}
which can be then reduced to rational functions of $N$ thanks to \eqref{eq:fu_fission}. Some examples for the fully contracted and totally symmetrised adjoint traces follow:
\begin{align}\label{eq:Adj_traces_explicit}
    &\Tr_{\text{Adj}}(T^aT^a) = N(N^2{-}1)~,~~~~~~  \Tr_{\text{Adj}}(T^{(a}T^aT^bT^{b)}) = \frac{5N^2}{2}(N^2{-}1)~, \\
    \Tr_{\text{Adj}}&(T^{(a}T^aT^bT^bT^cT^{c)}) = \frac{35N^3}{4}(N^2{-}1)~,~~~ \Tr_{\text{Adj}}(T^{(a}T^a\dots T^dT^{d)}) = \frac{21N^2}{4}(N^2{-}1)(7N^2{+}2)~.
\end{align}

\section{Two-loop four-point correlators from superspace formalism} \label{app:N=1SUSY}
In \cite{DAlessandro:2005fnh} a complete one- and two-loop analysis for four-point functions of half-BPS operators with some fixed values of charge was carried out using the $\cN=1$ superspace formalism.  In this appendix, we extend that analysis to the class of $\vev{(\cO_2)^p(\cO_2)^p\cO_2\cO_2}$ correlators for generic values of $p$ and $N$. We refer to \cite{DAlessandro:2005fnh} for the diagrammatic details, here we exploit those results and extend them to the general $p$ case. We find that the $\cN=1$ superspace formalism up to two-loop order is perfectly consistent with both our diagrammatic argument coming from chiral Lagrangian insertions (which holds at leading order in the large-$p$ limit) and with the integrated correlator results from supersymmetric localization (both at leading and subleading orders in large-$p$), see appendix \ref{app:integrated}. We concentrate on the simplest case, $\cH=(\cO_2)^p$, but the computation for different classes of heavy operators would be analogous.

The four-point function (with an explicit dependence on $p$) admits the following expansion at weak coupling 
\cite{DAlessandro:2005fnh}:
\begin{align}\label{eq:H_perturbative}
	\cT_{p|0}(u, v) = -\frac{\gym^2}{\pi^2}\,\cT_{p|0; \, 1}(u, v) + \frac{\gym^4}{8\pi^4} \, \cT_{p|0; \, 2}(u, v) + O(\gym^6)~,
\end{align}
where we have used the overall normalisation to be consistent with our convention and the one- and two-loop contributions read 
\begin{align}
	\cT_{p|0;\, 1}(u, v) &=  \widehat D^{(1)}(p,N)\, \frac{1}{u}\Phi^{(1)}(u,v)\,, \label{eq:1_loop}\\
	\cT_{p|0; \, 2}(u, v) &= \widehat D_{1}^{(2)}(p,N)\Bigg[\! \frac{1+v-u}{4u^2}\big(\Phi^{(1)}(u,v)\big)^2+{1 \over u}\Phi^{(2)}\left(\tfrac{1}{u},\tfrac{v}{u} \right) +\frac{1}{u\,v}\Phi^{(2)} \left(\tfrac{u}{v},\tfrac{1}{v} \right)+\frac{1}{u^2}\Phi^{(2)}(u,v)\Bigg]\notag \\ 
    &+\widehat D_{2}^{(2)}(p,N)\frac{1}{2u}\big(\Phi^{(1)}(u,v)\big)^2+ \widehat D_{3}^{(2)}(p,N) {1\over u}\Phi^{(2)}(u,v)~,\label{eq:2_loop}
\end{align}
and once again $\Phi^{(\ell)}(u,v)$ is the $\ell$-loop ladder integral as in \eqref{eq:ladderL}. The coefficients $\widehat D_{i}^{(L)}(p,N)$ with $i=1,2,3$ at $L$-th loop order are the colour factors encoding the $p$ and $N$ dependence of the one- and two-loop correlators, which is normalised by the two-point function $\vev{(\cO_2)^p (\cO_2)^p}$  as in \eqref{eq:tree_contraction}. In this appendix, we concentrate on computing the colour factors $\widehat D_{i}^{(L)}(p,N)$ with an explicit dependence on $p$ and $N$. This allows a direct comparison with both the chiral Lagrangian insertion method as described in section \ref{sec:integrand} and with the supersymmetric localization results.

As in the main text, the four operators are defined by the following tensorial structures \footnote{Since the results from this appendix can be seen as an additional check of our argument in the main text, here we mainly follow the conventions from \cite{DAlessandro:2005fnh}. We will point out the main differences with respect to the main text.}: 
\begin{equation}\label{eq:def_color_app}
\begin{split}
  (\cO_2)^p(x_1) &= X_{\cH}^{a_1 \dots a_{2p}} (\varphi_{a_1} \dots \varphi_{a_{2p}})(x_1)~, ~~~~~~(\cO_2)^p(x_2) = X_{\cH}^{b_1 \dots b_{2p}}(\varphi_{b_1} \dots \varphi_{b_{2p}})(x_2) \\
  \cO_2 (x_3) &= \frac{1}{2} \delta^{c_1c_2} (\varphi_{c_1}\varphi_{c_2})(x_3)~,~~~~~~~~~~ \qquad \cO_2 (x_4) = \frac{1}{2} \delta^{d_1d_2} (\varphi_{d_1}\varphi_{d_2})(x_4)  ~,
  \end{split}
\end{equation}
where the colour tensors for $(\cO_2)^p$ are defined as: 
\begin{equation}\label{eq:def_color_Op_app}
    X_{\cH}^{a_1 \dots a_{2p}} = \frac{1}{2^{p}} \delta^{(a_1a_2} \dots \delta^{a_{2p-1}a_{2p})}~,
\end{equation}
as in the main text, see \eqref{eq:def_color_Op}.

The colour factor of the $\cO_2$'s simply carry a single delta function, so the tensor structure that we need to take into account only comes from the $(\cO_2)^p$'s. In particular, we begin with the tree-level contraction of the two heavy operators as given in \eqref{eq:Gtree}, which we quote here:  
\begin{equation}\label{eq:Gtree_app}
    \cG_{\mathrm{free}}(N,p) =  p!  \frac{\Gamma(\alpha+p)}{\Gamma(\alpha)}~, \qquad \alpha= \frac{N^2-1}{2}~.
\end{equation}
This expression evaluated at $p=2$ and $p=3$ matches the tree level results of \cite{DAlessandro:2005fnh} (see their Table 2 for $p=2$, their Tables 6 and 7 for $p=3$), up to irrelevant overall normalisations.

We then consider the loop corrections.
Up to two-loop order, a limited number of traces can be involved in the insertion of possible subdiagrams coming from the $\cN=1$ superspace formalism, in order to close the four-point function in a connected way. Hence one can write two possible tensor structures, with 4 and 6 open indices, which read: 
\begin{align}
    Y^{a_1a_2b_1b_2} &=  X_\cH^{a_1 a_2 e_1 e_2\dots e_{2p-2}} X_\cH^{b_1 b_2 e_1 e_2\dots e_{2p-2}}\,,\label{eq:Y_tensor}\\
    Z^{a_1a_2a_3b_1b_2b_1} &=  X_\cH^{a_1 a_2 a_3 e_1 e_2\dots e_{2p-3}} X_\cH^{b_1 b_2 b_3 e_1 e_2\dots e_{2p-3}}~.\label{eq:Z_tensor}
\end{align}
These tensors have been explicitly computed in \cite{DAlessandro:2005fnh} for $p=2,3$. Here we extend them to the generic-$p$ case, where we must take into account the $p$-combinatorics analogously to what was done in section \ref{sec4:exampleO2}. In particular, the $Y$ tensor is exactly the same, and the $Z$ tensor is a slight generalisation with an additional pair of open indices:
\begin{align}
    Y^{a_1a_2b_1b_2} &= \frac{1}{2} \binom{p}{1}\,p! \;\frac{\Gamma \left(\alpha +p\right)}{\Gamma \left(\alpha +1\right)}\,\delta^{a_1 a_2} \delta^{b_1 b_2}\notag \\  &~~~+ \binom{p}{2}\,p! \;\frac{\Gamma \left(\alpha +p\right)}{\Gamma \left(\alpha+2\right)}\,\left(\delta^{a_1 a_2} \delta^{b_1 b_2}+\delta^{a_1 b_1} \delta^{a_2 b_2}+\delta^{a_1 b_2} \delta^{b_1 a_2}\right)~,\label{eq:Y_tensor_b}\\
    Z^{a_1a_2a_3b_1b_2b_3} &= \binom{p}{2}p! \, \frac{\Gamma \! \left(\alpha +p\right)}{\Gamma \! \left(\alpha+2\right)} \! \left(\delta^{a_1 a_2} \delta^{a_3 e}+\delta^{a_1 a_3} \delta^{a_2 e}+\delta^{a_1 e} \delta^{a_2a_3}\!\!\phantom{\Big|} \! \right) \! \left(\delta^{b_1 b_2} \delta^{b_3 e}+\delta^{b_1 b_3} \delta^{b_2 e} +\delta^{b_1 e} \delta^{b_2b_3}\! \right) .\label{eq:Z_tensor_b}
\end{align}
These tensors will be contracted with the colour vertices coming from the one- and two-loop subdiagrams. Such subdiagrams are fully listed in \cite{DAlessandro:2005fnh} using $\cN=1$ superspace. We have explicitly checked that the tensors \eqref{eq:Y_tensor_b} and \eqref{eq:Z_tensor_b} contracted with the subdiagrams listed in \cite{DAlessandro:2005fnh} for lower values of $p$  reproduce the colour factors reported in \cite{DAlessandro:2005fnh} (again for $p=2$ see their Table 2, for $p=3$ Tables 6 and 7), up to different overall normalizations, which anyway compensate the different normalisation from the tree-level contraction and thus in the ratio with \eqref{eq:Gtree_app}. 

Here we extend the colour factor computations to the generic $p$ case.
The crucial element of these colour tensors is the binomial combinatorial factor, which counts the number of traces from $(\cO_2)^p$ in which each subdiagram is inserted. This provides the correct scaling in $p$ when taking the large-charge 't Hooft limit.

\subsection{One-loop colour factor}
At the order $O(g_{_{\rm YM}}^2)$ only a single subdiagram contributes, and it has a colour tensor with four open indices: 
\begin{equation}\label{H1_1loop}
    H_1^{a_1b_1a_2b_2} =  f^{a_1b_1e} f^{a_2b_2e}~.
\end{equation}
After contracting this tensor
with the $Y$ tensor \eqref{eq:Y_tensor_b} and
normalizing by \eqref{eq:Gtree_app} we simply get:
\begin{equation}\label{eq:color_1loop_app}
    \widehat D^{(1)}(p,N) = \binom{p}{1} N  = p N~.
\end{equation}
Such colour factor comes with the combinatorial term $\binom{p}{1}$ and one can explicitly check that the contraction of \eqref{H1_1loop} with the second line of \eqref{eq:Y_tensor_b} (proportional to $\binom{p}{2}$) completely vanishes. This is consistent with the idea that only one trace from $(\cO_2)^p$ can be involved in this insertion (in order to have a connected diagram), and also (trivially) respects the double scaling limit at large charge, analogously to the one-loop case in section \ref{sec:integrand}.

\subsection{Two-loop colour factors}
The two-loop computation is more interesting and non-trivial, since we can explicitly see the different scaling in $p$ and $N$, as suggested in the main text.

The colour tensors of the two-loop subdiagrams fall into three categories, with 4 or 6 open indices (see \cite{DAlessandro:2005fnh} for more details):
\begin{align}
    &H_1^{a_1b_1a_2b_2} = f^{a_1 c d} f^{f c d} f^{f b_1e} f^{a_2b_2e}~,  ~~~~ H_2^{a_1b_1a_2b_2} =  f^{a_1 c d} f^{b_1 d e}f^{a_2 e f} f^{b_2 f c}~, \\
    &H_3^{a_1b_1a_2b_2a_3b_3} =  f^{a_1 c d} f^{b_1 e d}f^{a_2 b_3 e} f^{a_3 b_2 c}~.
\end{align}
After contracting $H_1$ and $H_2$ with \eqref{eq:Y_tensor_b}, $H_3$ with \eqref{eq:Z_tensor_b} and taking the ratio with \eqref{eq:Gtree_app}, we get:
\begin{align}\label{eq:2loop_color}
    \widehat D^{(2)}_1(p,N) = \binom{p}{1} N^2~, ~~~ \widehat D^{(2)}_2(p,N) = \binom{p}{1} N^2 + \binom{p}{2}  \frac{10N^2}{N^2+1}~, ~~~ \widehat D^{(2)}_3(p,N) = \binom{p}{2} \frac{10N^2}{N^2+1}~.
\end{align}
We first notice a special behaviour in the scaling of $p$ and $N$ that we mentioned in the main text and that is even more evident from the integrated results of appendix \ref{app:integrated}. The leading power in $N$ appears in $\widehat D^{(2)}_1(p,N)$ and in $\widehat D^{(2)}_2(p,N)$ and comes with the minimal scaling in $p$ (lying in the combinatorial factor $\binom{p}{1}$). On the contrary, the leading power in $p$ (coming from the combinatorial factors $\binom{p}{2}$ in $\widehat D^{(2)}_2(p,N)$ and $\widehat D^{(2)}_3(p,N)$) comes with the minimal power in $N$. Qualitatively, maximising $p$ selects the Feynman diagrams which are ``spread" among the maximal number of pairs of traces, and this process minimises the powers of $N$ in the colour space. This alternating behaviour in the roles of $p$ and $N$ naturally suggests that the large-charge limit selects the maximally non-planar insertions in the colour space.
An analogous diagrammatic analysis using $\cN=1$ superspace formalism was performed in the context of extremal correlators in $\cN=2$ SCFTs, see \cite{Beccaria:2020azj} for more details.

We now use the result \eqref{eq:2_loop} with the colour factors computed in  \eqref{eq:2loop_color} to match the expectations from the chiral Lagrangian insertion method as presented in section \ref{sec:integrand} and \ref{sec:colourfactor}.
Selecting only the leading large-charge colour factors from \eqref{eq:2loop_color} we can drop the first line of the two-loop correlator \eqref{eq:2_loop}, obtaining:
\begin{equation}
    \cT_{p|0; \, 2} (u,v) = \frac{5 N^2}{2^2(N^2+1)} \frac{1}{u}\left[ (\Phi^{(1)}(u,v))^2 + 2 \Phi^{(2)}(u,v) \right]+O(p^{-1})~.
\end{equation}
In the spacetime dependent part, the residual combinations of ladder diagrams perfectly match the expectations from the chiral Lagrangian insertion argument, see \eqref{eq:F2loop}. Furthermore, the resulting colour factor matches the result for $d_{p|0,\,N;2}$ from the main text, see eq. \eqref{eq:colour-factor1}.

As a further consistency check about the colour factor computations, we check \eqref{eq:2_loop} with the colour factors displayed in \eqref{eq:color_1loop_app} and \eqref{eq:2loop_color} against the results for integrated correlators for general $p, N$ reported in appendix \ref{app:integrated}. We combine the colour factors in \eqref{eq:2loop_color} for any charge $p$ with the spacetime integration \eqref{eq:def-integation} over all the ladder diagrams in \eqref{eq:2_loop}. After some simple algebra and using the general property of the integration measure \eqref{eq:def-integation2} when integrating ladder integrals, we perfectly reproduce the one and two-loop localisation results for general $p$ \cite{Paul:2023rka,Brown:2023why, Brown:2023cpz}, reported in the first line of \eqref{eq:C1}. One can also see that when $p=1$, the result reduces to the known answer for $\langle \cO_2 \cO_2 \cO_2 \cO_2 \rangle$ \cite{Eden:2000mv, Bianchi:2000hn}. 

Similar computations at the level of Feynman diagrams for higher perturbative orders are more challenging, and indeed the higher loop results using the $\cN=1$ superspace formalism Feynman diagrams are not available in the literature. However, as we showed in the main text, the chiral Lagrangian insertion method (combined with the supersymmetric non-renormalisation theorems) appears to be much more efficient for computing this class of observables, which allows us to determine the relevant Feynman integrals to all loop orders for the HHLL correlators in the large-charge 't Hooft limit.

\section{Results for integrated correlators with generic $p$ and $N$} \label{app:integrated}
In this appendix we first store some technical details about the organisation of the $\cO^{(i)}_{p|M}$ operators.
We then report some results for the integrated correlators in the presence of operators $\cO^{(i)}_{p|M}$ using the Laplace difference equation \eqref{eq:laplace}, which allows to write explicit result for any $p$ and any $N$. 

\subsection{Organising superconformal primary operators}
\label{sec:OpM}

As discussed in subsection \ref{sec:heavy_choice}, we organise the half-BPS superconformal primary operators as follows 
\begin{equation}
    \label{eq:def-tower2}
\cO^{(i)}_{p|M} = (\cO_2)^p \cO^{(i)}_{0|M}\, .
\end{equation}
Here we review the construction of $\cO^{(i)}_{0|M}$, following \cite{Gerchkovitz:2016gxx,Brown:2023cpz, Brown:2023why}. To define $\cO^{(i)}_{0|M}$ explicitly, we begin by considering a general operator $\cO_{p_1, \cdots, p_n}$ as given in \eqref{eq:sin-tr} with all $p_i >2$ (namely, $\cO_2$ is excluded). These operators are ordered according to their dimensions, which are labelled as $B^{(i)}_M$ with dimension $M$ and $i$ denoting possible degeneracy. The first few examples are $B_0 = \mathbb{I}, B_3 = \cO_3$, $B_4 = \cO_4$, and $B_5 = \cO_5$ (there is no degeneracy for $M<6$, so we have dropped the index $i$). For $M=0$, we define $\mathcal{O}_{0|0} = B_0 = \mathbb{I}$, which is the identity operator, and then $\mathcal{O}^{(i)}_{0|M}$ with $M \geq 3$ are defined recursively as
\ie \label{eq:defOm}
\mathcal{O}^{(i)}_{0|M}  = B_M^{(i)} - \sum_{ M'\leq M} \sum_j \frac{\vev{ B_M^{(i)} \, , \mathcal{O}^{(j)}_{0|M'} }} { \vev{\mathcal{O}^{(j)}_{\delta|M'}\, , \mathcal{O}^{(j)}_{0|M'} }} \,  \mathcal{O}^{(j)}_{\delta|M'} \, , 
\fe
where we have used the definition \eqref{eq:def-tower} for $\mathcal{O}^{(j)}_{\delta|M'}$, and the summation over $j$ is restricted to $j<i$ when $M'=M$. Here $\delta=(M-M')/2$ so that $\mathcal{O}^{(j)}_{\delta|M'}$ has the same dimension as $B^{(i)}_M$. The vevs $\vev{}$ in \eqref{eq:defOm} define the two-point functions that are protected for half-BPS operators in $\cN=4$ SYM theory (i.e they do not receive quantum corrections), and hence can be computed via free theory Wick contractions in terms of rational functions of $N$.

Below we list examples of $\mathcal{O}^{(i)}_{0|M}$ with some small values of $M$. As we mentioned, when $M\leq 5$ there is no degeneracy, so we will simply drop the degeneracy index $i$:
\begin{equation}
    \label{eq:example_SPOs}
\cO_{0|0} =\mathbb{I}\, , ~~~\cO_{0|3} = \cO_3\, , ~~~ \cO_{0|4} = \cO_4- \frac{2N^2-3}{ N(N^2+1)} \cO_{2,2}\, ,
~~~ \cO_{0|5} = \cO_5- \frac{5(N^2-2)}{N(N^2+5)} \cO_{2,3} \, .
\end{equation}
The degeneracy index $i$ becomes relevant starting at $M=6$, and in this case, we have two different operators, 
\begin{equation}
    \begin{split}
        \label{eq:SPO_dim6}
\cO^{(1)}_{0|6} &= \cO_6 - \frac{3N^4-11N^2+80}{N(N^4+15N^2+8)}\cO_{3,3}- \frac{6(N^2-4)(N^2+5)}{N(N^4+15N^2+8)}\cO_{4,2}+ \frac{7(N^2-7)}{N^4+15N^2+8}\cO_{2,2,2} \, ,\\
\cO_{0|6}^{(2)} &= \cO_{3,3} -\frac{9}{N^2+7} \cO_{4,2} +\frac{3 \left(5 N^2+1\right)}{N
   \left(N^2+3\right) \left(N^2+7\right)} \cO_{2,2,2}\, .
    \end{split}
\end{equation}
More generally, the degree of degeneracy increases with $M$.\footnote{As noted in \cite{Brown:2023cpz, Brown:2023why} that $\cO^{(1)}_{0|6}$ and the ones in \eqref{eq:example_SPOs}, as well as their higher-dimensional generalisations, are identical to the so-called single-particle operators \cite{Aprile:2020uxk}, which are defined by requiring the operators to be orthogonal to all the multi-trace operators.} 
This construction was introduced because the organisation of the half-BPS primary operators in terms of $\mathcal{O}^{(i)}_{p|M}$ is the most natural when computing the integrated correlators using supersymmetric localisation \cite{Gerchkovitz:2016gxx}. Since the localisation computation is done on $S^4$, one must perform a Gram-Schmidt orthogonalisation to map the observables to $R^4$. As defined in \eqref{eq:def-tower}, the operators $\cO_{0|M}^{(i)}$ are manifestly orthogonal to each other, so they greatly simplify the Gram-Schmidt procedure. For the purpose of this paper, half-BPS operators organised as $\mathcal{O}^{(i)}_{p|M}$ also provide a natural way of defining the heavy operators that we will consider.

\subsection{General expressions of integrated HHLL correlators}
We report the exact results in $p$ and $N$ for integrated correlators in presence of $\cO_{p|M}^{(i)}$ operators for fixed values of $M$.
The formulas are lengthy for generic values of $p$ and $N$, therefore we will only keep the first few loop orders in perturbation theory. They are enough to highlight their very special behaviour: the leading order in $p$ corresponds to the lowest order in $N$ and vice versa. This is the simplest way to see that the leading charge limit selects the class of Feynman diagrams inserted in a maximally non-planar way in the colour space.
Below we list the results for the integrated HHLL correlators for $\cH=\cO_{p|M}$ with $M=0,3,4,5$ and $\cH=\cO^{(i)}_{p|6}$ for $i=1,2$. All the results are written as functions of $\tau_2=\tfrac{4\pi}{\gym^2}$. 
The result for $\cG_{p|0}$ was originally obtained in \cite{Paul:2023rka}: 
\begin{align}
    {\cG}_{p|0}&= \frac{12 N p \zeta(3)}{\tau_2 \pi}-\left[\frac{p}{N^2+1}  \left(N^2-1 + 2 p \right)  \right] \frac{75 N^2 \zeta(5)}{ \tau_2^{2} \pi^{2}}\label{eq:C1} \\
&-\Bigg[ \frac{p}{(N^2+1)(N^2+3)}\left( 3N^4-3N^2+4+15(N^2-1)p +20 p^2 \right) \Bigg]\frac{245 N^3\zeta(7)}{2 \tau_2^{3} \pi^{3}} +\ldots  \nonumber
\end{align}
When $M=3,4,5$, the results have been obtained in \cite{Brown:2023cpz}, which we quote below, 
\begin{align}
{\cG}_{p|3}&=\left[p+ \frac{3}{2}\right] \frac{12 N \zeta(3)}{\tau_2 \pi}-\left[3+ { \left(5N^2+67\right) \over \left(N^2{+}5\right) \left(N^2{+}7\right)  } \left( \frac{N^2{+}5}{2}\,p+p^2 \right) \right] \frac{30 N^2 \zeta(5)}{\tau_2^{2} \pi^{2}} \\
&+\Bigg[ 1+ {\left(N^2+23\right) \over  \left(N^2 {+} 5\right)  \left(N^2 {+} 7\right) \left(N^2 {+} 9\right) } \left( \frac{ 3 N^4{+}33 N^2{+}94 }{3} \,p+5\left(N^2{+}5\right) p^2+\frac{20}{3} \,p^3 \right) \Bigg]\frac{735 N^3\zeta(7)}{2 \tau_2^{3} \pi^{3}} +\ldots  \nonumber
\end{align}

\begin{align} \nonumber
 {\cG}_{p|4}&=\left[p+2\right] \frac{12 N \zeta(3)}{\tau_2 \pi}-\left[ \frac{4N^4{+}6}{N^2{+}1}+\frac{\left(5 N^6+96 N^4-5 N^2+144\right)}{\left(N^2{+}1\right)\left(N^2{+}7\right)
   \left(N^2{+}9\right)}  \left(  p^2 + {N^2{+}7\over 2} p \right) \right] \frac{30 \zeta(5)}{\tau_2^{2} \pi^{2}}  \\ \nonumber
&+\Bigg[\frac{\left(2 N^4 {-} 3 N^2 {+} 9\right)}{N^2{+}1} + \frac{\left(7 N^6{+}240 N^4{-}367 N^2{+}1080\right)}{\left(N^2 {+} 1\right)\left(N^2 {+} 7\right) \left(N^2 {+} 9\right) \left(N^2 {+} 11\right)} \times \\ 
&  ~~ \left( \frac{3 N^4{+}45 N^2{+}172}{15 }\,p+\left(N^2{+}7\right)p^2 +\frac{4}{3} p^3 \right) \Bigg]  \frac{525 N \zeta(7)}{2\tau_2^{3} \pi^{3}} +\ldots \,.
\end{align}
\begin{align}   \nonumber
 {\cG}_{p|5}&=\left[ p+\frac{5}{2}\right] \frac{12 N \zeta(3)}{\tau_2 \pi} -\left[ \frac{N^6{+}2N^2{+}6}{N^2{+5}}+ \frac{\left(N^6+28 N^4+43 N^2+144\right)}{\left(N^2 {+} 5\right)\left(N^2 {+} 9\right) \left(N^2 {+} 11\right)}  \left( p^2+ \frac{N^2+9 }{2 }p \right) \right] \frac{150 \zeta(5)}{\tau_2^{2} \pi^{2}}    \\ \nonumber
&+\Bigg[ \frac{2\left(5N^4{-}8N^2{+}87\right)}{\left(N^2{+}5\right)}+ \frac{ \left(7 N^6{+}328 N^4{-}515 N^2{+}5220\right)}{\left(N^2{+}5\right) \left(N^2 {+} 9\right)
   \left(N^2 {+} 11\right) \left(N^2 {+} 13\right)} \times \\
&  ~~\left( \frac{\left(3 N^4{+}57 N^2{+}274\right)}{15}\,p + \left(N^2{+}9\right) p^2 +\frac{4}{3}p^3 \right) \Bigg]\frac{525 N \zeta(7)}{\tau_2^{3} \pi^{3}}+\ldots \, .
\end{align}
For $M=6$, the results have not appeared in the literature, but it is straightforward to compute them from localisation, and due to the lengthy expressions we will list only the first two loop results, 
\begin{align} \label{eq:M=61}
 {\cG}^{(1)}_{p|6}&=\left[p+3\right] \frac{12 N \zeta(3)}{\tau_2 \pi} -\left[\frac{4N^6+35 N^4+49 N^2+80}{N^4+15 N^2+8}  \right. \\
 & \left. + \frac{2 \left(5N^8+214N^6+1225N^4+1724
  N^2+2880\right)}{3 \left(N^2+11\right) \left(N^2+13\right)
   \left(N^4+15N^2+8\right)} \left( \frac{N^2+11 }{2 }p+p^2 \right) \right] \frac{45 \zeta(5)}{\tau_2^{2} \pi^{2}}  +\ldots \, . \nonumber
\end{align}
\begin{align} \label{eq:M=62}
 {\cG}^{(2)}_{p|6}&=\left[p+3\right] \frac{12 N \zeta(3)}{\tau_2 \pi} -\left[\frac{4N^8+103N^6+566N^4+1535
  N^2+288}{N^6+22N^4+113N^2+56}  \right. \\
 & \left. + \frac{2 \left(5 N^{10}+249 N^8+4163 N^6+20091
   N^4+54980 N^2+10368\right)}{3 \left(N^2+7\right)
   \left(N^2+11\right) \left(N^2+13\right) \left(N^4+15
   N^2+8\right)} \left( \frac{N^2+11 }{2 }p+p^2 \right) \right] \frac{45 \zeta(5)}{\tau_2^{2} \pi^{2}}  +\ldots \, . \nonumber
\end{align}
As we emphasised in the main text, all these examples indeed show that the terms dominate in the large-charge limit have the lowest power in $N$. Note in these examples, the charges are $2p+M$.

\providecommand{\href}[2]{#2}\begingroup\raggedright\endgroup

%\bibliography{biblio}

 \end{document}